\documentclass{aa}  

\usepackage{graphicx}
\usepackage{lscape}
\usepackage{longtable}
\usepackage{natbib}
\usepackage{color}
\usepackage{array} 
\usepackage{array} 
\usepackage{tikz,array}
\usetikzlibrary{calc}
\usepackage{txfonts}
\usepackage{color}
\newcommand{\troy}{{\fontfamily{pzc}\fontsize{10pt}{13pt}\fontseries{b}\selectfont TROY}}

\newcommand{\mearth}{~M$_{\oplus}$}
\newcommand{\rearth}{~R$_{\oplus}$}

\begin{document}

   \title{The {\fontfamily{pzc}\fontsize{19pt}{13pt}\fontseries{b}\selectfont TROY} project: II. Multi-technique constraints on exotrojans \\ in nine planetary systems 
   \thanks{Based on observations collected at the Centro Astron\'omico Hispano Alem\'an (CAHA) at Calar Alto, operated jointly by the Max-Planck Institut f\"ur Astronomie and the Instituto de Astrof\'isica de Andaluc\'ia (CSIC).}\fnmsep
   \thanks{Partly based on data obtained with the STELLA robotic telescopes in Tenerife, an AIP facility jointly operated by AIP and IAC.}\fnmsep
   \thanks{Based on observations collected at the European Organisation for Astronomical Research in the Southern Hemisphere under ESO programmes 297.C-5051, 098.C-0440(A), and 298.C-5009}
   }
   \subtitle{}

   \author{
   J.~Lillo-Box\inst{\ref{eso}}, 
   A.~Leleu\inst{\ref{cheops}},
   H.~Parviainen\inst{\ref{iac}}\fnmsep\inst{\ref{ull}}\fnmsep\inst{\ref{uoxford}},
   P.~Figueira\inst{\ref{eso}}\fnmsep\inst{\ref{iace}}, 
   M.~Mallonn\inst{\ref{postdam}}, 
   A.C.M.~Correia\inst{\ref{obsparis}}\fnmsep\inst{\ref{coimbra}}\fnmsep\inst{\ref{aveiro}},
   N.C.~Santos\inst{\ref{iace}}\fnmsep\inst{\ref{depfisporto}}, 
   P.~Robutel\inst{\ref{obsparis}},
   M.~Lendl\inst{\ref{austria}},
   H.M.J.~Boffin\inst{\ref{esogarching}},
   J.P.~Faria\inst{\ref{iace}}\fnmsep\inst{\ref{depfisporto}},
   D.~Barrado\inst{\ref{cab}},
   J. Neal\inst{\ref{iace}}\fnmsep\inst{\ref{depfisporto}}
}
\institute{
European Southern Observatory (ESO), Alonso de Cordova 3107, Vitacura Casilla 19001, Santiago 19, Chile \label{eso}\\
              \email{jlillobox@eso.org}
\and Physics Institute, Space Research and Planetary Sciences, Center for Space and Habitability - NCCR PlanetS, University of Bern, Bern, Switzerland \label{cheops}
\and Instituto de Astrof\'sica de Canarias (IAC), E-38200 La Laguna, Tenerife, Spain \label{iac} 
\and Dept. Astrof\'sica, Universidad de La Laguna (ULL), E-38206 La Laguna, Tenerife, Spain \label{ull} 
\and Sub-department of Astrophysics, Department of Physics, University of Oxford, Oxford, OX1 3RH, UK \label{uoxford} 
\and Instituto de Astrof\' isica e Ci\^encias do Espa\c{c}o, Universidade do Porto, CAUP, Rua das Estrelas, PT4150-762 Porto, Portugal \label{iace} 
\and Leibniz-Institut f\"ur Astrophysik Potsdam, An der Sternwarte 16, D-14482 Potsdam, Germany \label{postdam}
\and IMCCE, Observatoire de Paris - PSL Research University, UPMC Univ. Paris 06, Univ. Lille 1, CNRS, 77 Avenue Denfert- Rochereau, 75014 Paris, France \label{obsparis}
\and Department of Physics, University of Coimbra, 3004-516 Coimbra, Portugal \label{coimbra}
\and CIDMA, Departamento de F\'isica, Universidade de Aveiro, Campus de Santiago, 3810-193 Aveiro, Portugal \label{aveiro}
\and Departamento de F\'{i}sica e Astronomia, Faculdade de Ci\^{e}ncias, Universidade do Porto, Portugal   \label{depfisporto} 
\and Space Research Institute, Austrian Academy of Sciences, Schmiedlstr. 6, 8042 Graz, Austria \label{austria}
\and ESO, Karl Schwarzschild Strasse 2, 85748 Garching, Germany \label{esogarching}
\and Depto. de Astrof\'isica, Centro de Astrobiolog\'ia (CSIC-INTA), ESAC campus 28692 Villanueva de la Ca\~nada (Madrid), Spain\label{cab} 
            }

\titlerunning{The TROY project: II. Multi-technique constraints on exotrojans in nine planetary systems}
\authorrunning{Lillo-Box et al.}

   \date{In preparation}

% \abstract{}{}{}{}{} 
% 5 {} token are mandatory
 
  \abstract
   {Co-orbital bodies are the byproduct of planet formation and evolution, as we know from the Solar System. Although planet-size co-orbitals do not exists in our planetary system, dynamical studies show that they can remain stable for long periods of time in the gravitational well of massive planets. Should they exist, their detection is feasible with the current instrumentation.}
   {In this paper, we present new ground-based observations searching for these bodies co-orbiting with nine close-in ($P<5$~days) planets, using different observing techniques. The combination of all of them allows us to restrict the parameter space of any possible trojan in the system. }
   {We use multi-technique observations (radial velocity, precision photometry and transit timing variations), both newly acquired in the context of the \troy\ project and publicly available, to constrain the presence of planet-size trojans in the Lagrangian points of nine known exoplanets.}
   {We find no clear evidence of trojans in these nine systems through any of the techniques used down to the precision of the observations. However, this allows us to {constrain the presence} of any potential trojan in the system, specially in the trojan mass/radius versus libration amplitude plane. {In particular,} we can set upper mass limits in the super-Earth mass regime for {six} of the studied systems.}
   {}


   \maketitle
%

%------------------------------------------------------------------------------------------------
%						INTRODUCTION
%------------------------------------------------------------------------------------------------

\section{Introduction}

The development of state-of-the-art instrumentation and space-based facilities in the past decades boosted the discovery of extrasolar planets up to several thousands of detections\footnote{\url{http://exoplanet.eu}}. This plethora has shown the wide diversity of intrinsic and orbital properties that planets can have. Exoplanet research is currently focused on the deep understanding of the planet composition, structure and atmosphere, in parallel to the search for Earth analogues. From our own system, we know that extrasolar systems should also host other components that also played an important role in moulding the architecture and properties of the planets. In the Solar System, moons and more recently trojans (e.g., Lucy mission, \citealt{levison17}) are targets for \textit{in situ} exploration since they contain clues on the formation and early evolution of our planetary system (e.g., \citealt{morbidelli05,borisov17}). 

Trojan bodies co-rotate with planets in a wide variety of orbital configurations, mainly in tadpole (orbiting the gravity wells of the L$_4$ and L$_5$ Lagrangian points) and horseshoe (librating from L$_4$ to L$_5$ in a horseshoe-like orbit in the co-rotating frame) orbits (see, e.g., \citealt{laughlin02}). The formation of these bodies is still under debate and two main mechanisms are proposed. First, in the \textit{in situ} formation scenario, these bodies would form in the Lagrangian points throughout multiple inelastic collisions between the remnant dust particles of the protoplanetary disk trapped in the gravitationally stable regions. Indeed, swarms of particles trapped in the Lagrangian points are common outcomes of the hydrodynamical simulations used to explain the features observed in transition disks with potentially forming planets (e.g., \citealt{fuente17, laughlin02b}). Similarly to the core accretion process, this trapped material could have grown to larger bodies from kilometer- to moon- or planet-size by particles and pebble collisions. Interestingly, \cite{beauge07} demonstrate that co-orbital bodies up to 6 times the mass of Mars can be formed in these regions under certain conditions and remain in stable orbits. Another proposed mechanism is the capture of these bodies in the Lagrangian points of giant planets during their migration along the disk (see, e.g., \citealt{namouni17}). This would allow larger bodies to be trapped in these regions. Each of these mechanisms, occurring during the first satages of the planet formation and evolution processes, have different imprints in the physical and orbital properties of the co-orbitals. Hence, they contain primordial information about these first stages of the system. Consequently, the study of trojans in the Solar System and the search for trojans in extrasolar systems (and minor bodies in general, \citealt{lillo-box18b}) is key to understand the whole picture of the properties of planetary systems. In particular they can provide significant insight into the nature of formation processes by allowing us to understand if some exoplanetary properties, like the presence of Hot Jupiters, are a product of nature (formation) or nurture (evolution). 

Previous works have developed different techniques to search for these bodies but none has been found yet. Several techniques have been explored, namely radial velocity \citep{ford06,leleu15,leleu17}, transits \citep{janson13,hippke15}, or transit timing variations of the planet \citep{ford07, madhusudhan08, schwarz16}. In \cite{lillo-box18a}, we presented the \troy~project, a multi-technique effort to detect the first bodies co-orbiting to known extrasolar planets. We analyzed a combination of radial velocity and transit data with the methodology described in \cite{leleu17} (a generalization of the \citealt{ford06} technique, expanding the parameter space to horseshoe and large amplitude tadpole orbits). We used archive radial velocity data together with Kepler and ground-based light curves to constrain exotrojan masses in 46 single-planet systems in short-period orbits ($P<5$~days). We found no significant evidence of trojan bodies in any of the studied systems, but we could place upper limits to masses in both Lagrangian points and start populating the parameter space towards a definition of the trojan occurrence rate at different mass regimes. For instance, we could discard Jupiter-mass (or more massive) trojans in 90\% of the systems, which might indicate difficulties in forming, capturing, or keeping trojans in stable orbits during the inward migration of the planet (as theoretically predicted in, e.g., \citealt{rodriguez13}). 

Some of the systems studied in \cite{lillo-box18a} show hints for a mass imbalance between the two Lagrangian points, at the $2\sigma$ level. Despite not being significant detections, they certainly deserve additional follow-up. In this paper we present an extensive amount of dedicated radial velocity and light curve data of nine of these systems to look for the planet-mass co-orbital candidates found in our previous work. In Section~\ref{sec:observations} we describe the multi-technique observations and data reduction. In Section~\ref{sec:methodology} we present the methodology followed to constrain the different regions of the parameter space to restrict the possible presence of trojan planets in these systems; and Section~\ref{sec:results} present the results for each individual system. In Section~\ref{sec:discussion} we discuss these results. {In Appendix~\ref{app:figures} and \ref{app:tables} we show the large figures and long tables (respectively).  }

%------------------------------------------------------------------------------------------------
%						OBSERVATIONS
%------------------------------------------------------------------------------------------------

\section{Observations}
\label{sec:observations}

We have used several datasets from previous publications as well as newly acquired data. In this section we summarize these observations and briefly describe the target sample.

%=======
\subsection{Target sample}

{Nine systems were selected for further follow-up from our radial velocity analysis in \cite{lillo-box18a} because they presented hints for some mass imbalance between the two Lagrangian points. The selected systems are shown in the first column of  Table~\ref{tab:observations}. The planets in all nine systems studied here have been confirmed and characterized through both transits and radial velocity observations. Among them, seven are gas giants with masses above one Jupiter mass, HAT-P-12\,b is a Saturn-mass planet, and GJ\,3470\,b is a 13.7\mearth\ planet. All of them transit main-sequence stars of spectral types FGK (and M in the case of GJ\,3470) with orbital periods below 3.5 days.}

%=======
\subsection{Radial velocity}

We used the archive radial velocity data presented in \cite{lillo-box18a} and newly acquired high-resolution spectra from HARPS \citep{mayor03}, HARPS-N \citep{cosentino12} and CARMENES \citep{quirrenbach14}. In Table~\ref{tab:observations} we present a summary of the new radial velocity observations obtained for each of the nine systems, distinguishing between archive and new observations.  

\begin{table*}[]
\centering
\setlength{\extrarowheight}{5pt}
\caption{Summary of the data used in this paper for the nine targets analyzed, including the number of datapoints for each technique, both new (N$_{\rm new}$) and from archive data (N$_{\rm arch}$).}
\label{tab:observations}
\begin{tabular}{l|llll|ll|ll}
\hline
\hline
 						& \multicolumn{4}{c|}{{Radial Velocity}}                          & \multicolumn{2}{c|}{{Lagrangian transit}}     & \multicolumn{2}{c}{{TTV}}  \\ \hline
System                  & {N$_{\rm arch}$} & {N$_{\rm new}$} & {Inst.}\tablefootmark{a} & {N$_{\rm tot}$} & {N$_{\rm tr}$} & {Inst.}         & {N$_{\rm TTVs}$} & {Ref.} \\
\hline
\object{GJ 3470}                &   110            &      6/10       &  HN/C     &  126            &   1            & CAFOS           &   25               &  [1] \\
\object{HAT-P-12}               &   23             &      6/5        &  HN/C     &  34             &   1            & CAFOS           &   60               &  [2] \\
\object{HAT-P-20}               &   45             &      4/9/15     &  H/HN/C   &  73             &   3            & CAFOS           &   33               &  [3] \\
\object{HAT-P-23}               &   36             &          0      &  -        &  36             &   3            & CAFOS/WiFSIP    &   54               &  [4] \\
\object{HAT-P-36}               &   16             &      7/7        &  HN/C     &  30             &   3            & CAFOS/WiFSIP    &   116              &  [5] \\
\object{WASP-2}                 &   64             &        0        &  -        &  64             &   1            & CAFOS           &   114               & [6] \\
\object{WASP-36}                &   36             &     8/12/24     &  H/HN/C   &  80             &   2            & CAFOS/WiFSIP    &   36               &  [7] \\
\object{WASP-5}                 &   43             &      17         &  H        &  60             &   1            & FORS2           &   33               &  [8] \\
\object{WASP-77}                &   16             &       7         &  C        &  23             &   4            & CAFOS/WiFSIP    &   21               &  [9] \\
\hline
\hline
\end{tabular}
\tablefoot{
\tablefoottext{a}{Instruments are H = HARPS, HN = HARPS-N, C = CARMENES.}
[1] \cite{Awiphan16}, \cite{Bonfils12}, \cite{poddany10}                       
[2] \cite{poddany10}, \cite{mallonn15}, \cite{lee12}, \cite{hartman09}                       
[3] \cite{Bakos10}, \cite{Basturk15}, \cite{poddany10}, \cite{Granata14}, \cite{sun17}                       
[4] \cite{Bakos10}, \cite{poddany10}                      
[5] \cite{Bakos12}, \cite{poddany10}                      
[6] \cite{Charbonneau07}, \cite{Collier-Cameron07}, \cite{Hrudkova09}, \cite{Southworth10}, \cite{poddany10}]                    
[7] \cite{smith12_wasp36}, \cite{poddany10}                      
[8] \cite{Hoyer12}                      
[9] \cite{poddany10}                                        
}

\end{table*}

All three instruments are high-resolution fiber-fed \'echelle spectrographs with resolving powers $R=120\,000$ (HARPS, 3.6\,~m telescope at La Silla Observatory, ESO, Chile), $R=120\,000$ (HARPS-N, TNG telescope at ORM observatory, La Palma, Spain), and $R=81\,200$ (CARMENES, 3.5\,~m telescope at Calar Alto Observatory, Almer\'ia, Spain). These instruments are all located in temperature and pressure controlled vacuum vessels inside isolated chambers to improve their stability. Also, all three instruments are equipped with a second fiber for simultaneous wavelength calibration. In the case of HARPS-N and CARMENES, we fed the second fiber with a Fabry-P\'erot, while HARPS was fed by a simultaneous ThAr lamp. 

In the case of HARPS and HARPS-N, the data were reduced by using the corresponding pipelines available at each observatory. In both cases, the pipeline also determines precise radial velocities using the cross-correlation function method \citep[CCF, ][]{baranne96,pepe02} with a binary template of a similar spectral type as the target star.  In the case of CARMENES, we use the CARACAL pipeline \citep{caballero16} to perform the basic reduction, wavelength calibration and extraction of the spectra. Then, we used our own code \textit{carmeneX}, partly based on the SERVAL pipeline \citep{zechmeister18}\footnote{Publicly available at \url{www.github.com/mzechmeister/serval}.}, to compute the radial velocities by using the CCF technique with a solar-type binary mask with more than 3\,000 spectral lines (adapted form the mask developed for CAFE, see \citealt{lillo-box15b} and \citealt{aceituno13}).

We used HARPS  in a five-night campaign\footnote{Programme ID: ESO 098.C-0440(A), PI: J. Lillo-Box.} on 21-25 January 2017 unfortunately affected by poor weather conditions, and only allowing us to use 1.5 nights due to high humidity, and in a monitoring campaign\footnote{Programme ID: ESO 297.C-5051, PI: J. Lillo-Box.} where we could gather 13 datapoints for WASP-5 over 4 months. In the case of HARPS-N, we had a three-night run\footnote{Programme ID: 18-TNG4/16A, PI: D. Barrado.} on 6-8 February 2017 with  successful observations during the whole campaign. Finally, we used CARMENES in a four-night run\footnote{Programme ID: CAHA F17-3.5-007, PI: Lillo-Box.} on 28-31 January 2017 with a success rate of 60\%, mostly due to thick clouds and high humidity, and in another four-night run on 12-15 December 2017\footnote{Programme ID: CAHA H17-3.5-024, PI: Lillo-Box.} with a 48\% success rate. 

In Tables~\ref{tab:rv1} to \ref{tab:rv7}, we show the derived radial velocities and their uncertainties for the new radial velocity observations.

%=======
\subsection{Differential photometry}

We have used CAFOS \citep{meisenheimer94} at the 2.2\,m telescope in Calar Alto Observatory, WiFSIP at the STELLA1 1.2\,m robotic telescope \citep{strassmeier10} of the Teide Observatory (Tenerife, Spain), and FORS2 \citep{appenzeller92} at the Very Large Telescope (VLT, Paranal Observatory, ESO, Chile)  to photometrically explore the regions around one of the Lagrangian points of the exoplanets studied in this work. The selection of the particular Lagrangian region (either L$_4$ or L$_5$) was done on the basis of  previous radial velocity analysis in \cite{lillo-box18a} for most of the targets. In Table~\ref{tab:photometry}, we present the detailed characteristics of these observations for each of the targets. In all cases, the small eccentricities of the planets allow us to compute the transit times of the Lagrangian points as $T_{0, \rm LP}=T_0\pm1/6\times P$, where $T_0$ is the planet mid-transit time, $P$ is the orbital period, and the plus (minus) sign represents the $L_5$ ($L_4$) location. Based on this, we used the orbital properties from NASA Exoplanet Archive to compute the transit times by using the Transit Ephemeris Service\footnote{\url{https://exoplanetarchive.ipac.caltech.edu/cgi-bin/TransitView/nph-visibletbls?dataset=transits}} at the corresponding location and custom phase. Since the radial velocity analysis is not sensitive to possible librations, the wider range of time we can observe around the Lagrangian point the better we will constrain the parameter space. 

% CAFOS
We used CAFOS (Calar Alto Faint Object Spectrograph) in imaging mode\footnote{Programme IDs: H17-2.2-018 and F18-2.2-004, PI: J. Lillo-Box} to obtain high-cadence relative photometry of eight of the targets. The field of view is reduced down to $7\times7$ arcmin (i.e., around $800\times$800 pixels for a plate scale of 0.53 arcsec/pix) to decrease the readout time and so increase the observational cadence. The exposure times for each target depend on their magnitudes and atmospheric conditions, ranging between 10-30\,s in the SDSS\,i filter. The total time span for the observations is typically four hours centered on the Lagrangian point mid passage. We slightly defocused the telescope in order to increase the signal-to-noise ratio of the source, reaching an average of 50\,000 counts per pixel. In the case of HAT-P-20, we defocused the telescope (both in CAFOS and WiFSIP observations) just up to 2 arcsec in order to avoid contamination from a close companion at 6.2~arcsec \citep{Bakos10}. On the contrary, for WASP-77 we strongly defocused the telescope (up to donut-shaped point spread function) in order to merge the light from a close companion at 3~arcsec  \citep{maxted12} and so avoiding flux fluctuations inside the aperture due to possible seeing variations along the observation. 

The CAFOS photometry was reduced with a custom aperture photometry pipeline\footnote{The pipeline is written in Python and built on top of Astropy \citep{Astropy2013}, Photutils \citep{Photutils2017}, xarray \citep{xarray2017}, StatsModels \citep{statsmodels2010}, NumPy \citep{NumPy2011}, and SciPy}. The pipeline applies first the basic data reduction steps of bias subtraction and flat field division, after which it calculates the aperture photometry for the target star and a set of potential comparison stars using five aperture sizes. The final relative light curve is generated by finding a combination of comparison stars and aperture sizes that minimizes the relative light curve point-to-point scatter. A slightly modified version of the pipeline is used when the target star is accompanied by a bright nearby star (close enough for the PSFs to blend). In this case the photometry is calculated for a set of circular apertures and a set of elliptical apertures that contain the target and the contaminant. We calculate the relative target-contaminant flux based on the frames with best seeing, and remove the fractional contaminant flux from the combined flux (and use the circular aperture photometry centred on the target and contaminant to test that the contaminant is stable).

% WiFSIP
We also used WiFSIP (Wide Field STELLA Imaging Photometer) to explore the Lagrangian points of 3 exoplanets\footnote{Programme IDs: 52-Stella9/17B and 49-Stella4-18A, PI: D. Barrado}. The WiFSIP field of view is 22$\times$22 arcmin and the plate scale corresponds to 0.322 arcsec/pix. In this case, given the robotic nature of the telescope and in order to increase the execution probability of the program, we asked observations of three times three hours around the Lagrangian point mid transit time, starting at random phases between two hours and half an hour before the start of the expected transit. The typical exposure times range between 10 and 80\,s in the SDSSr band. In the same way as explained before, we also defocused the telescope for some targets (see Table~\ref{tab:photometry}) in order to reach the maximum precision possible.  The data have been reduced with the software tools used previously for high-precision exoplanet transit photometry with STELLA/WiFSIP (e.g., \citealt{mallonn16}). Bias and flat field correction was done with the standard STELLA pipeline. For aperture photometry we employed the publicly available tool SExtractor \citep{bertin96}. Our software tools choose the selection of reference stars that minimize the standard deviation in the light curves, and by the same criterion also choose the best aperture size. 

% FORS2
Finally, we also used FORS2 on 2016-10-26 in imaging mode with the z\_SPECIAL filter to explore the L$_5$ region of WASP-5b during 3.68 hours\footnote{Programme ID: ESO 298.C-5009, PI: J. Lillo-Box}. An individual exposure of 12\,s was set, providing photon-noise limited images without need to defocus the telescope\footnote{Note that at the VLT, this can be done by doing bad-AO. However, this can introduce additional systematics in the relative photometry}.
The FORS2 data was reduced and extracted following the same principles described above for the CAFOS data, using a similar (adapted) pipeline. 

In all cases, regardless of the instrument used, the extraction procedure gathers a set of covariates (airmass, median sky level, full width at half maximum, and centroid shift of the target) that are used next to detrend the light curve (see \S~\ref{sec:LCfitting}).

%=======
\subsection{Transit timing variations}

We have collected the transit times of all targets studied in this paper through the Exoplanet Transit Database (ETD, \citealt{poddany10}). We only used transit epochs with data quality better or equal than 3, as flagged at the ETD. In the last two columns of Table~\ref{tab:observations}, we show the number of epochs for each target together with the corresponding references.

%------------------------------------------------------------------------------------------------
%						METHODOLOGY
%------------------------------------------------------------------------------------------------

\section{Methodology: Constraining the parameter space}
\label{sec:methodology}

We use all the above described data to constrain the parameter space of a potential co-orbital planet. While the combination of the radial velocity and planet transit data (\S~\ref{sec:rvtr}) can constrain the mass of the trojan (see \citealt{lillo-box18a,leleu17}), the dedicated multi-epoch photometric exploration of the Lagrangian points (\S~\ref{sec:tr}) and the measured transit timing variations of the planet (\S~\ref{sec:ttv}) can constrain other regions of the parameter space characterizing the co-orbital and its orbit. Here we explain each of these approaches. 

%=======
\subsection{Radial velocity + Planet transit}
\label{sec:rvtr}
In all cases where new RV data from our own monitoring campaigns or from public data are available, we apply the same technique as in \cite{lillo-box18a} based on the theoretical approach described in \cite{leleu17} to obtain an upper limit to the mass of the trojan and to decide on the Lagrangian point to be explored photometrically. This analysis is based on the determination of the $\alpha$ parameter which corresponds to $m_t/m_p\sin{\zeta}$ to first order in eccentricity, where $m_t$ is the mass of the trojan, $m_p$ is the mass of the planet, and $\zeta$ is the resonant angle representing the difference between the mean longitudes of the trojan and the planet. If $\alpha$ is significantly different from 0, the system is hence a strong candidate to harbor co-orbitals.  Consequently, for a known planetary mass and an assumed resonant angle, an upper limit on $\alpha$ can be directly translated into an upper limit for the mass of trojans at the Lagrangian points. We will then obtain $m_{\rm t,RV}^{\rm max}$ as the maximum mass (95\% confidence level) that a potential trojan could have on the average location of the co-orbital along the observation timespan. 

\begin{figure}
\centering
\includegraphics[width=0.5\textwidth]{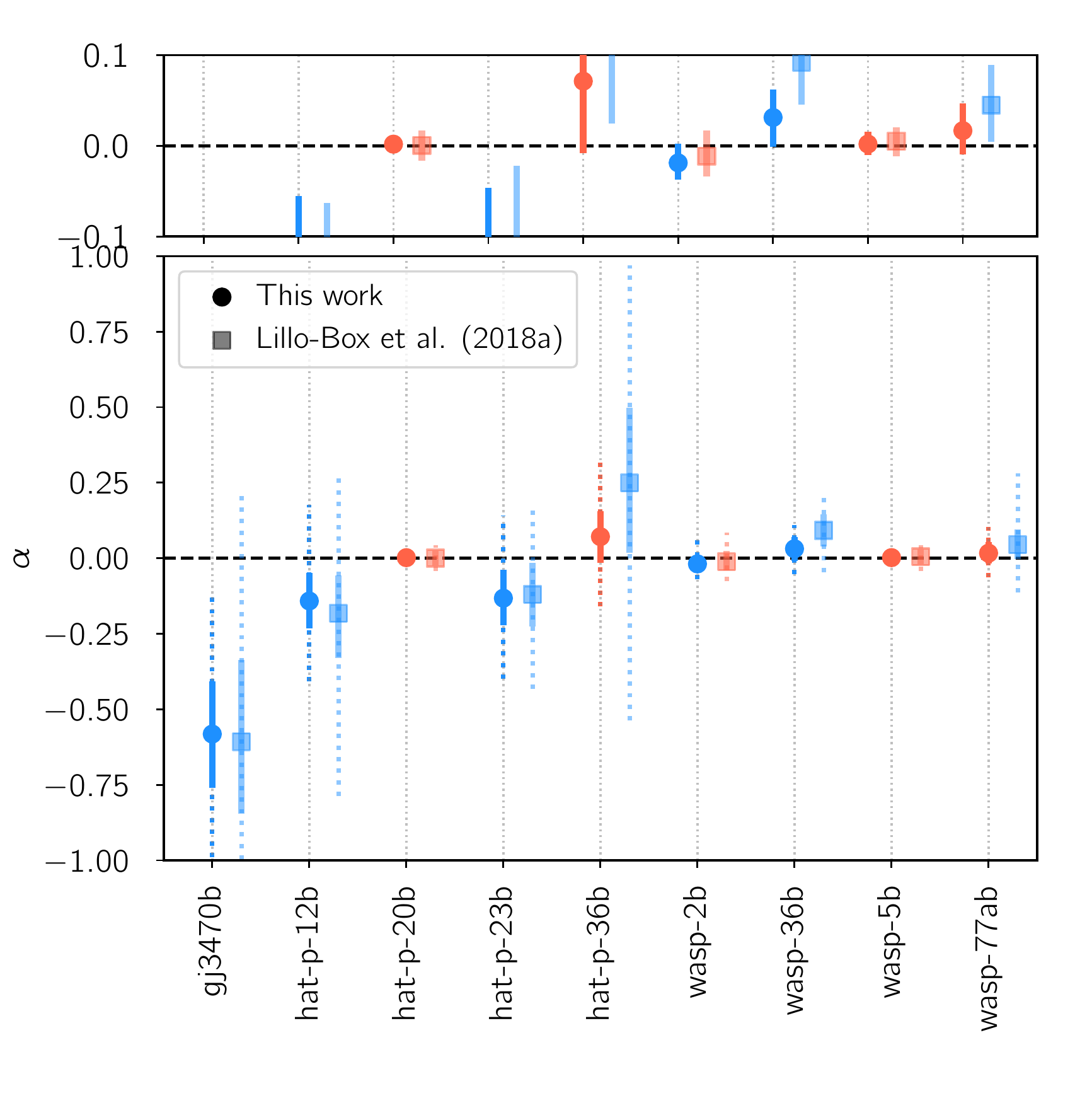}
\caption{Comparison between the $\alpha$ values obtained with the new data (circles) and the values published in \cite{lillo-box18a}. The colored dotted lines represent the $3\sigma$ uncertainties. Blue colors indicate $\alpha\pm\sigma>0$ (i.e., candidate) and red colors indicate $\alpha-\sigma< 0 < \alpha+\sigma$ (i.e., no detection). The top panel is a close view of the region around $\alpha=0$}
\label{fig:uppermasslim}
\end{figure}

\begin{table*}[]
\setlength{\extrarowheight}{7pt}
\caption{Derived parameters for the nine planetary systems analyzed. \label{tab:RVresults}}
\small
\begin{center}
\begin{tabular}{llllllllll}
\hline
\hline

Object & $V_{\rm sys}$ & $P$    & $T_0$          & $K$    & $\alpha$ & $c$ & $d$ \\ 
       & (km/s)       & (days) & (BJD-2450000) & $(m/s)$ &          &     &  \\ \hline

gj3470b              & 	$26.51640^{+0.00053}_{-0.00051}$ & 	$3.33666^{+0.00011}_{-0.00010}$   & 	$6090.47687^{+0.00047}_{-0.00044}$ & 	$8.28^{+0.59}_{-0.60}$ & 	$-0.58^{+0.16}_{-0.17}$ & 	$-0.174^{+0.071}_{-0.074}$ & 	$0.176^{+0.073}_{-0.072}$ 	\\
hat-p-12b            & 	$0.0002^{+0.0035}_{-0.0034}$ & 	$3.2130599^{+0.0000063}_{-0.0000065}$ & 	$4419.19555^{+0.00061}_{-0.00057}$ & 	$38.6^{+1.9}_{-2.0}$ & 	$-0.142^{+0.082}_{-0.080}$ & 	$-0.062^{+0.039}_{-0.040}$ & 	$-0.004^{+0.039}_{-0.036}$ 	\\
hat-p-20b            & 	$0.0874^{+0.0052}_{-0.0057}$ & 	$2.8753186^{+0.0000053}_{-0.0000050}$ & 	$6708.35627^{+0.00026}_{-0.00027}$ & 	$1252.0^{+3.9}_{-4.2}$ & 	$0.0018^{+0.0047}_{-0.0048}$ & 	$-0.0140^{+0.0017}_{-0.0017}$ & 	$0.0092^{+0.0035}_{-0.0034}$ 	\\
hat-p-23b            & 	$-0.007^{+0.017}_{-0.019}$ & 	$1.212862^{+0.000042}_{-0.000039}$ 	  & 	$4852.26469^{+0.00055}_{-0.00054}$ & 	$359^{+15}_{-15}$ & 	$-0.132^{+0.082}_{-0.079}$ & 	$-0.001^{+0.018}_{-0.017}$ & 	$-0.098^{+0.034}_{-0.039}$ 	\\
hat-p-36b            & 	$0.007^{+0.032}_{-0.028}$ & 	$1.3273435^{+0.0000074}_{-0.0000071}$ & 	$5565.18152^{+0.00053}_{-0.00057}$ & 	$327^{+8.1}_{-10}$ & 	$0.071^{+0.072}_{-0.075}$ & 	$0.020^{+0.024}_{-0.020}$ & 	$-0.037^{+0.022}_{-0.023}$ 	\\
wasp-2b              & 	$-27.864^{+0.011}_{-0.011}$ & 	$2.1522215^{+0.0000012}_{-0.0000013}$ & 	$3991.51537^{+0.00048}_{-0.00053}$ & 	$152.9^{+3.7}_{-3.5}$ & 	$-0.019^{+0.018}_{-0.015}$ & 	$-0.0016^{+0.0036}_{-0.0035}$ & 	$0.027^{+0.024}_{-0.027}$ 	\\
wasp-36b             & 	$-13.211^{+0.011}_{-0.010}$ & 	$1.5373595^{+0.0000044}_{-0.0000043}$ & 	$5569.83733^{+0.00028}_{-0.00027}$ & 	$375.8^{+2.8}_{-2.8}$ & 	$0.031^{+0.027}_{-0.029}$ & 	$-0.0029^{+0.0094}_{-0.0088}$ & 	$0.0290^{+0.0090}_{-0.0090}$ 	\\
wasp-5b              & 	$20.0174^{+0.0075}_{-0.0079}$ & $1.6284261^{+0.0000020}_{-0.0000021}$ & 	$4375.62495^{+0.00073}_{-0.00075}$ & 	$267.4^{+1.3}_{-1.4}$ & 	$0.0019^{+0.0096}_{-0.0088}$ & 	$-0.0036^{+0.0026}_{-0.0027}$ & 	$-0.0020^{+0.0055}_{-0.0062}$ 	\\
wasp-77ab            & 	$1.6604^{+0.0086}_{-0.0081}$ & 	$1.3600332^{+0.0000048}_{-0.0000049}$ & 	$5870.44975^{+0.00043}_{-0.00043}$ & 	$325.3^{+6.7}_{-6.4}$ & 	$0.017^{+0.027}_{-0.023}$ & 	$0.00165^{+0.00099}_{-0.00095}$ & 	$0.047^{+0.031}_{-0.030}$ 	\\

\hline
\hline
\end{tabular}
%\\tablefoot{}
\end{center}
\end{table*}%

We followed the same procedure as in our previous work, modeling the radial velocity with the equation described in \cite{leleu17} and including a Gaussian Process to account for the presence of active regions in the stellar surface that can lead to correlated noise in the data. To this end we used a quasi-periodic kernel (described in \citealt{faria16}). We set Gaussian priors for the orbital period, time of mid-transit of the planet, and $c\approx e\cos{\omega}$ when this parameter can be constrained from the detection of the secondary eclipse. These priors are centered on the values from the literature and we set a width equal to three times the estimated uncertainties provided in the literature. Log-uniform priors are set to the GP hyperparameters and uniform priors are used for the rest of the parameters, including the systemic velocity ($V_{\rm sys}$), the radial velocity semi-amplitude ($K$), $\alpha$, and $d\approx e\sin{\omega}$. 
The results of the RV fitting are shown in Fig.~\ref{fig:RV1} and the median and 68.7\% confidence values of the main fitted parameters are shown in Table.~\ref{tab:RVresults}, and will be discussed individually in \S~\ref{sec:results}. Also, in  Fig.~\ref{fig:uppermasslim}, we compare the $\alpha$ values for these nine systems between the current work and our first analysis in \cite{lillo-box18a}. The main conclusion is that we can reduce the uncertainty in systems where a significant amount of new data points were obtained compared to the previous data. This allows to set lower upper mass limits. For instance, in the case of HAT-P-36 and WASP-77A, the $\alpha$ parameter is now compatible with null within $1\sigma$.

%=======
\subsection{Photometric exploration of the Lagrangian points}
\label{sec:tr}
The photometric exploration of the Lagrangian points can constrain: i) the trojan size, if we assume small librations around the Lagrangian point; and ii) the orbital parameters of the co-orbital body, specially the libration amplitude assuming coplanarity. In the following section we describe the different analysis of the precise photometry obtained during our ground-based campaigns. Each subsection focuses on particular aspects, namely individual transit searches on each epoch assuming coplanarity and hence similar transit duration as the planet (\S~\ref{sec:LCfitting}), the implications of non detection on the trojan libration amplitude (\S~\ref{sec:PSconstrain}), and transit search on the combined light curve assuming no libration (\S~\ref{sec:LCcombine}).

%+++++++
\subsubsection{Individual epochs (coplanar case): search for transits}
\label{sec:LCfitting}
We used the different epochs for a single system independently to look for transits in the observed timespans. The search is done through transit fitting of the light curve, which also allows us to estimate the maximum object size in case no transit is detected. This fit is done by using the \textit{batman}\footnote{\url{http://astro.uchicago.edu/~kreidberg/batman/}} python module \citep{kreidberg16}, where we assume a Gaussian prior on the orbital period ($P$), semi-major-axis to stellar radius ($a/R_{\star}$) and inclination ($i$) parameters. This implies that we are looking for coplanar (or very close to coplanar) trojans, thus having transit durations similar to that of the planet. But, we note that not completely fixing these values still allows for some freedom in the trojan transit duration. We then leave the mid-transit time of the trojan ($T_{0,t}$), the trojan-to-stellar radius ($R_t/R_{\star}$), the zero level (out-of-transit flux level, $F_0$), and a photometric jitter to account for white noise ($\sigma_{\rm jit}$), as free parameters with uniform priors. The specific priors and ranges are shown in Table~\ref{tab:LCpriors}. We also include a baseline model simultaneously to account for possible correlations with airmass ($\chi$), seeing ($s$, full width at half maximum of the target point spread function), time ($t$), position of the target on the detector ($xy$) and background ($b$). In a first stage, we use linear dependencies for these parameters, $\mathcal{B}(t, \chi, s, x, y, b)$. {Mathematically, the baseline model can be represented as $\mathcal{B} = \sum_i a_i p_i $, where $a_i$ are the coefficients to be determined and $p_i$ are each of the parameters described above. The baseline and transit models are fitted simultaneously, assuming that each datapoint is a realization of $d(t) = \mathcal{B}(p) + \mathcal{T}(t) + \sigma_{\rm jit}$.}

\begin{figure*}
\centering
\includegraphics[width=1.\textwidth]{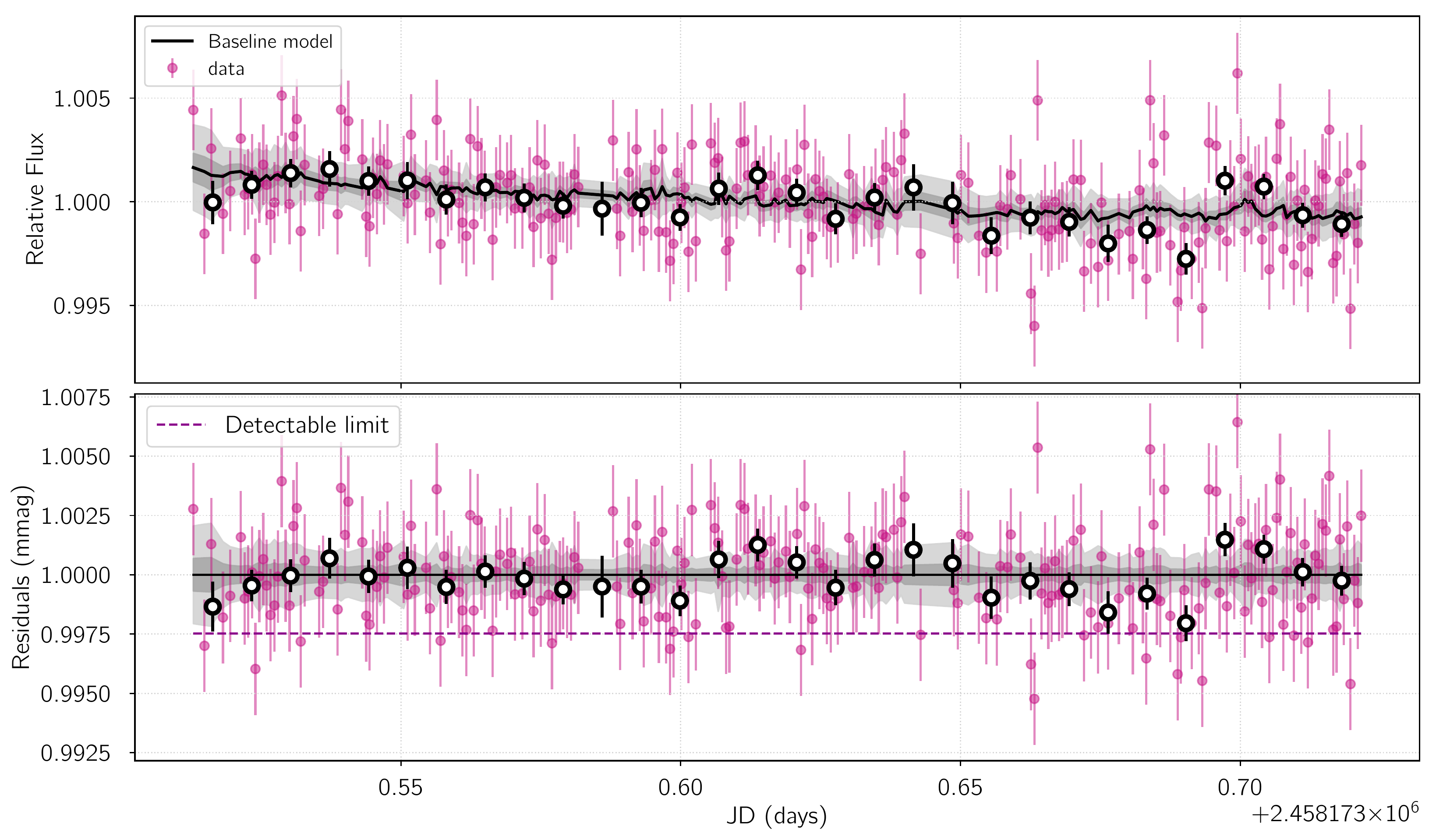}
\caption{Example of the light curve analysis {of the L$_5$ region of HAT-P-36\,b (observed with CAFOS on 2018-03-22) including the linear baseline model (see Sect.~\ref{sec:LCfitting})}. The upper panel shows the raw light curve (violet symbols) together with the 15-min binned data (black open circles, for visualization purposes) and the baseline model fitted. In this case, we assume to trojan (null hypothesis). The bottom panel shows the baseline-corrected light curve with the detectable limit calculated as the 95\% interval for the $R_t/R_{\star}$ parameter in the trojan model. In both panels, the shaded regions correspond to the 68.7\% and 99.7\% confidence intervals.}
\label{fig:LCshow}
\end{figure*}

\begin{table}
\small
\setlength{\extrarowheight}{5pt}
\caption{Priors for the light curve analysis and transit fitting. \label{tab:LCpriors}}
\begin{tabular}{l l l}
%\hline
 Parameter  & Prior & Units\\
\hline \hline

$^{\dagger}T_{0,t}$		& $\mathcal{U}\Big({\rm min}(t)-\frac{T_{\rm dur}}{4},{\rm max}(t)+\frac{T_{\rm dur}}{4}\Big)$ & days \\
$^{\dagger}P_{\rm orb}$			& $\mathcal{G}(\mu,\sigma)$ & days\\
$^{\dagger}a/R_{\star}$			& $\mathcal{G}(\mu,\sigma)$ & -\\
$^{\dagger}R_t/R_{\star}$	& $\mathcal{U}(0,R_p)$   & - \\
$^{\dagger}i$						& $\mathcal{G}(\mu,3\sigma)$ & degrees\\
$F_0$					& $\mathcal{U}(-0.3, 0.3)$  & - \\
$\sigma_{\rm jit}$		& $\mathcal{U}(0, 5)$ &  mmag\\

\hline
\end{tabular}
\tablefoot{
\tablefoottext{$^{\dagger}$}{These parameters are only included in the models with trojan (i.e., not in the null-hypothesis tests).
$\mathcal{U}(a,b)$ stands for uniform priors between $a$ and $b$, while $\mathcal{G}(\mu,\sigma)$ represents a Gaussian prior with mean $\mu$ and standard deviation $\sigma$.}
}
\end{table}

We use \textit{emcee}\footnote{See \url{http://dan.iel.fm/emcee} for further documentation.} \citep{foreman-mackey13} with 50 walkers and 50\,000 steps per walker to explore the posterior distribution of the transit parameters and baseline coefficients. We use the last half of each chain to compute the final posterior distributions and parameter-parameter dependencies. We test a model with trojan (trojan-hypothesis) and without trojan (null-hypothesis). In the latter case the number of free parameters reduces to $F_0$, $\sigma_{\rm jit}$ and the baseline parameters. The bayesian evidence ($E$) of  these two models are estimated using the \textit{perrakis} code\footnote{\url{https://github.com/exord/bayev}. {This code is a \textit{python} implementation by R. D\'iaz of the formalism explained in \cite{perrakis14}}.}. {Based on this Bayesian evidence, we can estimate the Bayes factor ($BF$) between the two models as the ratio between the evidence of the trojan hypothesis ($E_t$) and the null hypothesis ($E_0$), so that $BF=\ln{E_t}-\ln{E_0}$. A positive BF would favor the trojan hypothesis against the null hypothesis, with $BF>6$ considered as a strong evidence.}

Additionally, the posterior distribution of the trojan-to-star radius ratio ($R_t/R_{\star}$) is checked. If this parameter is significantly different from zero (i.e., the median value is larger than the 95\% confidence interval)  and the Bayes factor favors the trojan-hypothesis against the null-hypothesis, then we consider that we have a candidate trojan transit. In this case, we proceed by testing models (both with and without trojan) with quadratic dependencies for the baseline parameters, $p(t^2)+ p(\chi^2) +p (s^2) + p(xy^2)+ p(b^2)$. More specifically, we first run all parameters with quadratic dependencies, then we check the relevant parameters producing significant non-zero coefficients in the fit and we finally run a last model only including those relevant parameters. We then re-estimate the Bayesian evidence and look for the model with the largest evidence. If that model corresponds to a trojan-hypothesis model and keeps the significance of the $R_t/R_{\star}$ parameter, then we consider the transit as a strong trojan transit candidate. 

In the cases where no significant transit was detected, we can determine a maximum radius of the trojan per each epoch ($R_{\rm{t,TR}_i}^{\rm max}$). This is determined as the 95\% confidence interval of the posterior distribution of the $R_t/R_{\star}$ parameter. We also establish a time interval  (which can be translated to a phase interval) where we can assure that there is no transit of a body larger than $R_{\rm{t,TR}_i}^{\rm max}$. This time interval is considered as the entire observing window, which implies that we only consider a transit detection if we have more than half of the transit. We note that this is a conservative assumption for constraining the parameter space. An example of this approach is shown in Fig.~\ref{fig:LCshow} for the transit of the L$_5$ Lagrangian point of HAT-P-36\,b observed with CAFOS. The figure shows the modeling with the null hypothesis and the detectable limit by this observations in the bottom panel (dashed horizontal line). In Fig.~\ref{fig:LCfitting1}-\ref{fig:LCfitting2} we show for all nine systems the results of the null-hypothesis models once the fitted baseline contribution has been removed.

%+++++++
\subsubsection{Individual epochs (coplanar case): parameter space constraint}
\label{sec:PSconstrain}

We can then use the time range of non-detected transit for each epoch to constrain the parameter space of the trojan orbit.  We can do this by determining the parameter range where the trojan would not transit during these time ranges (assuming the trojan is larger than the above estimated maximum radius). To this end, we apply a Monte-Carlo Markov-Chain (MCMC) using a modified likelihood function ($\mathcal{L}$) which increases towards the edges of the observed time ranges and is flat and maximum outside of them:
~
\begin{equation}
\mathcal{L}_i = -0.5 \ln{2\pi} + \ln(\sigma^2)+\frac{r^2}{\sigma^2}
\end{equation}
~
\noindent where
~
\begin{equation}
  r =
  \begin{cases}
    \phi_{\rm mid}-x 				& \text{if $ (\phi_{\rm in}<x< \phi_{\rm out})$ \& $ (R_t>R_t^{\rm max,i})$}\\
    \phi_{\rm mid}-\phi_{\rm in} \equiv \mathcal{L}_i^{\rm max} 	& \text{if $ (\phi_{\rm in}<x< \phi_{\rm out})$ \& $ (R_t<R_t^{\rm max,i})$}\\
    \phi_{\rm mid}-\phi_{\rm in} \equiv \mathcal{L}_i^{\rm max}   		& \text{if $ x<\phi_{\rm in}$ or $ x> \phi_{\rm out}$}
  \end{cases}
\end{equation}
~
\noindent where $\phi_{\rm in}$ and $\phi_{\rm out}$ are the orbital phases corresponding to the earliest and latest edge of the time range (i.e., $t_{\rm in}$ and $t_{\rm out}$). $\phi_{\rm mid}$ represents the mid time of this time range and $\sigma$ is one fourth of this time span. Consequently, the likelihood is minimum at the mid time of the observations and is maximum and constant outside of this range. The total likelihood ($\mathcal{L}$) is the sum of the likelihoods ($\sum_i \mathcal{L}_i$) calculated as above for each transit epoch observed. 

\begin{figure*}
\centering
\includegraphics[width=1.\textwidth]{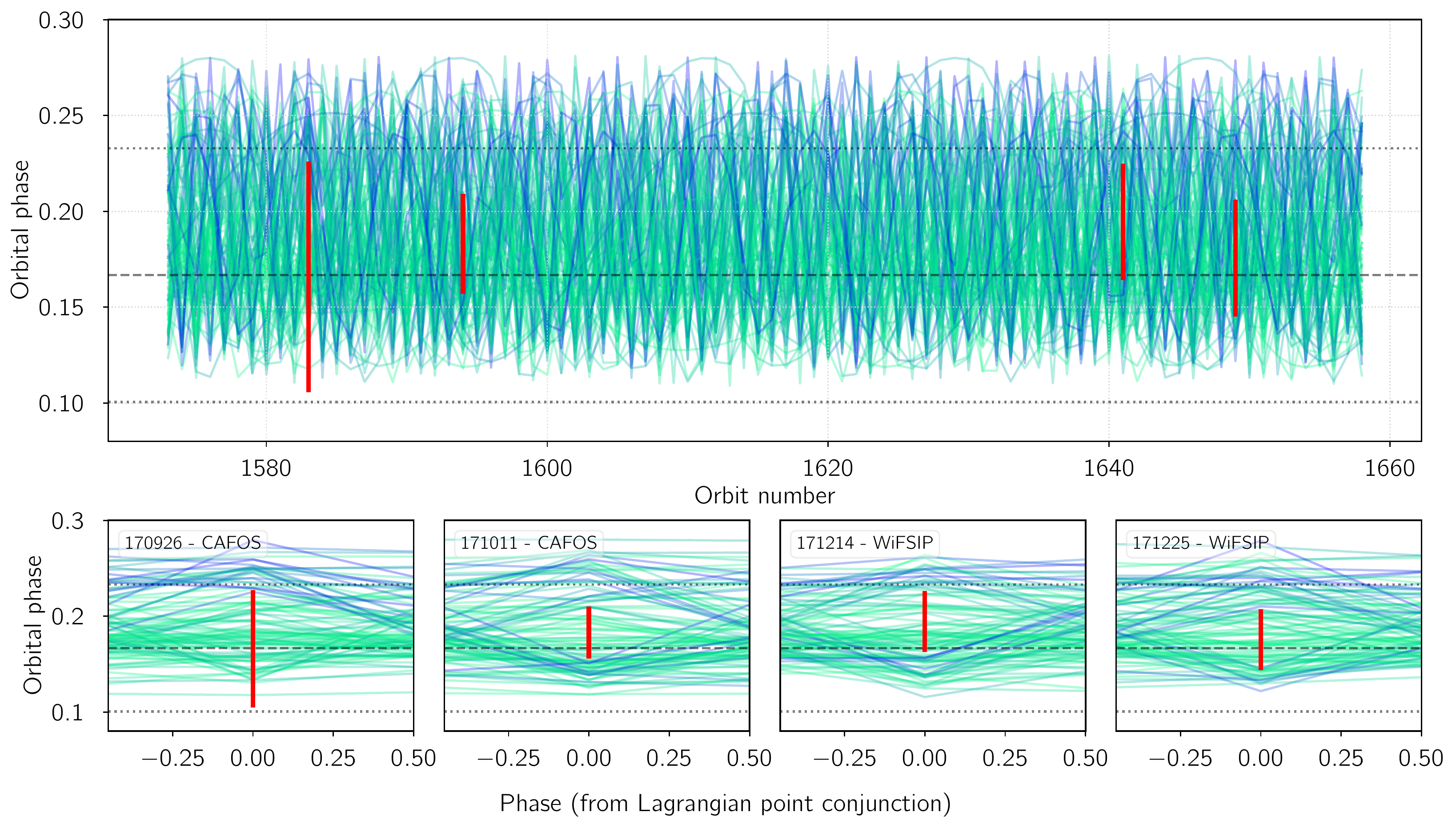}
\caption{Example of the analysis of the parameter space constrain based on the non-detection of transits in our data for the system WASP-77A. The upper panel shows the orbit number (X-axis) versus the orbital phase covered by our observations (red vertical lines). The transit of the Lagrangian point (L$_5$ in this case) is expected to happen between the dotted horizontal lines. The colored lines represent a small sample of the accepted models. For a model to be accepted it has to either not cross the red lines if the trial trojan radius is larger than the light curve detectability limit (blue lines) or if the trojan transits during our observations but it is smaller than the detectability limit (green lines). The bottom panels are just close views on each of the four epochs, with the X-axis being the phase from the Lagrangian point conjunction.}
\label{fig:PSshow}
\end{figure*}

The model calculates the expected time (or orbital phase) of the mid-transit based on the projected position of the trojan in the sky, which can be simplified from  Eqs.~53-55 in \cite{murray10} for small eccentricities as

\begin{equation}
\begin{split}
	X_t = &\ a/R_{\star}\Big[\cos{\lambda_t} + e_t/2  \Big(-3\cos{\omega_t} + \cos(2\lambda_t-\omega_t)\Big) \Big] \\
	Y_t = &\ a/R_{\star}\Big[\sin{\lambda_t} + e_t/2  \Big(-3\sin{\omega_t} + \sin(2\lambda_t-\omega_t)\Big) \Big]\cos{i} \\
	Z_t = &\ a/R_{\star}\Big[\sin{\lambda_t} + e_t/2  \Big(-3\sin{\omega_t} + \sin(2\lambda_t-\omega_t)\Big) \Big]\sin{i}, \\
\end{split}
\end{equation}

\noindent where $e_t$ and $\omega_t$ are the trojan eccentricity and argument of the periastron; $\lambda_t$ is the mean longitude of the trojan;
$i$ is the orbital inclination; and $a/R_{\star}$ is the semi-major axis to stellar radius ratio. Based on these equations and the reference frame described in \cite{leleu17}, the trojan transit occurs when $X_t=0$ (conjunction), $|Y_t|<1$ (the object does transit the star),  and $Z_t<0$ (primary eclipse). Hence, we can calculate the times at which these conditions are fulfilled (i.e., the mid-times of the trojan transit on each orbit) and, in particular, at the orbits that we observed. 

By comparing these values with the time ranges for each epoch we can constrain the parameter space of the parameters involved so that transits do not occur during these time ranges.  To that end, we use 20 walkers with 50\,000 steps per walker in our MCMC and we only select the steps with $\mathcal{L}$ equal to the sum of the maximum likelihoods on each epoch, i.e., $\sum_i \mathcal{L}_i^{\rm max}$. With these selected steps (usually around 80\% of the original chain) we can then construct the corner plot diagram with the parameter-parameter dependencies of the orbital models not having trojan transits during the observed time ranges. This diagram provides the constrains of the parameter values based on the observed photometric data in the absence of trojan transits and can be used, for instance, to provide a minimum libration amplitude for the trojan. In case several epochs were observed, we include the possibility of an eccentric orbit for the trojan. In case only one epoch is available, we assume circular orbit for the trojan ($e_t=0$).  

As a matter of example, we show in Fig.~\ref{fig:PSshow} the results for the analysis of WASP-77A. The red vertical lines represent the orbital phase interval that we observed, finding no transits. The expected transit is marked by the horizontal lines (dotted line being the expected ingress and egress phases). The colored lines represent a sample of accepted models from the whole MCMC chain. For a model to be accepted it has to either not cross the red lines if the trial trojan radius is larger than the light curve detectability limit (blue colored lines), or if the trojan transits during our observations but it is smaller than the detectability limit (green lines). In this particular example, our data allows us to reject trojans larger than 4\rearth with librations amplitudes shorter than $25^{\circ}$.

%+++++++
\subsubsection{Combined epochs}
\label{sec:LCcombine}

\begin{figure}
\centering
\includegraphics[width=0.5\textwidth]{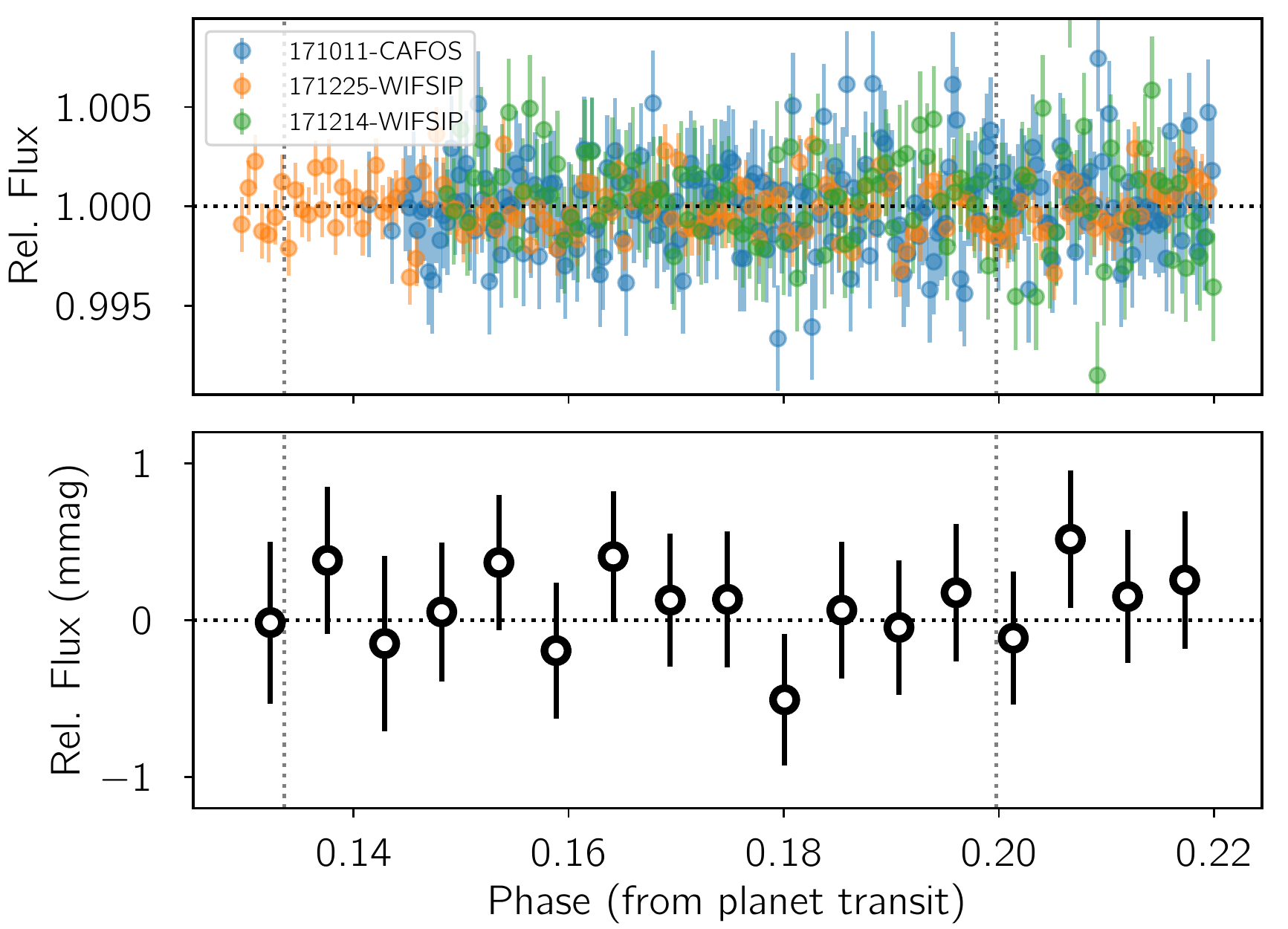}
\caption{Example of the light curve combination in phase for the case of WASP-77A. The individual epochs are shown in the upper panel (after removing the linear baseline model) and the combined light curve is shown in the bottom panel with 10-min bins having 225 ppm rms. The vertical dotted lines mark the expected transit ingress and egress of the Lagrangian point.}
\label{fig:LCcombine_show}
\end{figure}

Finally, we can also combine all epochs for the same object and Lagrangian point by assuming that the would-be transits are achromatic, given that we observed with different filters on different telescopes. This provides a smaller maximum radius of the trojan in case of no libration or in the case that the libration period is much larger than the time span of the observations. To do this we remove the median baseline model for each epoch in the null hypothesis and then  combine all epochs in orbital phase (see Fig.~\ref{fig:LCcombine_show}). We now try to search for a transit in this combined lightcurve in order to get a maximum radius for a stationary trojan (i.e., with no libration). To this end we follow the same procedure as in \S~\ref{sec:LCfitting}.

We can translate the maximum trojan radius into a maximum trojan mass by using the \textit{forecaster} module \citep{chen17}. According to the documentation of this module, it provides accurate estimates of the masses in the case of terrestrial and Neptune-like worlds, which are the typical cases for the radius limits found in this paper.

%=======
\subsection{Transit timing variations}
\label{sec:ttv}

As presented in \cite{ford07}, the amplitude of the variations in the transit times of the hosting planet ($K_{tt}$) due to the libration of a trojan with amplitude\footnote{Note that $\zeta=\lambda_p-\lambda_t$ is the angular difference between the mean longitude of the planet ($\lambda_p$) and the mean longitude of the trojan ($\lambda_t$).} $K_{\zeta}$ are well represented by the Eq.1 in that paper, i.e., 

\begin{equation}
K_{tt} = 60\,s \left(\frac{P_{\rm orb}}{4~{\rm days}}\right) \left(\frac{m_t}{m_{\oplus}}\right) \left(\frac{0.5 M_{\rm Jup}}{m_p+m_t}\right) \left(\frac{K_{\zeta}}{10^{\circ}}\right).
\end{equation}

If no periodic variations are found, we can estimate an upper limit for this amplitude by simply fitting a sinusoidal function to the measured TTVs, with amplitude $K_{tt}$. This way, we can find a direct relation between the trojan mass and the libration amplitude, thus constraining this parameter space. We can compute the maximum TTV amplitude ($K_{tt, \rm max}$) as the 95\% confidence level from the marginalized posterior probability of $K_{tt}$. In tadpole low eccentricity orbits, we expect the libration of the trojan to introduce sinusoid-like variations in the planet's time of transit. Hence, under this assumption, we can use the simple model $Z+K_{tt}\sin{(\nu_{\rm lib}t + \phi_{tt})}$ , being $\nu_{\rm lib}$ the libration frequency \citep{leleu15}, $\phi_{tt}$ a phase offset, and $Z$ a zero level to account for the uncertainties in the orbital period and mean mid-transit time. We assume a uniform prior for all parameters involved in the model, with $Z\in \mathcal{U}(-0.1,0.1)$ hours, $K_{tt}\in \mathcal{U}(0.0,5.0)$ hours, $\nu_{\rm lib}\in \mathcal{U}(\nu_{\rm Lp}/5, 5\nu_{\rm Lp})$ with $\nu_{\rm Lp}=2\pi/P_{\rm orb}\sqrt{27/4~m_p/(m_p+M_{\star})}$, and $\phi_{tt} \in \mathcal{U}(0,2\pi)$.

In order to sample the posterior distribution we use an MCMC algorithm by means of the \textit{emcee} code. We use 50 walkers and 100\,000 steps per walker. We remove the first half of the steps and compute the marginalized posterior distribution for the $K_{tt}$ parameter. We check that the posterior distribution is actually truncated at $K_{tt}=0$ (i.e., no detection of periodic TTVs). Then, the 95\% percentile is computed to obtain $K_{tt, \rm max}$. Given this value, we can get a contour in the $m_t(K_{\zeta})$ function to constrain the parameter space.

%=======
\subsection{Constraints to the trojan mass vs. libration amplitude parameter space}
\label{sec:mtlib}

\begin{figure}
\centering
\includegraphics[width=0.5\textwidth]{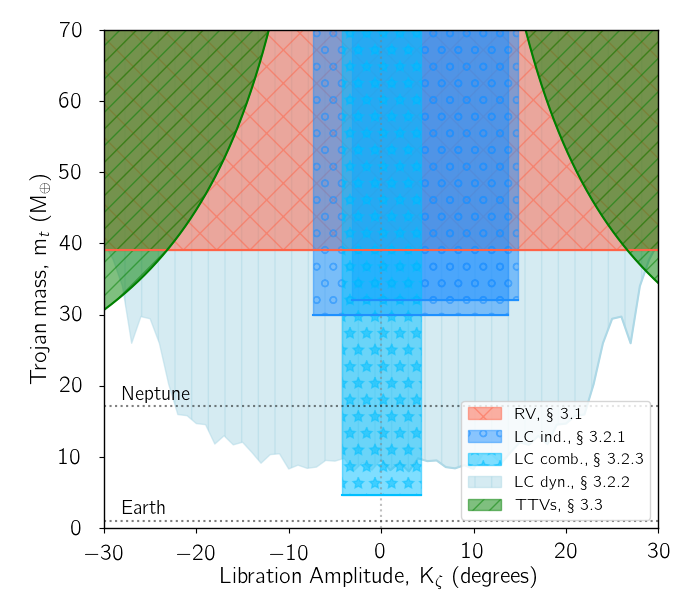}
\caption{Detailed example of the constraining of the trojan mass versus libration amplitude parameter space in the case of WASP-77A. Negative (positive) values for the libration amplitude are used to represent constraints on transits before (after) the Lagrangian point passage. Only the white region is not explored by our data. }
\label{fig:mtlib_show}
\end{figure}

The analysis described above for the the multi-technique data produces different constraints on several planes of the parameter space. One interesting plane is the trojan mass versus the libration amplitude, because it provides both physical and dynamical information relevant for detection purposes. In Fig.~\ref{fig:mtlib_show}, we show an example of the constraints provided by our analysis on this plane assuming coplanarity between the trojan and planet orbital planes. The analysis of the radial velocity provided in \S~\ref{sec:rvtr} constraints the mass of the trojan regardless of the libration amplitude, since the data was taken during a long timespan (much longer than the libration period). This is shown by the red shaded region in the example Fig.~\ref{fig:mtlib_show}. In order to translate the results from the light curve analysis (from the individual  analysis - \S~\ref{sec:LCfitting}-, the dynamical analysis combining the individual epochs - \S~\ref{sec:PSconstrain}-, and from the combined light curve - \S~\ref{sec:LCcombine}), we convert the maximum trojan radius to maximum trojan masses by using the \textit{forecaster} module \citep{chen17}. The libration parameter in this case is constrained by the time range of our observations in the case of individual light curves and from the expected transit duration for the combined light curves. Finally, the TTVs can constrain this parameter space as already described in \S~\ref{sec:ttv} (shown in green in the example Fig.~\ref{fig:mtlib_show}). The diagram corresponding to each of the nine systems can be found in Fig.~\ref{fig:mtlib}.

%------------------------------------------------------------------------------------------------
%						RESULTS
%------------------------------------------------------------------------------------------------
\section{Results per system}
\label{sec:results}

Based on the analysis of the multi-technique data presented in section \ref{sec:methodology}, we can now constrain the presence of co-orbital bodies in the surroundings of the Lagrangian points in these nine systems. Here we summarize the results for each of the analyzed planetary systems.

%=======
\subsection{GJ\,3470}

Only one Lagrangian transit is available for this system. The analysis of the light curve shows no significant detection of any transit up to $3.7~R_{\oplus}$. The analysis of the radial velocity data provides a value for the $\alpha$ parameter significantly different from zero. However, the eccentricity fitted is too large and thus the radial velocity equations are out of the validity range (i.e., $e<0.1$). Consequently, the result is degenerate and we can neither discard the trojan scenario nor confirm it. In order to solve this dichotomy, a measurement of the secondary eclipse of the planet would be needed\footnote{The estimated eclipse depth for this planet is smaller than 25 ppm. This is at the limit precision that will be achievable by the \textit{James Webb Space Telescope}.}, together with the development of new equations with a wider eccentricity validity range. Also, the TTVs do not show a significant periodic variation up to 3.2 minutes. This allows us to put important constraints on the mass of any potential librating trojan and confirms that should it exist with a moon- or planet-size, the libration amplitude should be smaller than 4$^{\circ}$ for trojans with masses larger than Earth and smaller than 10$^{\circ}$ for sub-Earth-mass trojans. 

%=======
\subsection{HAT-P-12}

The photometric exploration of $L_4$ in HAT-P-12 did not show any significant dim down to $4.8~R_{\oplus}$. However, only half of the Lagrangian point passage could be covered with CAFOS due to bad weather conditions. The radial velocity analysis provides $\alpha=-0.141\pm0.082$, corresponding to a trojan mass of $10.8\pm6.3~M_{\oplus}$ at L$_4$. This sets an upper limit of 21\mearth\ in L$_4$ and rejects any trojan more massive than 4~M$_{\rm Moon}$ at L$_5$. The TTVs in this system restrict the libration amplitude importantly, leaving only sub-Earth masses to libration amplitudes larger than $\sim7^{\circ}$. Hence, if larger bodies are present in this Lagrangian point, low libration amplitudes are expected and so sub-mmag precision light curves could explore this regime (an Earth-size body would induce $\sim$170 ppm transit depth in this system). This system illustrates that since TTVs are proportional to the trojan-to-planet mass, small co-orbitals should be more easily found co-rotating with low mass planets.

%=======
\subsection{HAT-P-20}

The large amount of radial velocity data that we gathered for HAT-P-20\,b added to the archive spectra allows now to estimate $\alpha=0.0017\pm0.0048$. This very small value allows us to set an upper limit on the mass of any potential trojan of 25\mearth. The TTVs do not show any clear periodic sinusoidal signal. However, the data show variations up to 2.5 minutes peak-to-peak. 

A total of three observations of the L$_5$ transit passage have been observed with CAFOS for this target. The quality of the observations is very variable, going from very bad quality (on 2017-04-03 with flux variations up to 5 mmag) to very good quality data at the $\sim$1.5 mmag level on individual measurements. We have tested the null hypothesis and the trojan hypothesis in all three epochs. The results show that for two epochs (2017-04-03 and 2017-11-16) the null hypothesis has a significantly larger evidence than the trojan hypothesis. They both allow us to set upper limits on the trojan size of 8.3\rearth\ and 2.9\rearth\ (respectively) in the phases covered by these epochs. 

\begin{figure}
\centering
\includegraphics[width=0.5\textwidth]{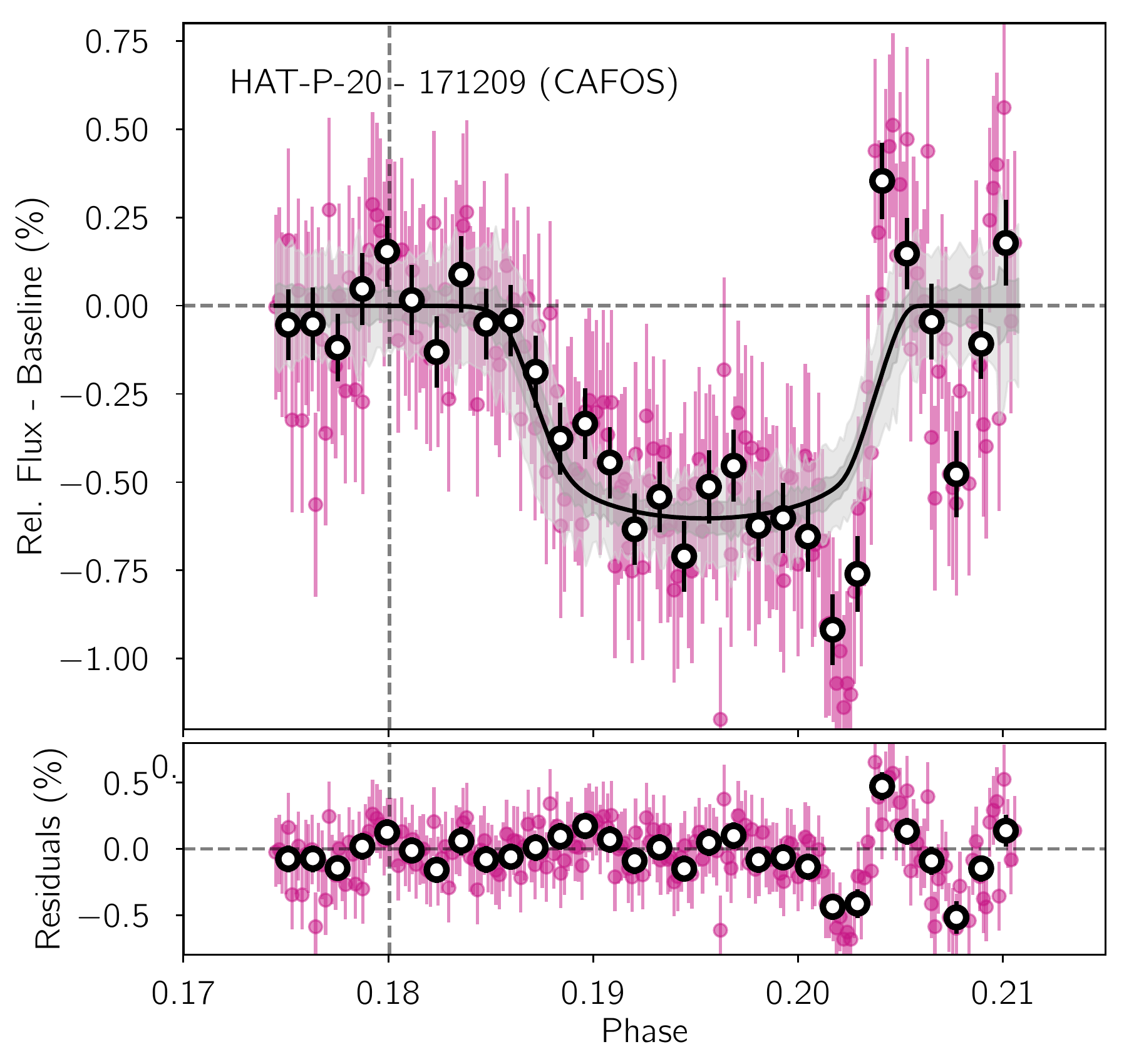}
\caption{Trojan hypothesis model after removing the linear baseline model for HAT-P-20. This trojan model is only favored against the null hypothesis if we include the whole data set. However by removing the feature at phases $\phi>0.20$, the null hypothesis is favored. The black shaded vertical line represents the expected egress of the Lagrangian point, showing the large time lag of the would-be trojan transit. This light curve corresponds to the data from CAFOS at CAHA on 2017-12-09.}
\label{fig:hatp20fit}
\end{figure}

However, the results on the night of 2017-12-09 provide a larger evidence to the trojan model against the null hypothesis, showing a Bayes factor of 12 in favor of the trojan model. These two models (with and without trojan) are built on the basis of linear dependencies with the baseline parameters. We have also tested different quadratic dependencies for the null hypothesis but all of them show Bayes factors even larger in favor of the trojan linear-baseline model. This favored model reveals a dim in the light curve at phase 0.195 that would correspond to a 5.8\rearth object. In Fig.~\ref{fig:hatp20fit}, we show the CAFOS light curve with the baseline model removed and the median fitted transit model. Despite the large significance of the trojan model, the large radius of the possible co-orbital is puzzling (although it could also be a compact trojan swarm). Also, the baseline models without trojan are able to reproduce qualitatively well the would-be ingress, removing the transit signal. In Fig.~\ref{fig:LCfitting1} (in the panel labeled as "HAT-P-20 - 171209"), this non trojan model is shown. 

As easily pointed out in both figures, at phases larger than 0.20 there is a clear modulation that any baseline model fails to reproduce. In order to test if this might be the only reason why the trojan model is favored against the non-trojan model we performed the same analysis by removing all datapoint after phase $\phi>0.20$ (after Julian date 2458097.2121). The results of this analysis, however, increase the significance of the non-trojan model providing an evidence 12 times larger now in favor of the null hypothesis. We then conclude that the strong modulation at the end of our observations strongly affects the result of the analysis and so we cannot conclude on the presence of any large body in this transit with the current data. Since this deep event (that would correspond to a gas giant) is not seen neither in the other epochs nor in the radial velocity data, we assume for the subsequent discussion that no trojan is present in this system.

Under this assumption, the analysis of the combined light curve produces an upper limit on the trojan mass of 2\rearth\ at the exact Lagrangian point.

%=======
\subsection{HAT-P-23}

In the case of HAT-P-23, we found $\alpha=-0.132\pm0.082$, corresponding to a 95\% upper limit of 205 $M_{\oplus}$. We photometrically explored the $L_4$ region during three different passages of this region in front of its star. We can discard coplanar transits in these observations of objects up to $5.1-6.2~R_{\oplus}$. Furthermore, the combined light curve allows discarding bodies larger than $1.39~R_{\oplus}$ laying exactly at the Lagrangian point (or experiencing very small librations). The analysis of the parameter space of the trojan properties that still remain plausible despite of the non transit detection, just leaves the possibility of large libration amplitudes for circular orbits of the trojan, which are avoided by the lack of  TTV modulations. Hence, should it exist, the trojan co-orbiting to HAT-P-23\,b must be either a planet smaller than Neptune with a large libration amplitude or a low mass ($<4~M_{\oplus}$) planet in a low libration amplitude but highly-eccentric orbit.

%=======
\subsection{HAT-P-36}

Photometric variability at the 1.5 mmag level appears {in the CAFOS and WiFSIP light curves around the L$_5$ region}, which might be due to the moderate activity of the host star ($\log{R}^{\prime} = -4.65$~dex, \citealt{mancini15}). No significant dim is detected down to 6\rearth\ at L$_5$ in the CAFOS light curve {(the WiFSIP data being of slightly worse quality in this case)}. The new radial velocity measurements presented in this paper allow us now to decrease the $\alpha$ parameter to $\alpha=0.071^{+0.072}_{-0.076}$ (from the previous $0.25^{+0.22}_{-0.24}$), {thus being now fully compatible with no trojan at 1-$\sigma$ level}. This now corresponds to a maximum mass of the trojan of 128\mearth\ at L$_5$ (around four times smaller than our previous upper limit) and 38\mearth\ at L$_4$.  {However, the combination of all three available transit observations allows us to set small upper radius of 2\rearth\ to any trojan body located at the exact Lagrangian point. }

%=======
\subsection{WASP-2}

In this case we find $\alpha=-0.018^{+0.018}_{-0.015}$, so that we can set an upper limit to the trojan mass of 15.1\mearth, similar to our previous measurement. The analysis of the TTVs also do not show any significant periodic variation and the single transit we could observe with CAFOS do not show any significant dim. The upper limit that we can impose based on this non-detection of the transit is 4.3 \rearth. The TTVs also indicate that trojans more massive than the Earth should have libration amplitudes smaller than $25^{\circ}$.

%=======
\subsection{WASP-36}

The 14 new radial velocity measurements obtained for this target with HARPS-N and CARMENES allows as to decrease the significance in the $\alpha$ parameter with respect to \cite{lillo-box18a}. We find now $\alpha=0.031\pm0.028$ (compared to the previous $0.092\pm0.043$). We have now decreased by two the uncertainty in this parameter. However, due to the large mass of the planet, we can only set an upper limit to its trojan mass at L$_5$ of $m_t<63$\mearth\ (compared to the previous 146\mearth).  Additionally, the TTVs do not show variations larger than 320\,s. Finally, the two WiFSIP observation of the Lagrangian transit do not show any dim down to 8\rearth\ and 5.4\rearth, respectively. The combined lightcurve, however, provides a much smaller upper limit to the size of any potential trojan of 2.1\rearth. This allows us to reject any possible trojan with $m_t>10$\mearth\ at the exact position of the Lagrangian point.

%=======
\subsection{WASP-5}

The new radial velocity analysis presented in this work (including new HARPS data) now provides $\alpha = 0.002 \pm 0.010$, corresponding to an upper limit of 10.4\mearth\ at L$_5$ and 7.6\mearth\ at L$_4$. The periodogram of the TTVs in this case shows a peak around 18~days although this peak is not statistically significant. The modeling of the TTVs consistently provides a possible periodic solution with an amplitude of $K_{tt}=32\pm18$\,s and a periodicity of $P_{tt}=18.06^{+0.17}_{-0.96}$\,days. If we neglect the mass of the trojan, the expected libration period would be 15.7 days in this system. But, this periodicity can be elongated due to different factors like mutual inclination or eccentricity.

Using Eq. 1 in \cite{ford07} we can use the estimated $K_{tt}$ to get a relation between the trojan mass and the libration amplitude. Given the upper limit on the mass provided by the radial velocity analysis, and assuming the TTV modulations, the possible trojan should have a minimum libration amplitude of 4$^{\circ}$. The lower the trojan mass the larger libration amplitude is needed to produce the observed TTV modulation. 

Such low libration amplitude implies that the trojan could transit the star very close to the Lagrangian point passage on every orbit. The upper mass limit of 10.4\mearth\ provided by the radial velocity would correspond to an object of $3.1^{+1.4}_{-0.8}$\rearth. Our FORS2 observations of the L$_5$ passage do not show any statistically significant dim, providing an upper limit on the trojan radius of 4.5\rearth. Hence, we would need at least two more transits to test the regime allowed by the radial velocity.

\begin{figure}
\centering
\includegraphics[width=0.5\textwidth]{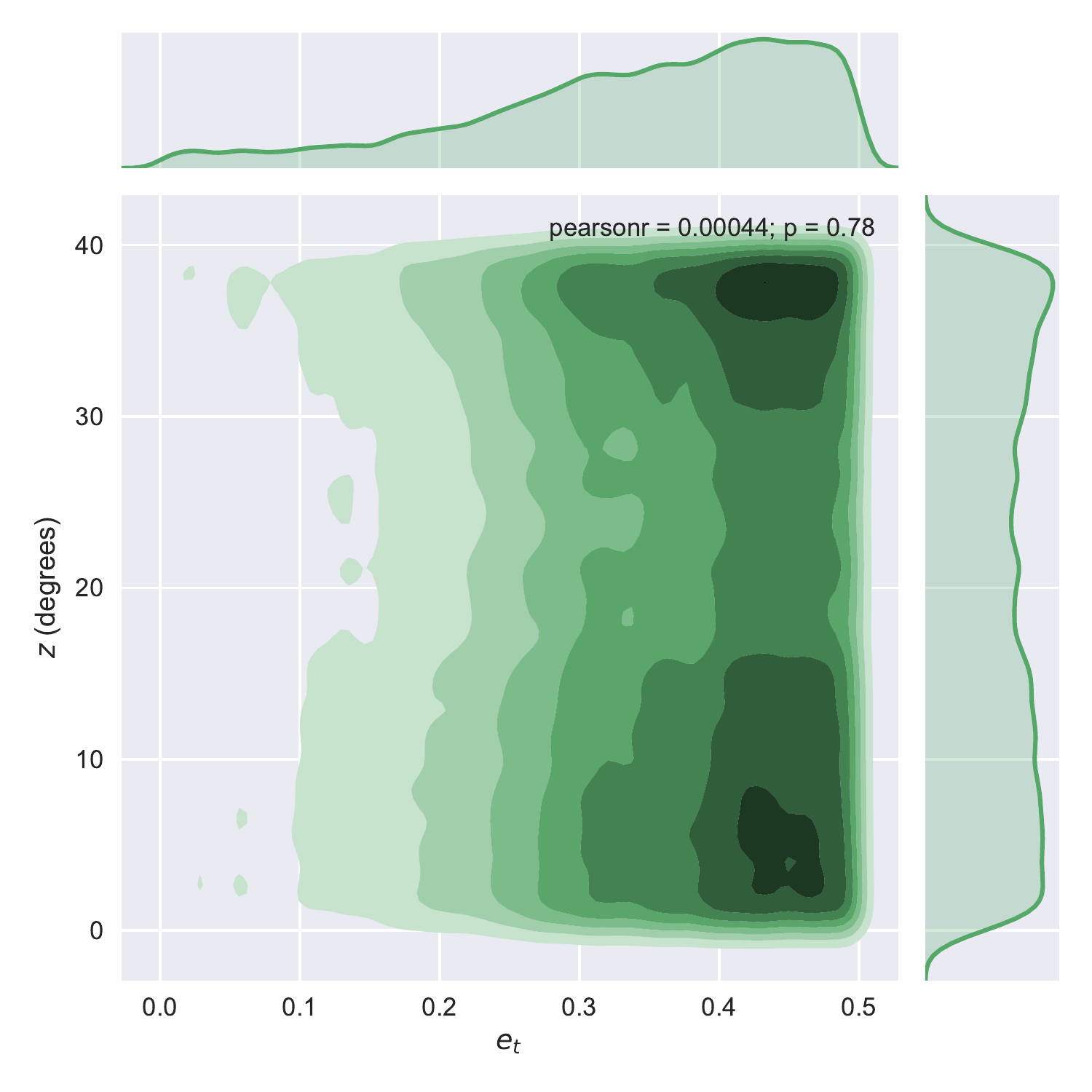}
\caption{Parameter space of trojan's eccentricity versus libration amplitude constrained by the non-detection of transits in the four Lagrangian passages observed for WASP-77A\,b. The results show that only highly eccentric orbits are possible for any libration amplitude and that low eccentric orbits are still allowed for libration amplitudes larger than $35^{\circ}$.  }
\label{fig:lib_et}
\end{figure}

%=======
\subsection{WASP-77}

The new RV data from CARMENES allows us now to decrease the upper limit in the mass of any potential trojan by a factor of four, now being $\alpha=0.017\pm0.027$, corresponding to an upper mass limit of $39~M_{\oplus}$ (compared to the $92~M_{\oplus}$ obtained in our previous analysis, \citealt{lillo-box18a}). The photometric data from the four epochs do not show statistically significant dims. Having four epochs allows us to set important constraints on the orbital properties of any potential trojan detectable with our photometric data but not transiting during our observing time ranges (see \S~\ref{sec:PSconstrain}). In this case, we can neglect trojans larger than 5\rearth\ with libration amplitudes smaller than $25^{\circ}$ in the case of circular orbit for the trojan (otherwise they would have been detected in our data). If we assume non-zero eccentricity for the trojan, Fig.~\ref{fig:lib_et} shows how our observations constrain this eccentricity as a function of the libration amplitude. In a nutshell, only a highly eccentric orbit would allow small libration amplitudes. 

The combination of the four epochs acquired for this target provides an upper limit for a trojan orbiting at the exact Lagragian point of $1.39~R_{\oplus}$. The combination of all three techniques discards trojans with masses larger than 4\mearth\ at the exact position of the L$_4$ Lagrangian point. Trojans up to $\sim$30\mearth\ with moderate libration amplitudes ($\zeta<25^{\circ}$) are still not rejected by our observations.

%------------------------------------------------------------------------------------------------
%						DISCUSSION
%------------------------------------------------------------------------------------------------

\section{Discussion and conclusions}
\label{sec:discussion}

We have used information from the radial velocity, transit and TTVs techniques to constrain the presence of co-orbital bodies in nine planetary systems previously showing hints for their presence \citep{lillo-box18a}. These systems correspond to short-period ($P<5$ days) mainly massive planets, where high-precision measurements from the three techniques can be obtained and found in the archive. The three techniques complement each other in the parameter space composed by the trojan mass versus libration amplitude, allowing us to progressively discard trojans at different regimes. 

\begin{figure}
\centering
\includegraphics[width=0.5\textwidth]{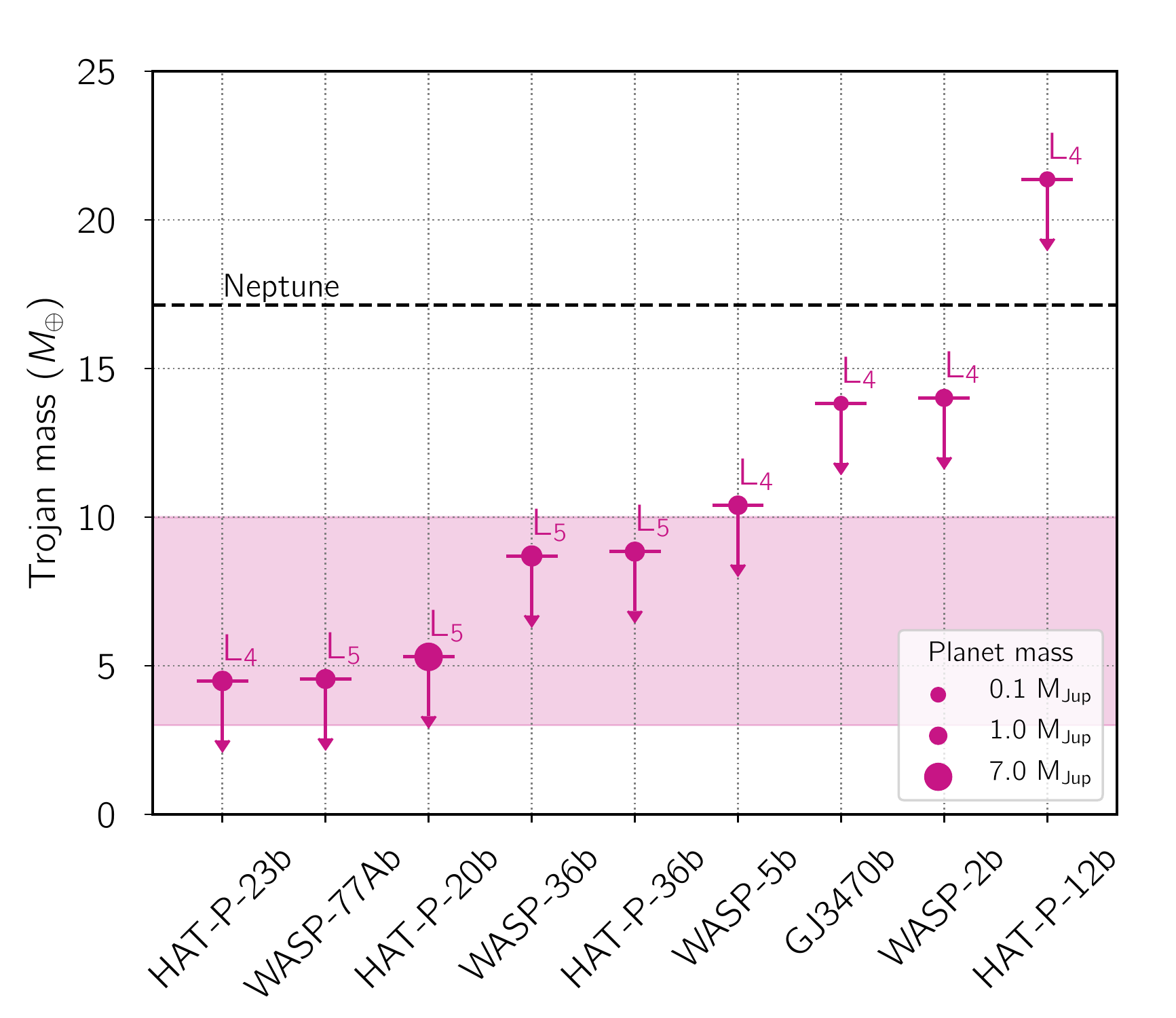}
\caption{Upper limits on the masses of trojan bodies located at the exact Lagrangian points of the nine systems studied. The symbol size scales as the mass of the planet. The mass of Neptune is marked for reference and the shaded regions represents the super-Earth mass regime. }
\label{fig:mtupper}
\end{figure}

For instance, in Fig.~\ref{fig:mtupper}, we show the upper mass limits for co-orbitals exactly at the Lagrangian points of eight of the studied systems combining all three techniques. As we can see, we can discard trojans more massive than 10\mearth\ in {six} of the systems and we can go down to 5\mearth\ regime in the case of HAT-P-23, WASP-77A, and HAT-P-20. The key observations {in these three systems} have been the combination of more than three transit observations of the Lagrangian passage with photometric precisions at the $\sim$1~mmag level with CAFOS and WiFSIP. In the case of WASP-5, the key technique was the radial velocity follow-up with HARPS combined with the accurate measurements of the planet's mid-transit time and mid secondary eclipse time (breaking the degeneracy with the eccentricity). 

We here show that ground-based photometric exploration of the Lagrangian points of known planets can provide constraints on trojan bodies in these regions down to the Earth-size regime when combining $>3$ epochs with mmag precision for solar-like stars. It is important to note that intensive ground-based monitoring from small robotic telescopes would strongly increase the sample of explored systems in an efficient way.

The exploration of Kepler/K2 data was partly done by \cite{janson13} with no positive detections. However, \cite{hippke15} found clear dims at L$_4$ and L$_5$ of the combined Kepler light curve obtained by stacking all planet candidates, showing that on average all Kepler planets have co-orbitals of few hundreds of kilometers size (or equivalently swarms of small trojans with an equivalent cross-section of this size). Hence, a dedicated exploration of this data (taking into account possible large libration amplitudes) should reveal the presence of the individual co-orbitals (Lillo-Box et al., 2018c, in prep.). In the future, TESS photometry \citep{ricker14} will also help to find these bodies in many systems with precisions similar to Kepler and a 2-min cadence for many of them, which is critical in case of large libration amplitudes. Additionally, the CHEOPS mission \citep{broeg13} will be a unique opportunity to follow-up best candidates, reaching lower trojan radii down to the rocky regime.

%-------------------------------------------------------------------

\begin{acknowledgements}
We thank Mathias Zechmeister for his useful advise and support with the SERVAL pipeline for CARMENES, which is partly used in our new pipeline to obtain the RVs with the CCF technique (carmeneX). J.L-B acknowledges financial support from the European Southern Observatory (ESO). Parts of this work have been carried out within the frame of the National Centre for Competence in Research PlanetS supported by the SNSF. DB acknowledges financial support from the Spanish grant ESP2015- 65712-C5-1-R. NCS, PF and JF were supported by Funda\c{c}\~ao para a Ci\^encia e a Tecnologia (FCT, Portugal) through the research grant through national funds and by FEDER through COMPETE2020 by grant PTDC/FIS-AST/1526/2014 \& POCI-01-0145-FEDER-016886, as well as through Investigador FCT contracts nr. IF/01037/2013CP1191/CT0001 and IF/00169/2012/CP0150/CT0002. J.J.N. also acknowledges support from FCT though the PhD:Space fellowship PD/BD/52700/2014. A.C. acknowledges support from CIDMA strategic project (UID/MAT/04106/2013), ENGAGE SKA (POCI-01-0145-FEDER-022217), and  PHOBOS (POCI-01-0145-FEDER-029932), funded by COMPETE 2020 and FCT, Portugal
H.P. has received support from the Leverhulme Research Project grant RPG-2012-661.

\end{acknowledgements}

%______________________________________________________________
\bibliographystyle{aa} % style aa.bst
\bibliography{/Users/lillo_box/11_MyPapers/biblio2} % your references Yourfile.bib

%----------------------------------------------------------------------------------------
%	FIGURES
%----------------------------------------------------------------------------------------
\newpage

\appendix

\section{Figures}
\label{app:figures}

% ========= Radial velocity
\begin{figure*}
\centering
\includegraphics[width=0.3\textwidth]{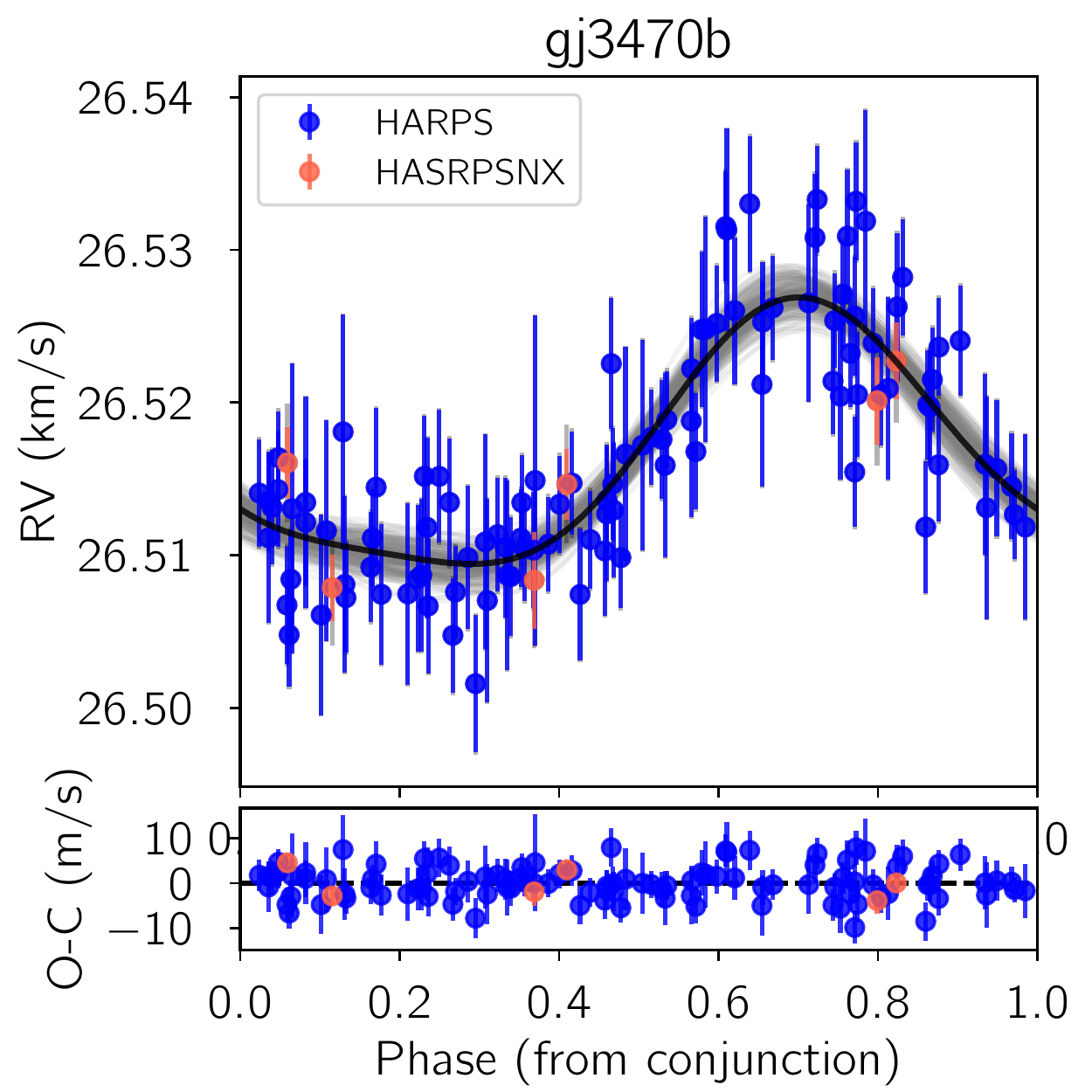}
\includegraphics[width=0.3\textwidth]{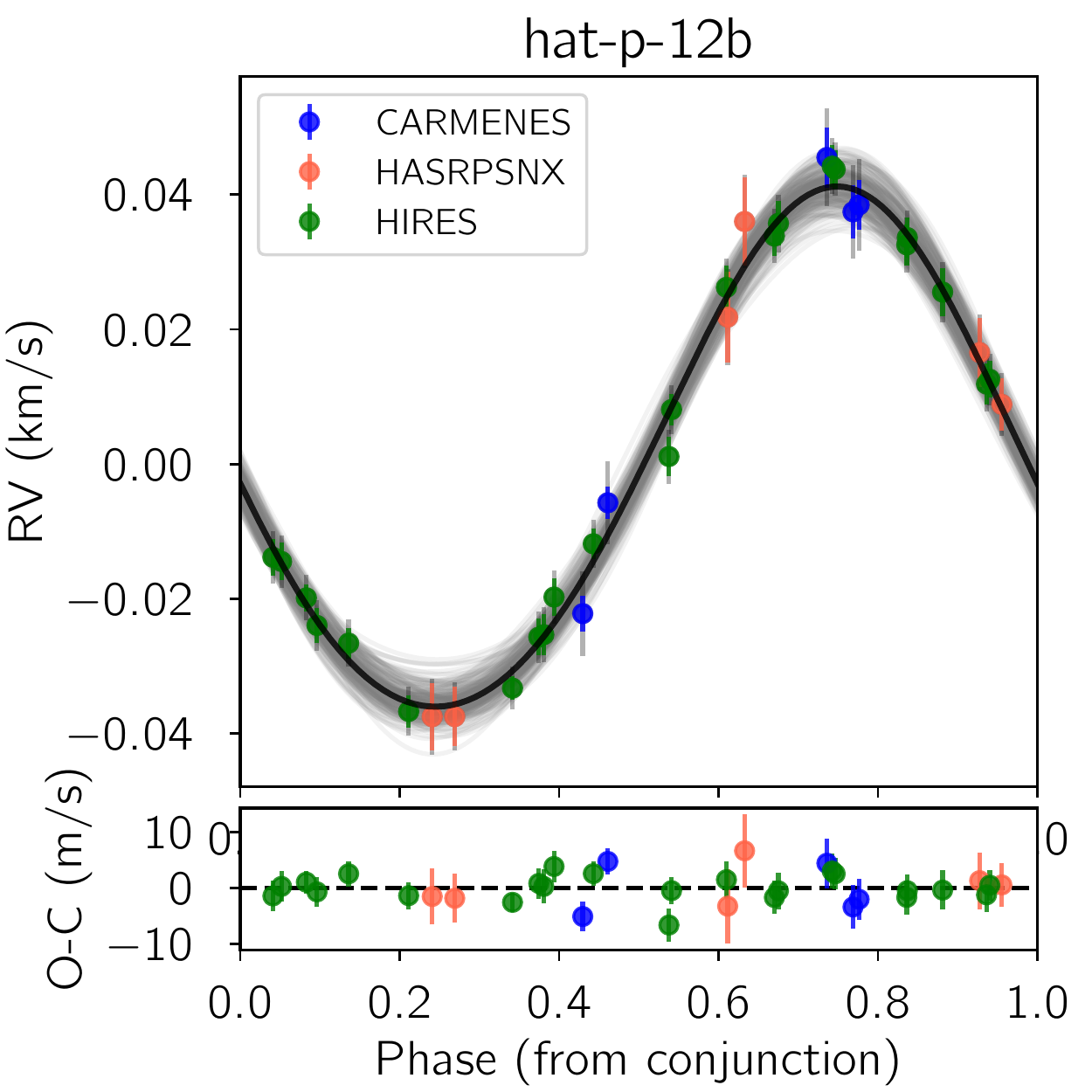}
\includegraphics[width=0.3\textwidth]{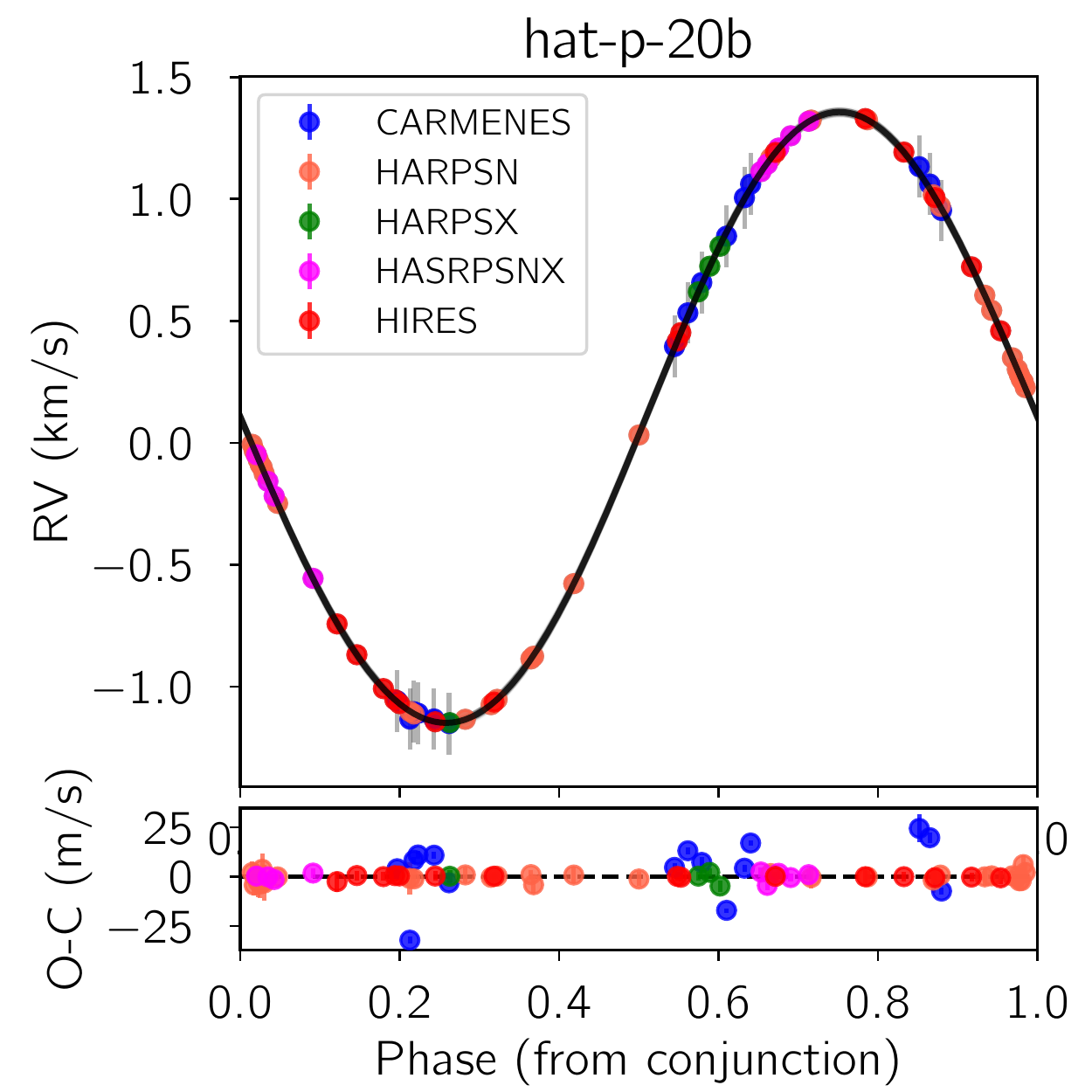}

\includegraphics[width=0.3\textwidth]{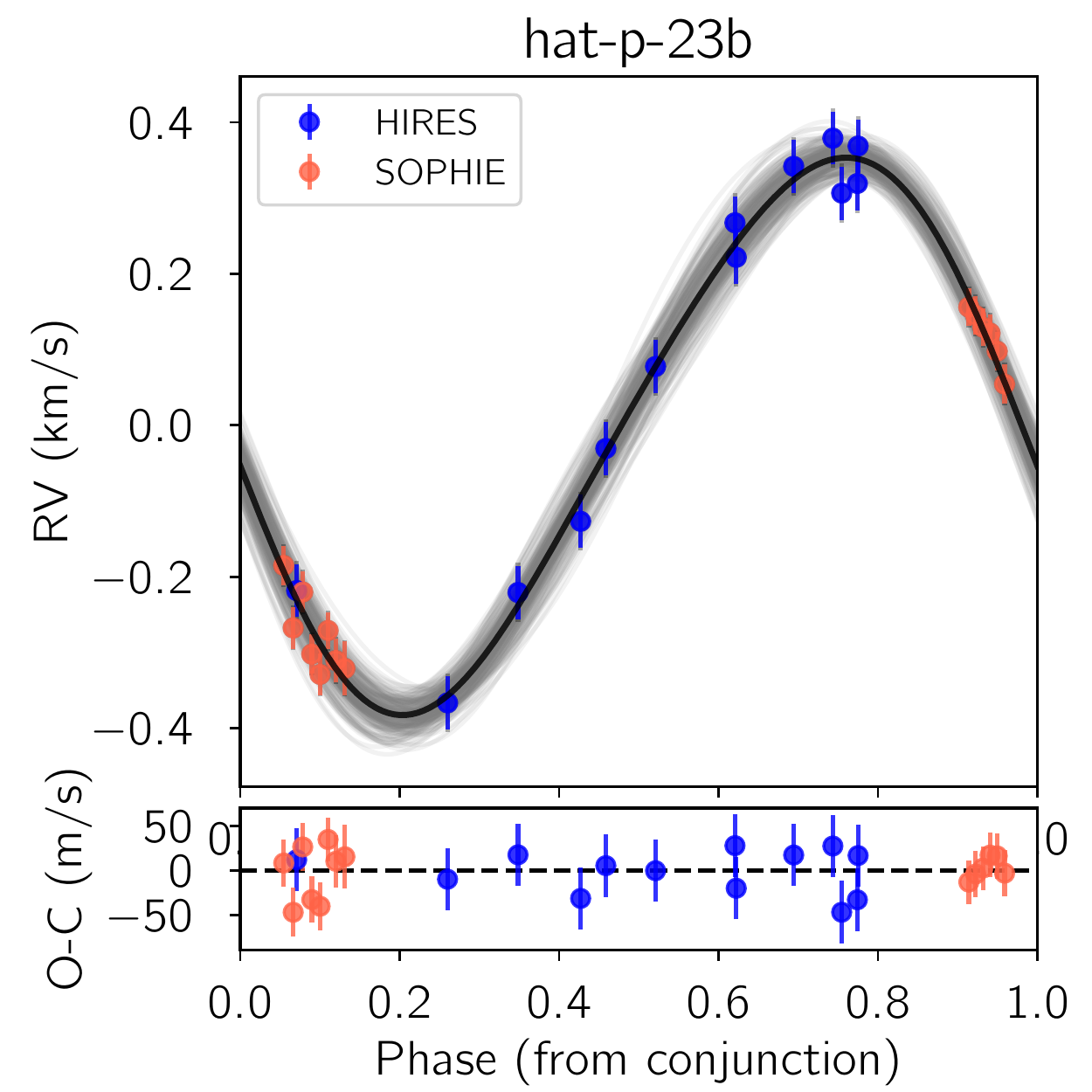}
\includegraphics[width=0.3\textwidth]{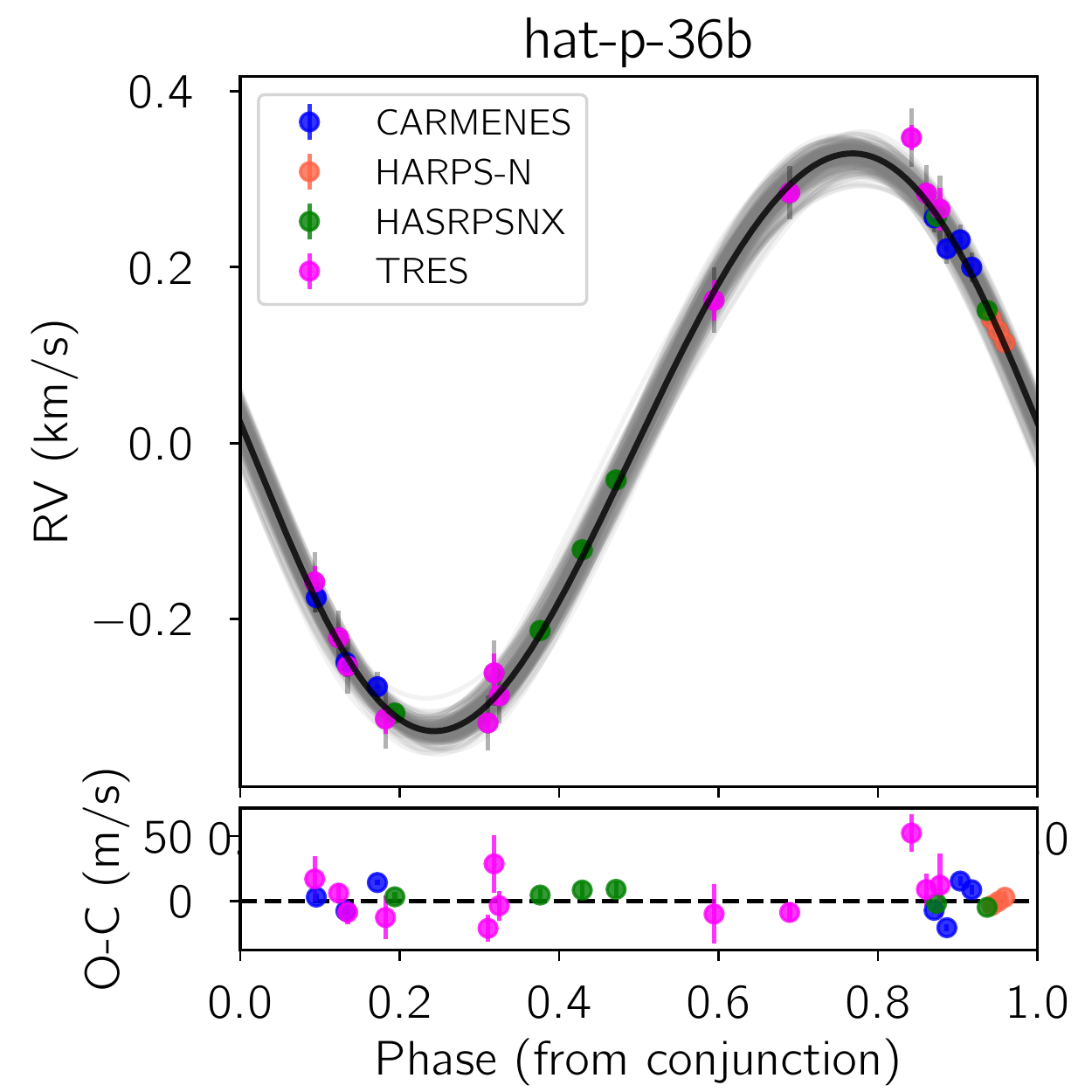}
\includegraphics[width=0.3\textwidth]{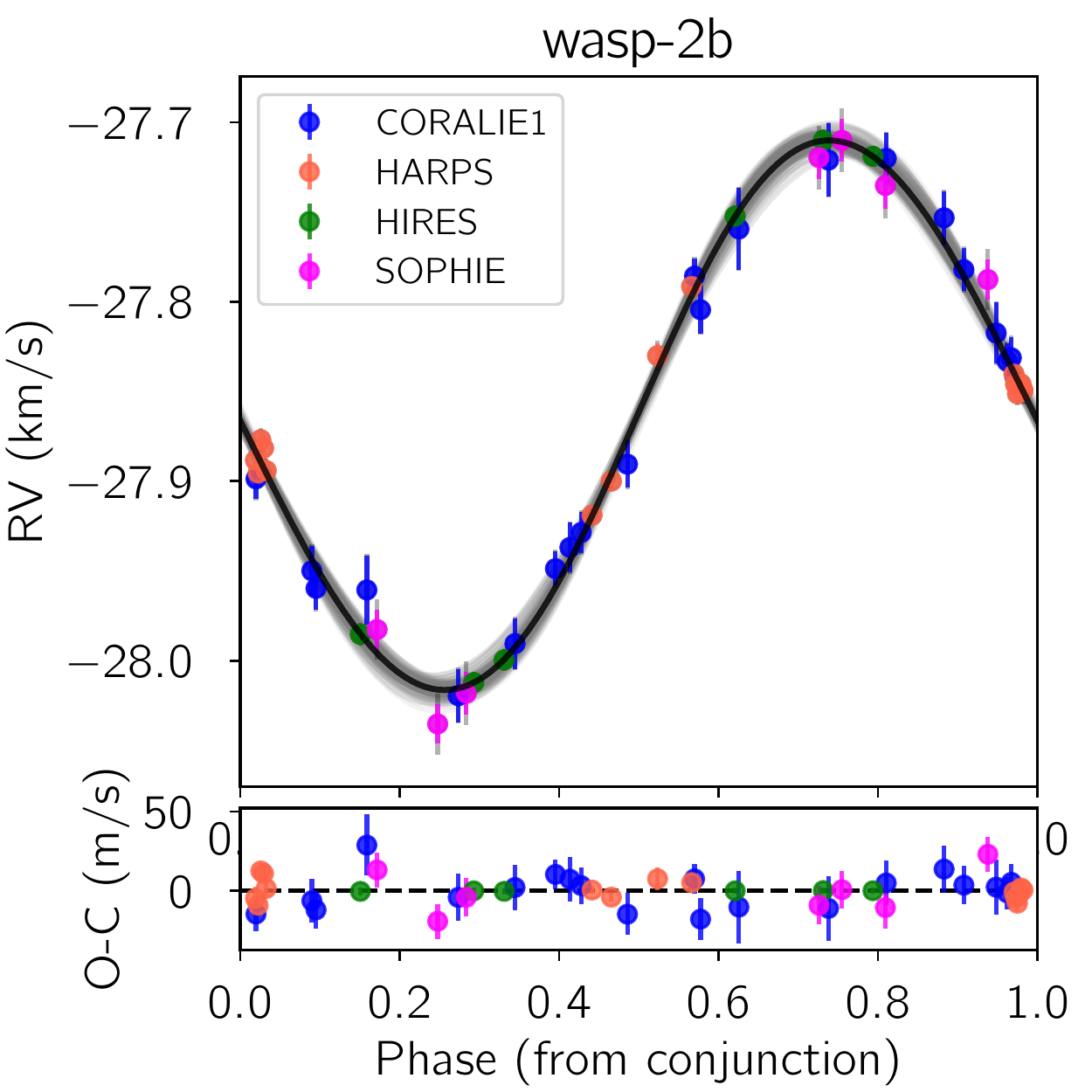}
\includegraphics[width=0.3\textwidth]{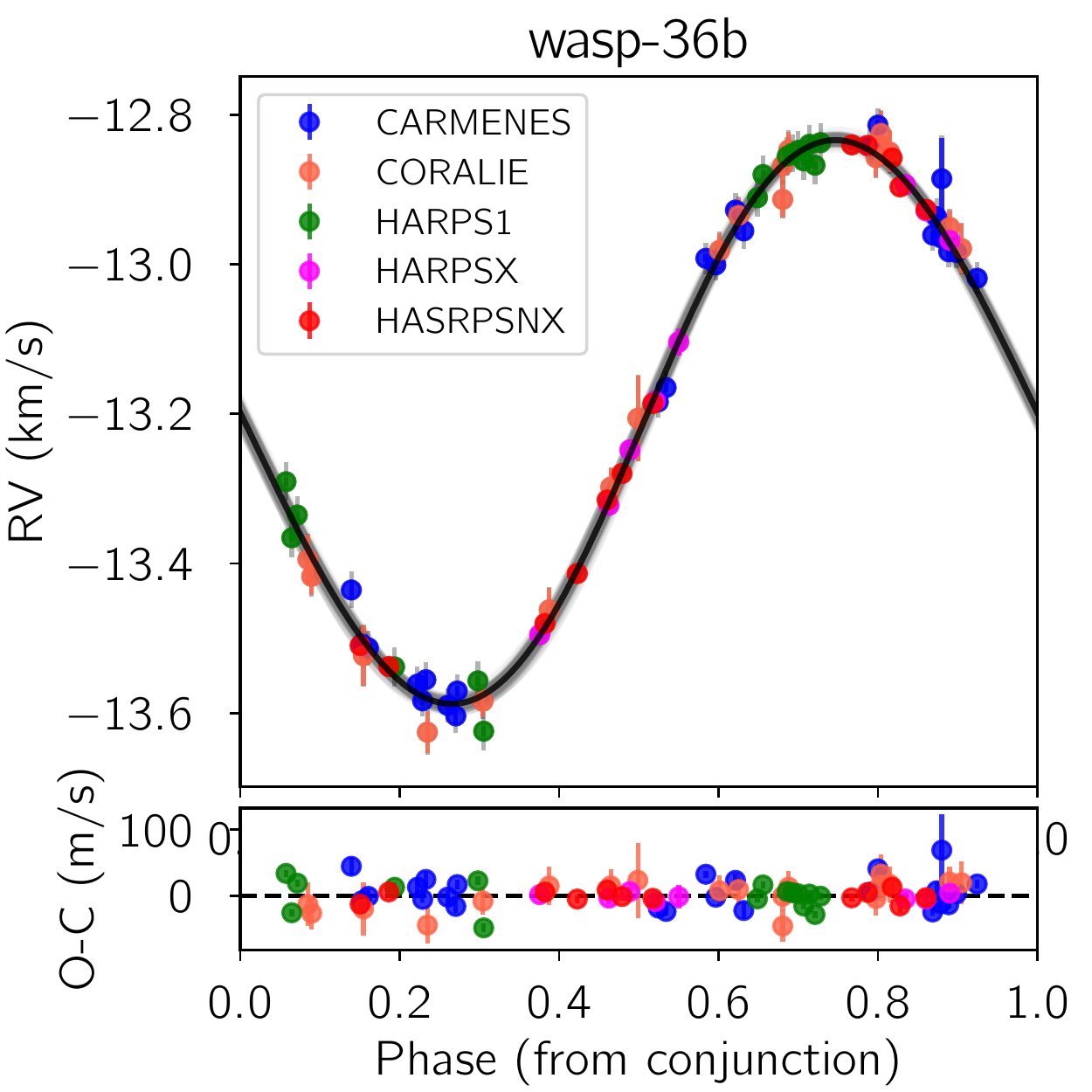}
\includegraphics[width=0.3\textwidth]{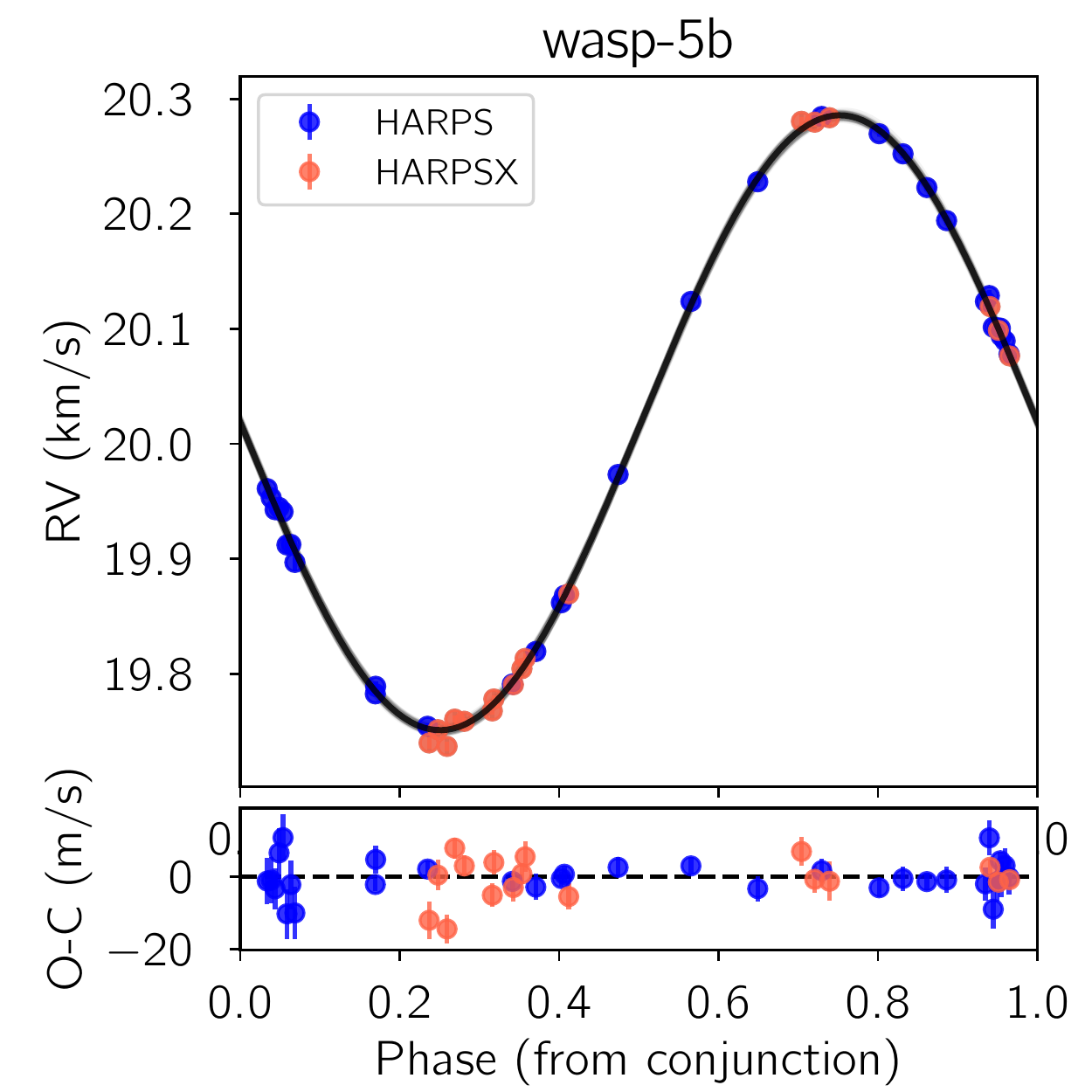}
\includegraphics[width=0.3\textwidth]{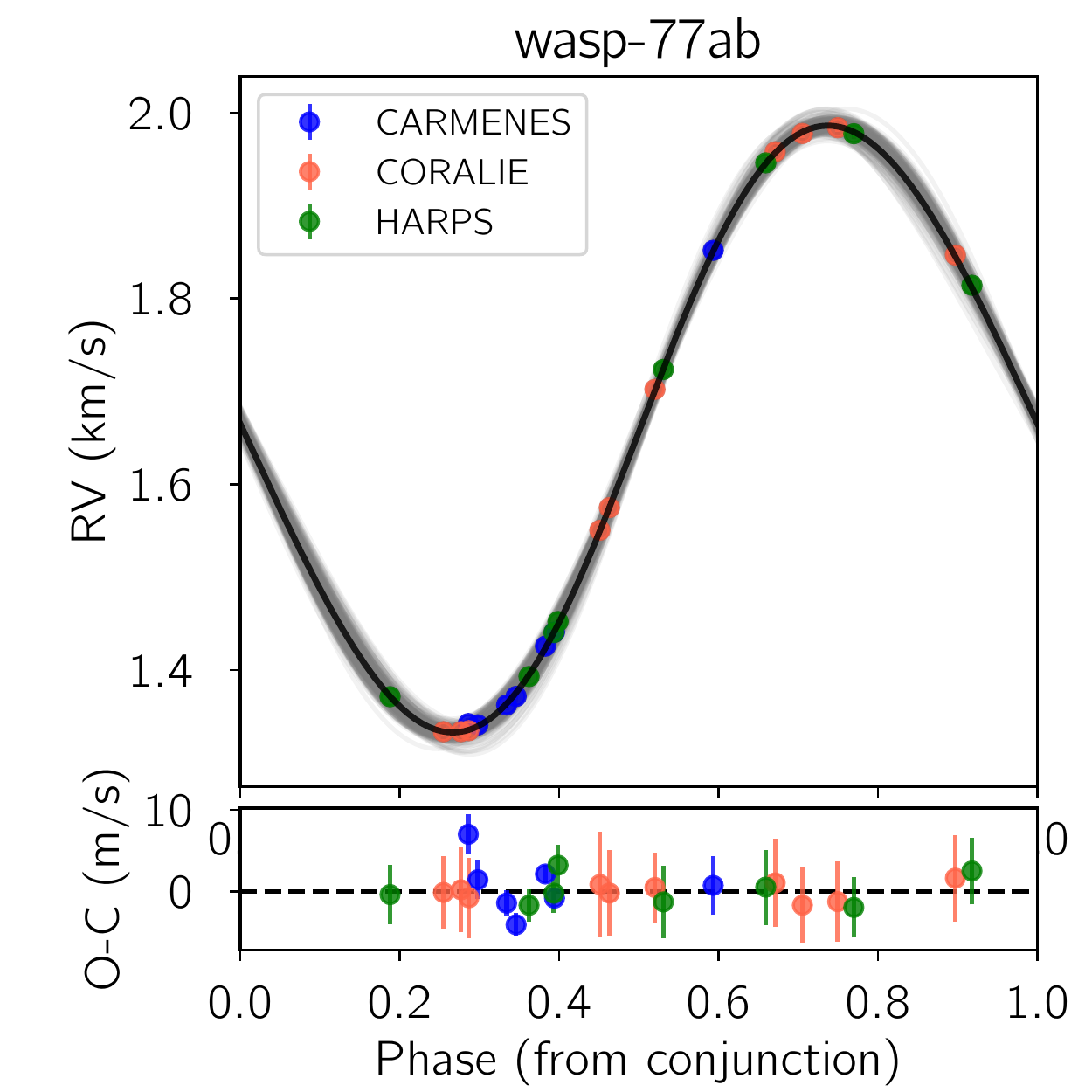}

\caption{Radial velocity analysis of the nine studied systems. The colors of the symbols represent the instrument used, with HARPSN being HARPS-N data from archive, HARPSNX and HARPSX being newly acquired data with HARPS-N and HARPS (respectively) in the context of \troy\ project. All CARMENES data were also obtained for this project.}
\label{fig:RV1}
\end{figure*}

% ========= LC fitting
%
\begin{figure*}
\includegraphics[width=0.48\textwidth]{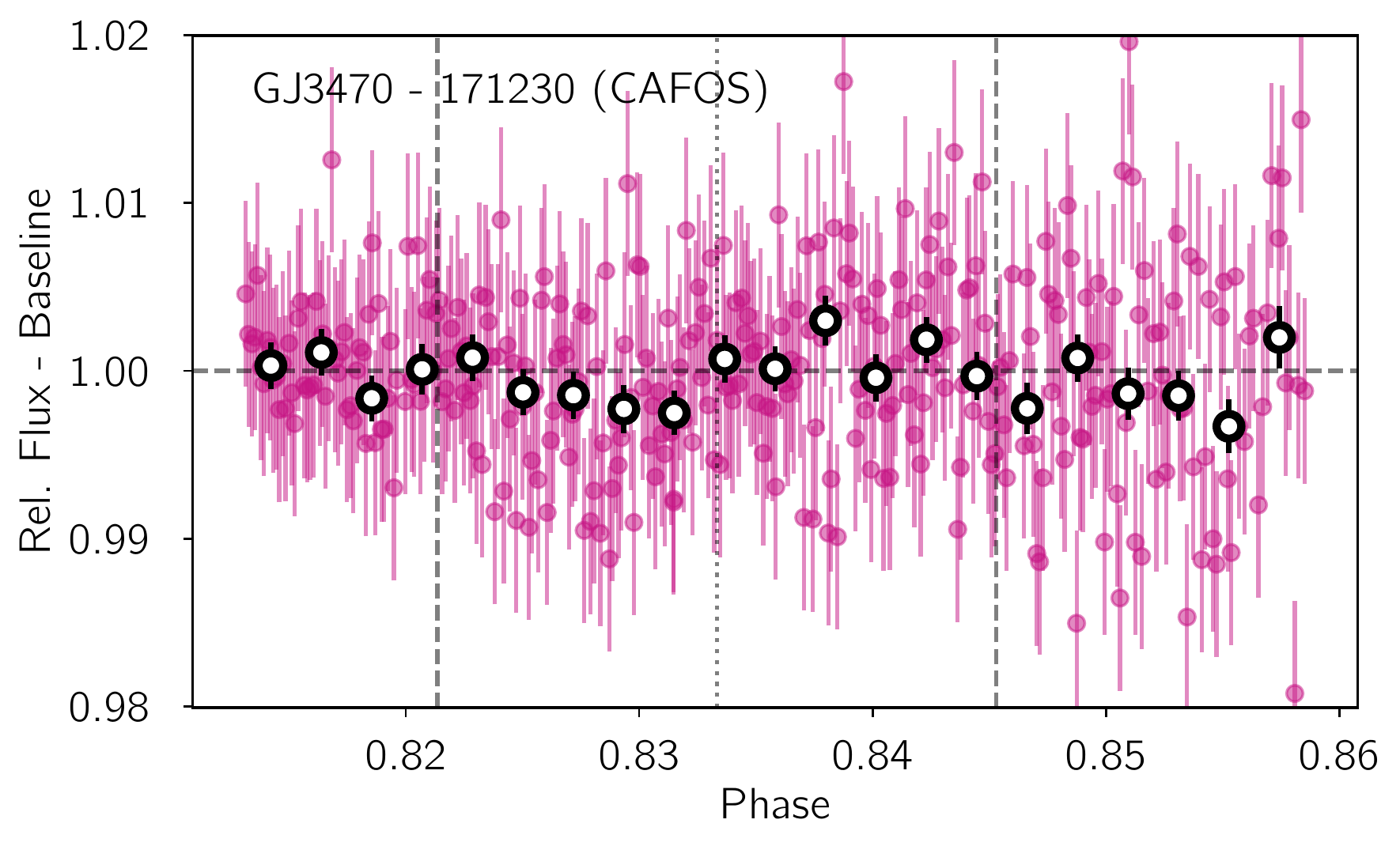} 
\includegraphics[width=0.48\textwidth]{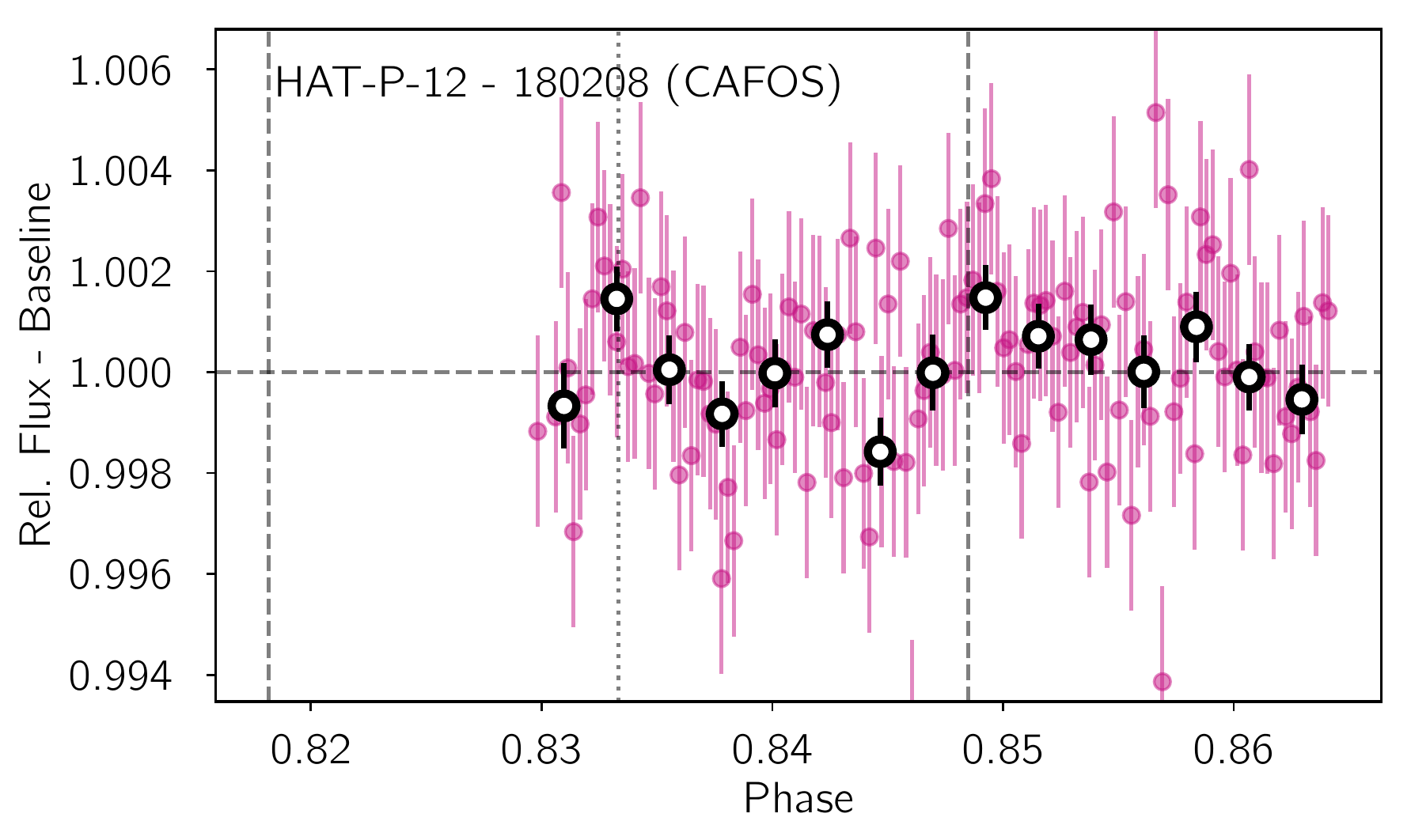} 

\includegraphics[width=0.48\textwidth]{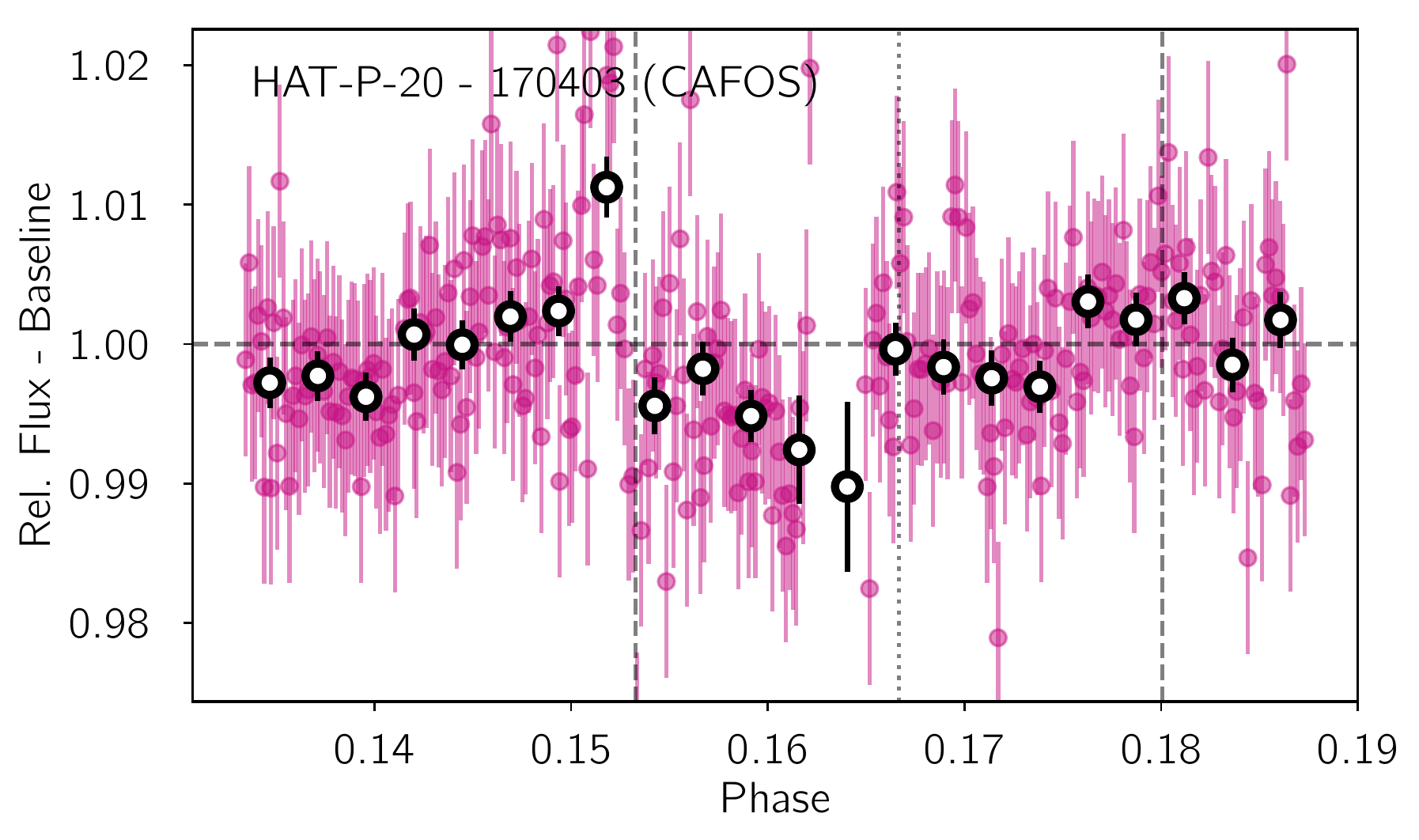} 
\includegraphics[width=0.48\textwidth]{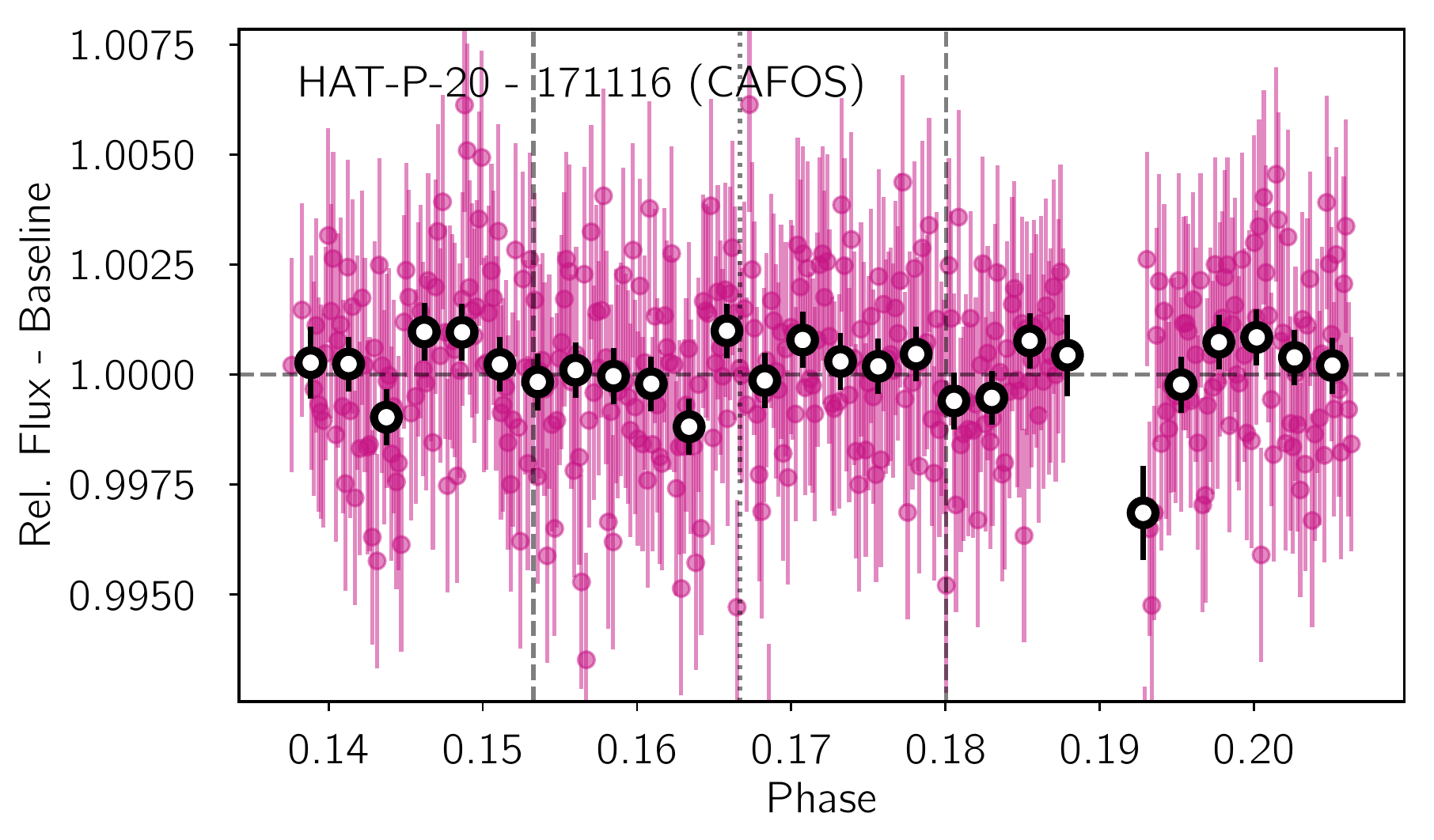} 

\includegraphics[width=0.48\textwidth]{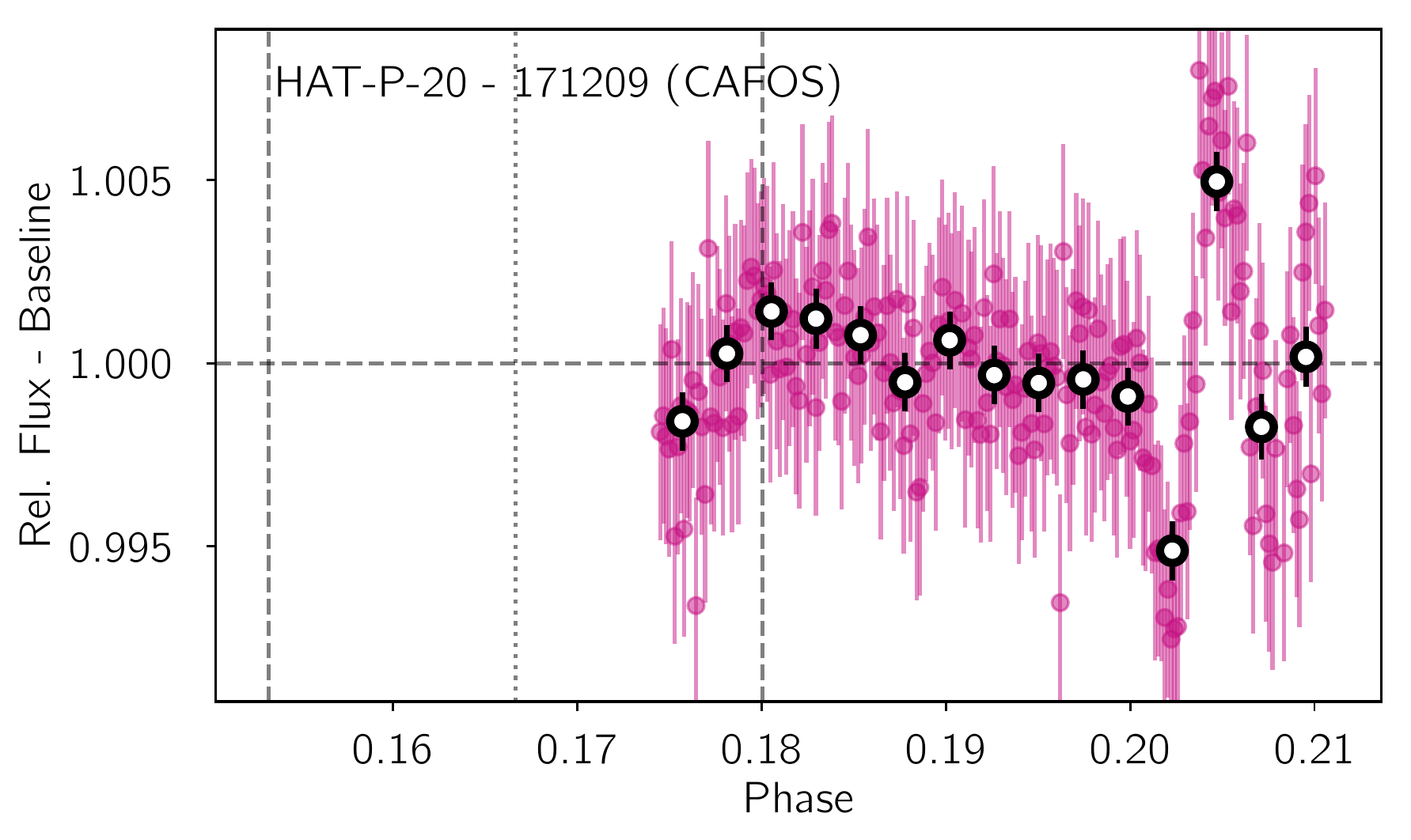} 
\includegraphics[width=0.48\textwidth]{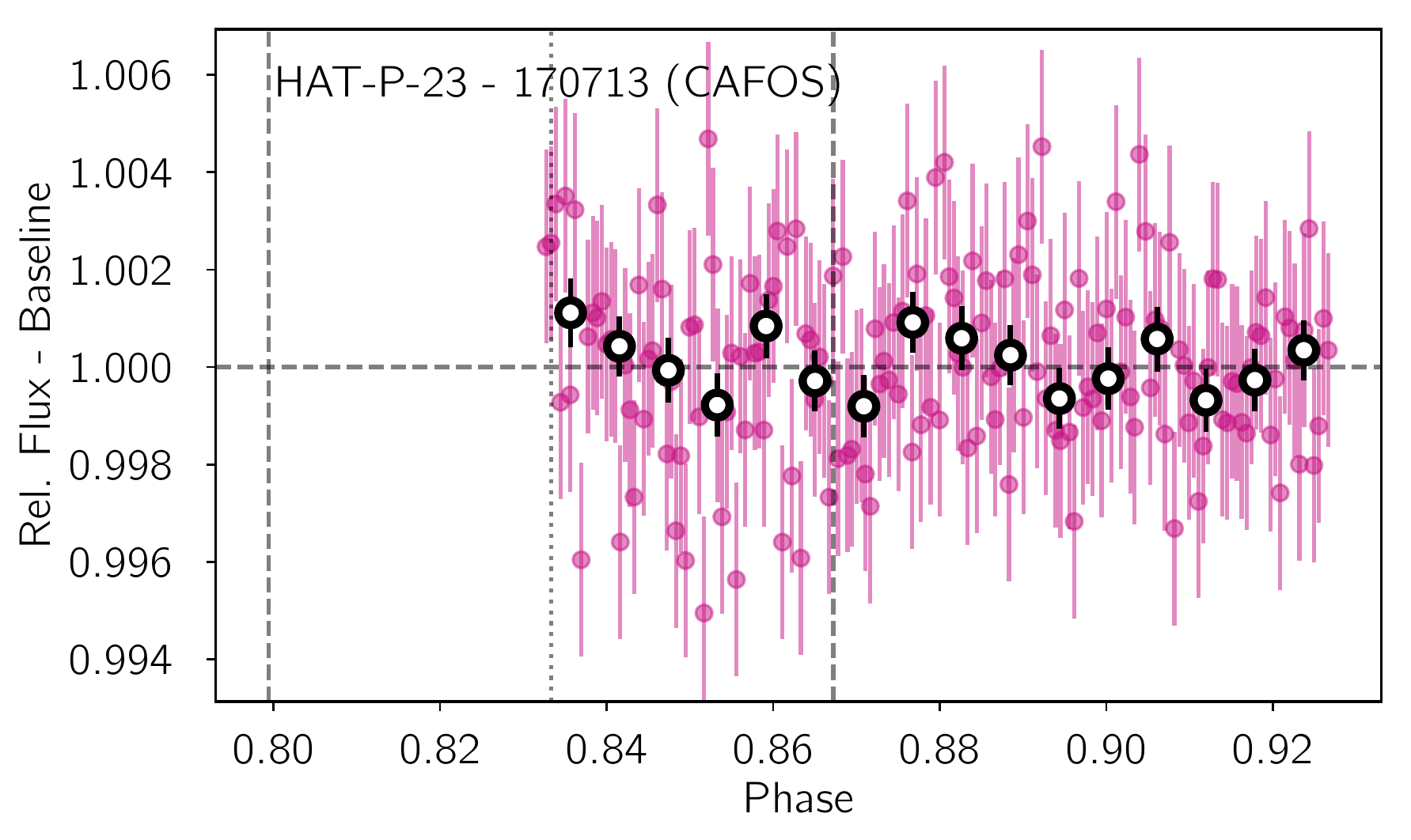} 

\includegraphics[width=0.48\textwidth]{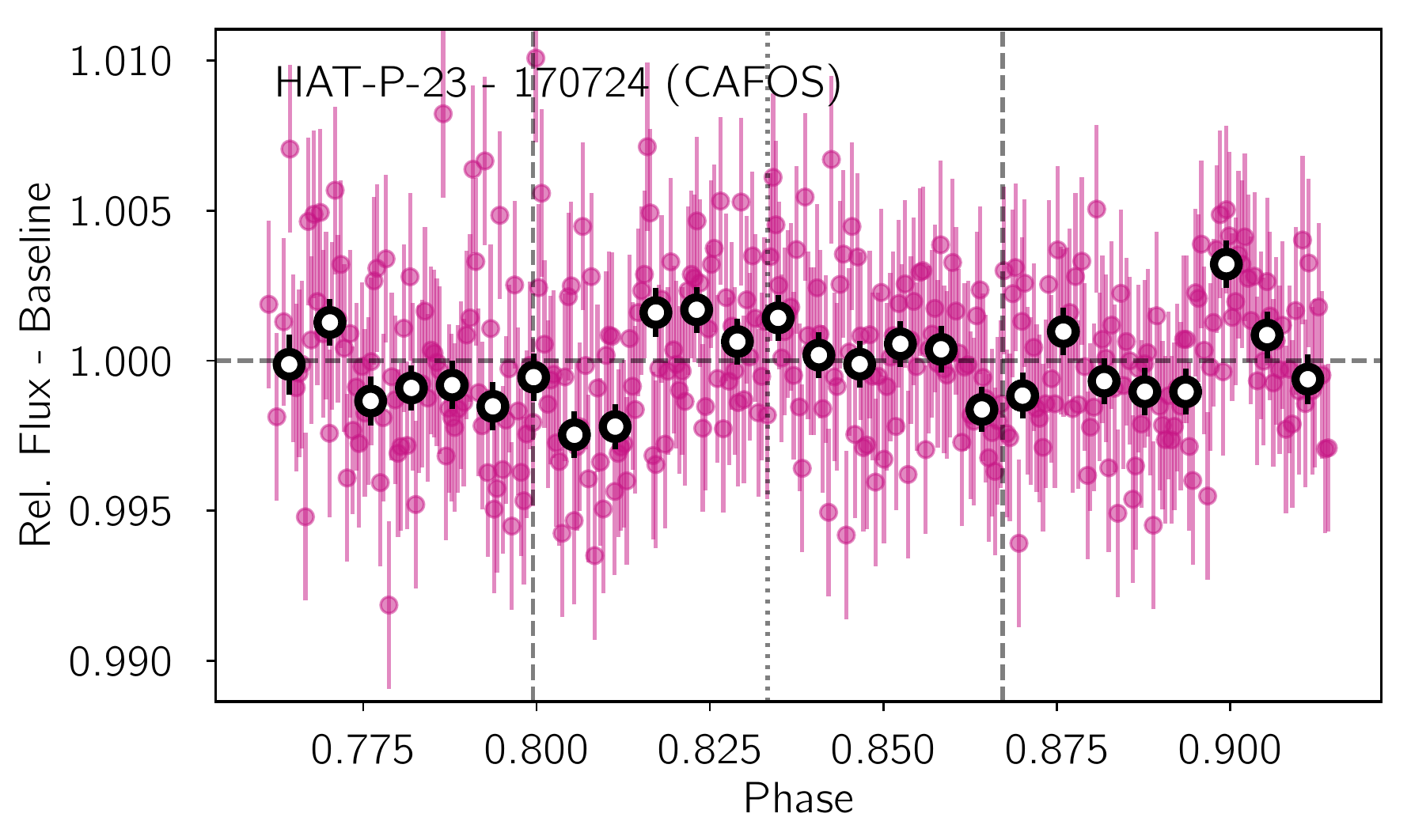} 
\includegraphics[width=0.48\textwidth]{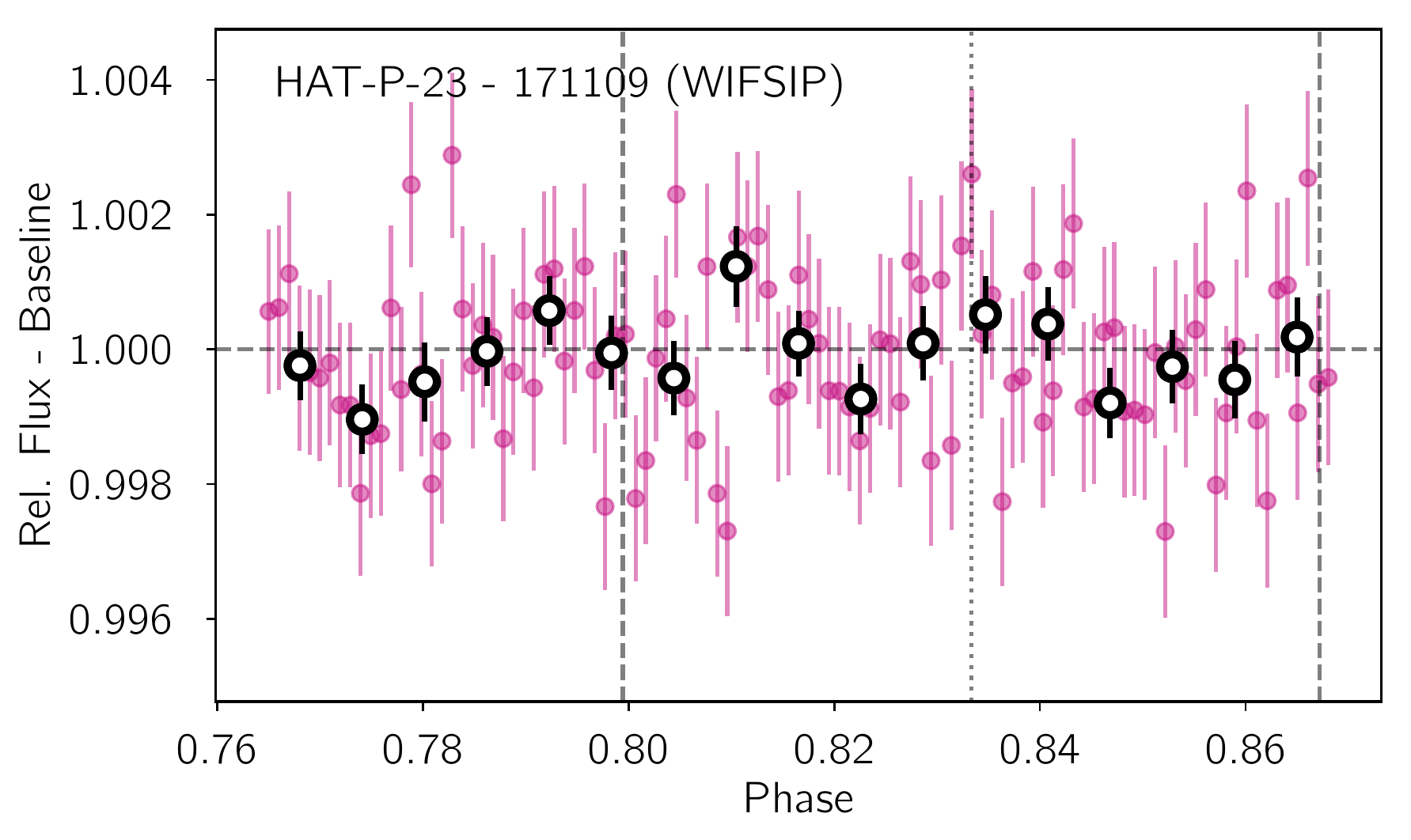}

\caption{Light curves of all Lagrangian point transits observed with ground-based facilities. A linear baseline model accounting for time, seeing, airmass, XY positions on the detector, and background has been removed. Purple data points are the individual measurements while big black open symbols represent 10-minute bins. The system and observing date (in YYMMDD format) are shown in each panel. The vertical dashed line indicates the mid-transit passage of the Lagrangian point and the dotted vertical lines indicate the total duration of the transit assuming the same as the planet.}
\label{fig:LCfitting1}
\end{figure*}

\begin{figure*}

\includegraphics[width=0.48\textwidth]{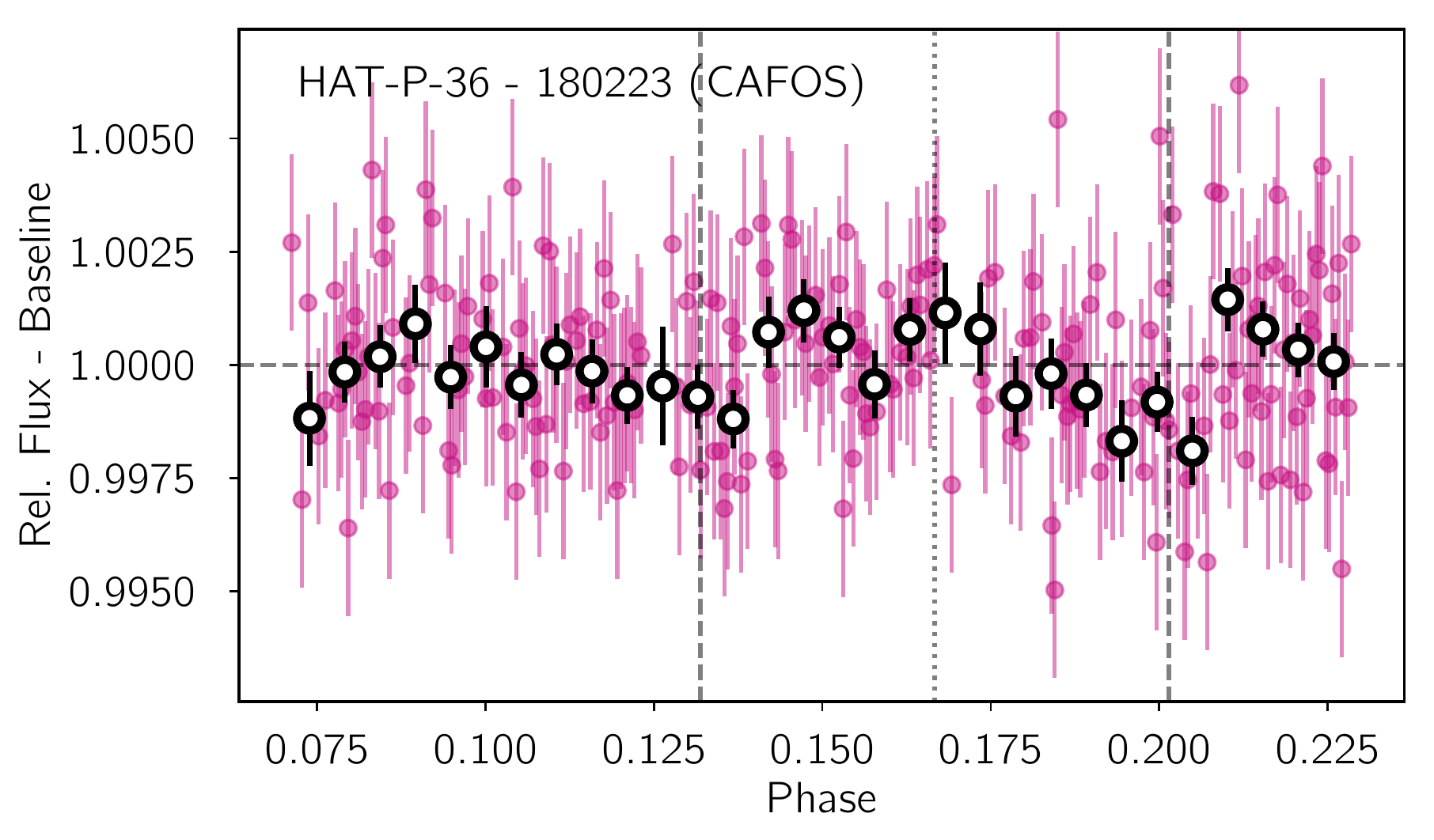} 
\includegraphics[width=0.48\textwidth]{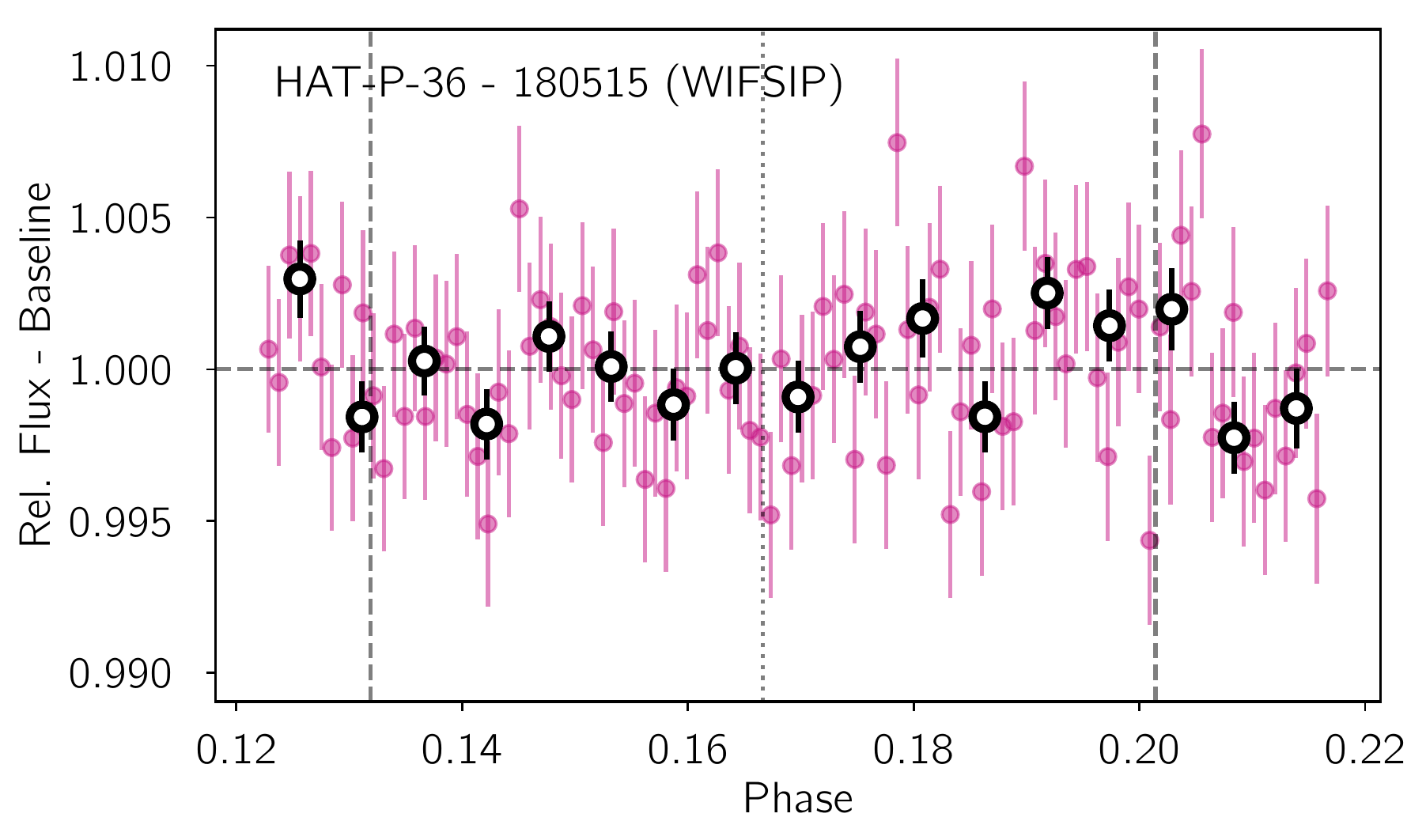} 

\includegraphics[width=0.48\textwidth]{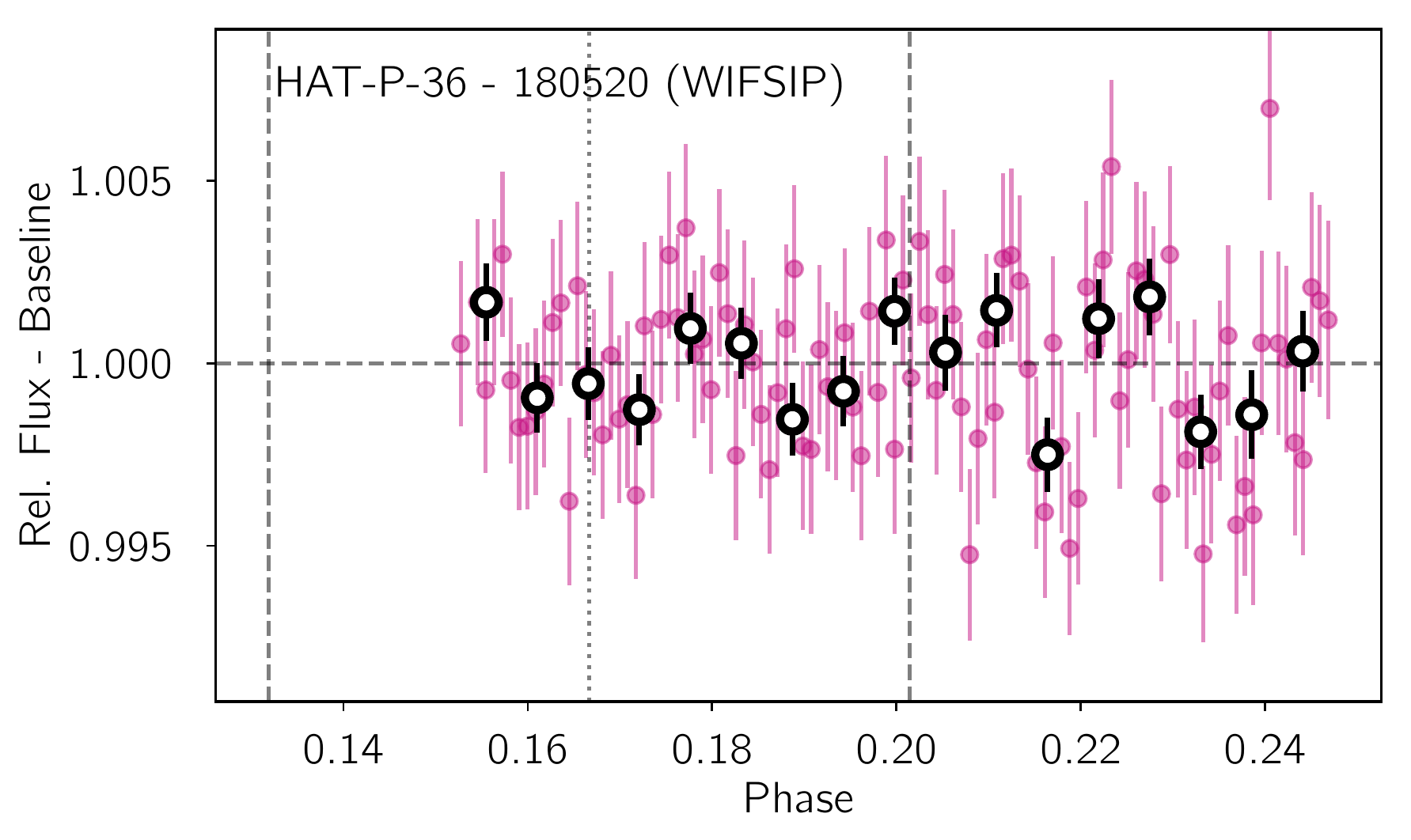} 
\includegraphics[width=0.48\textwidth]{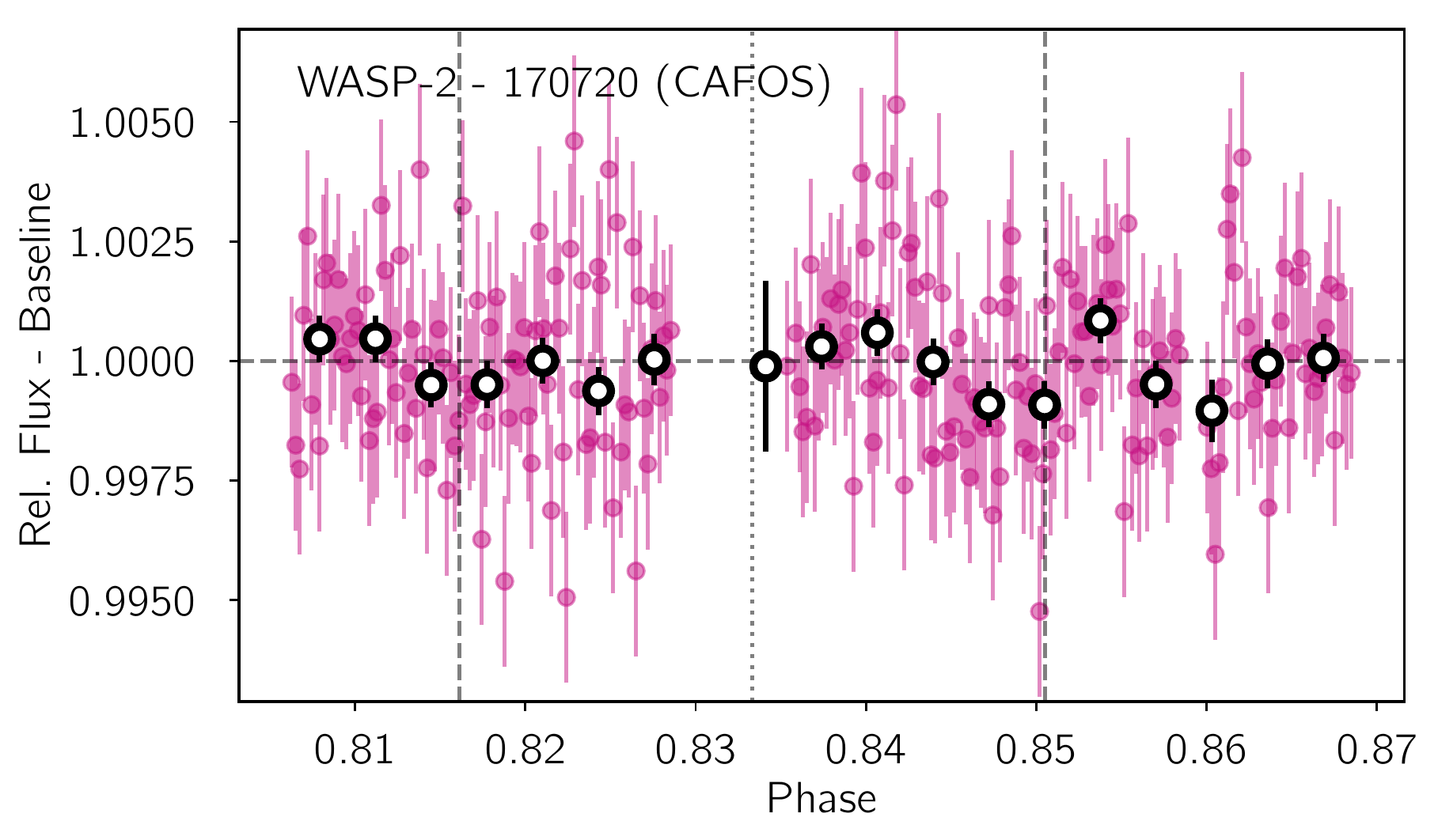} 

\includegraphics[width=0.48\textwidth]{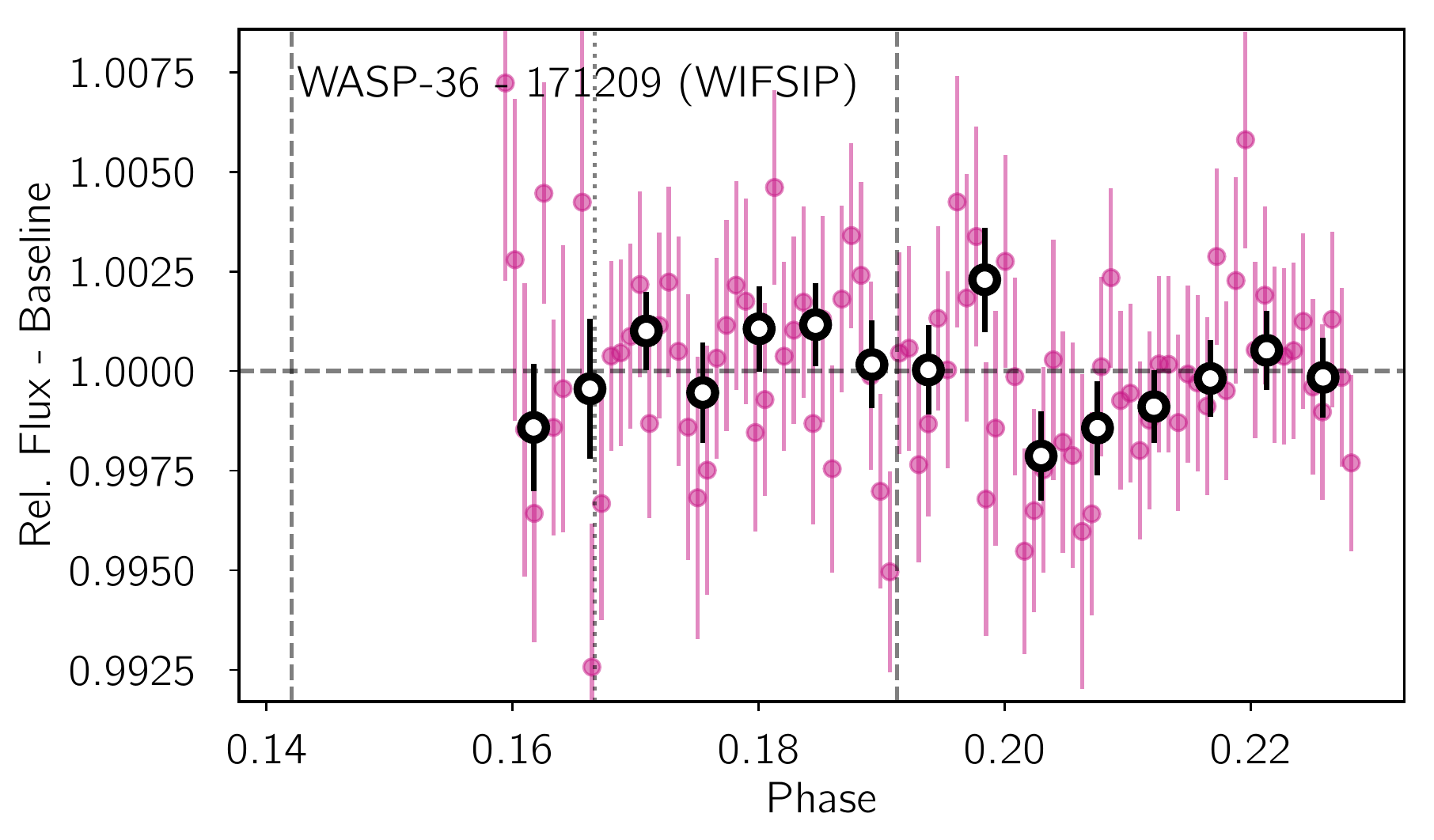}
\includegraphics[width=0.48\textwidth]{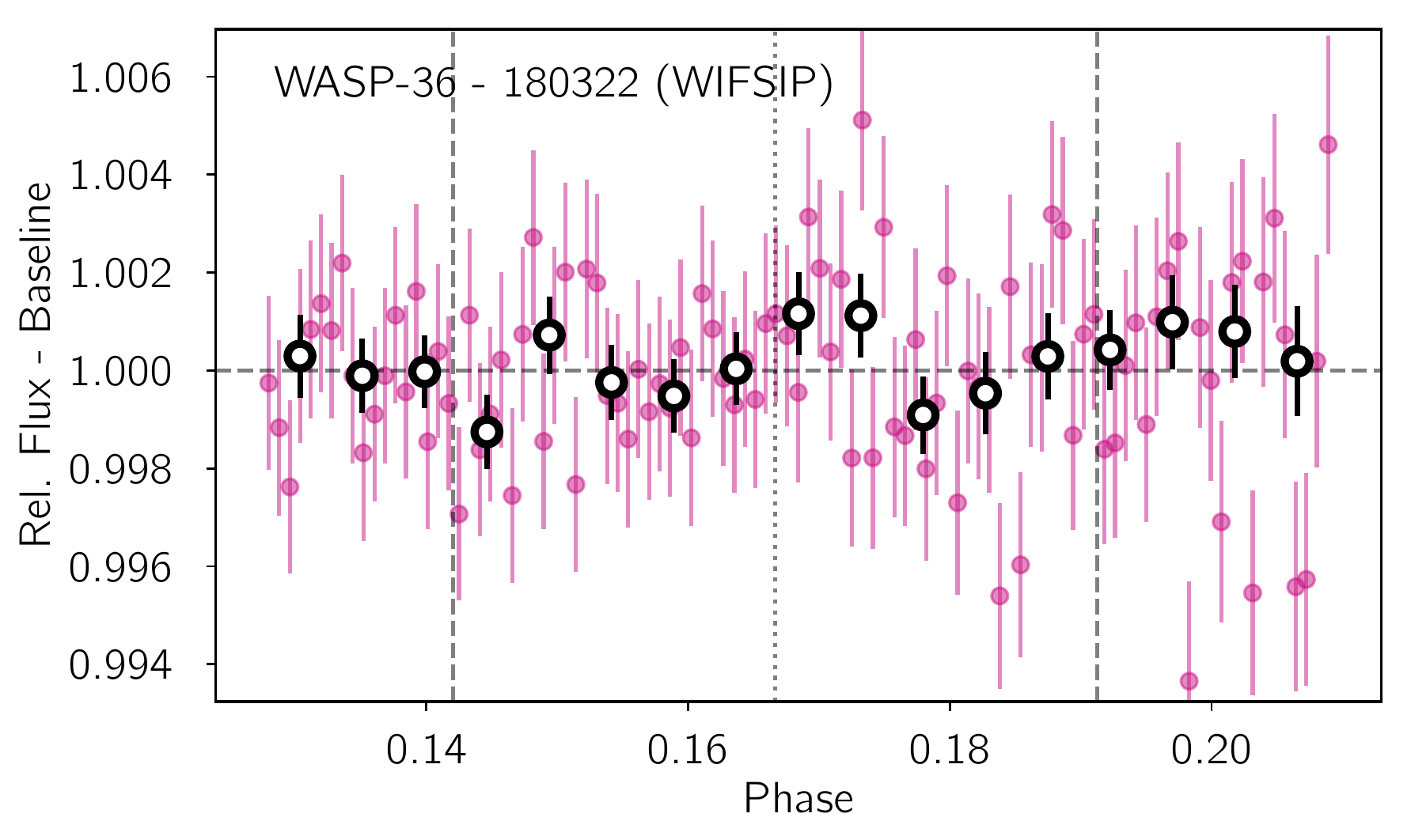}

\includegraphics[width=0.48\textwidth]{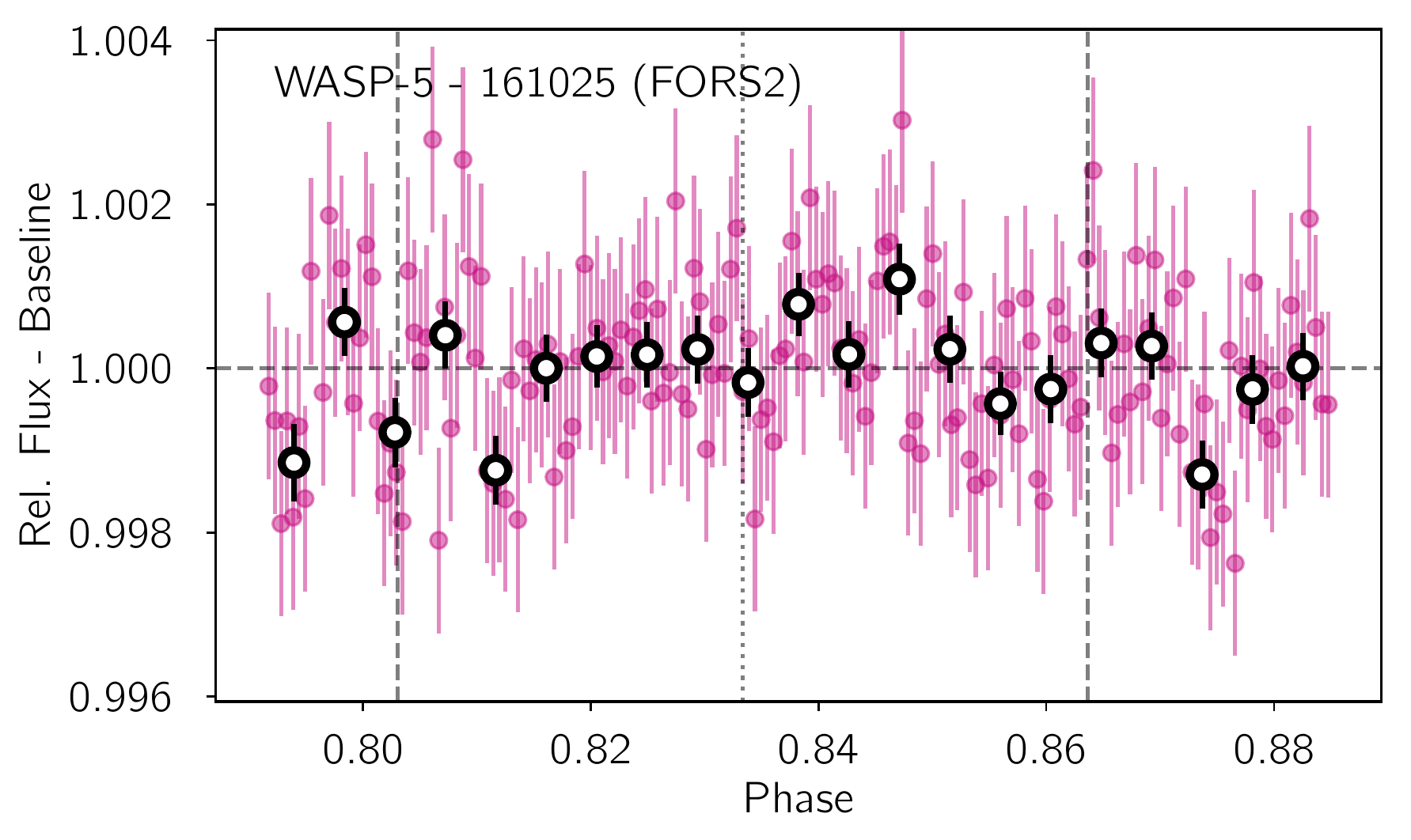}
\includegraphics[width=0.48\textwidth]{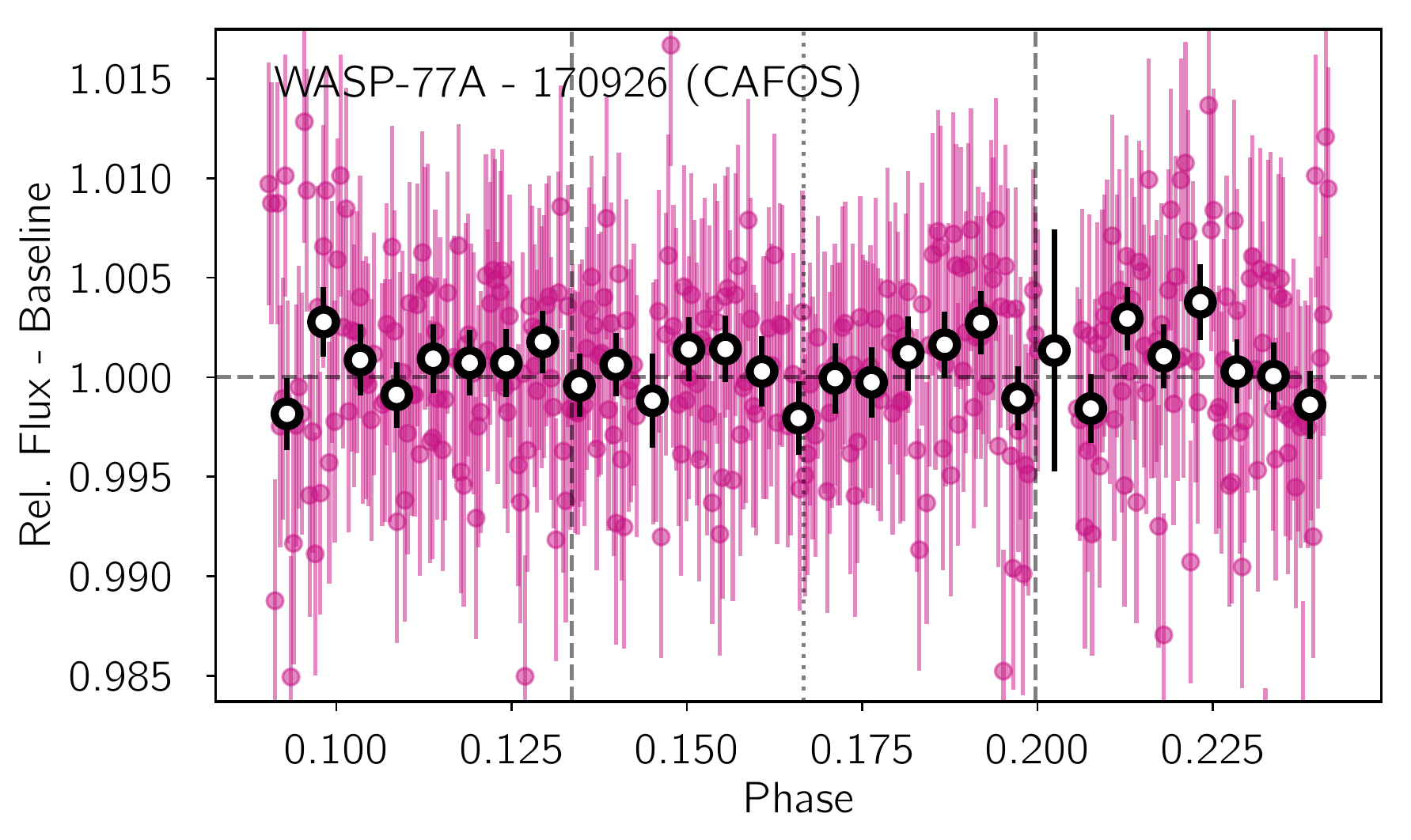}

\caption{Continuation of Fig~\ref{fig:LCfitting1}.}
\label{fig:LCfitting2}
\end{figure*}

\begin{figure*}

\includegraphics[width=0.48\textwidth]{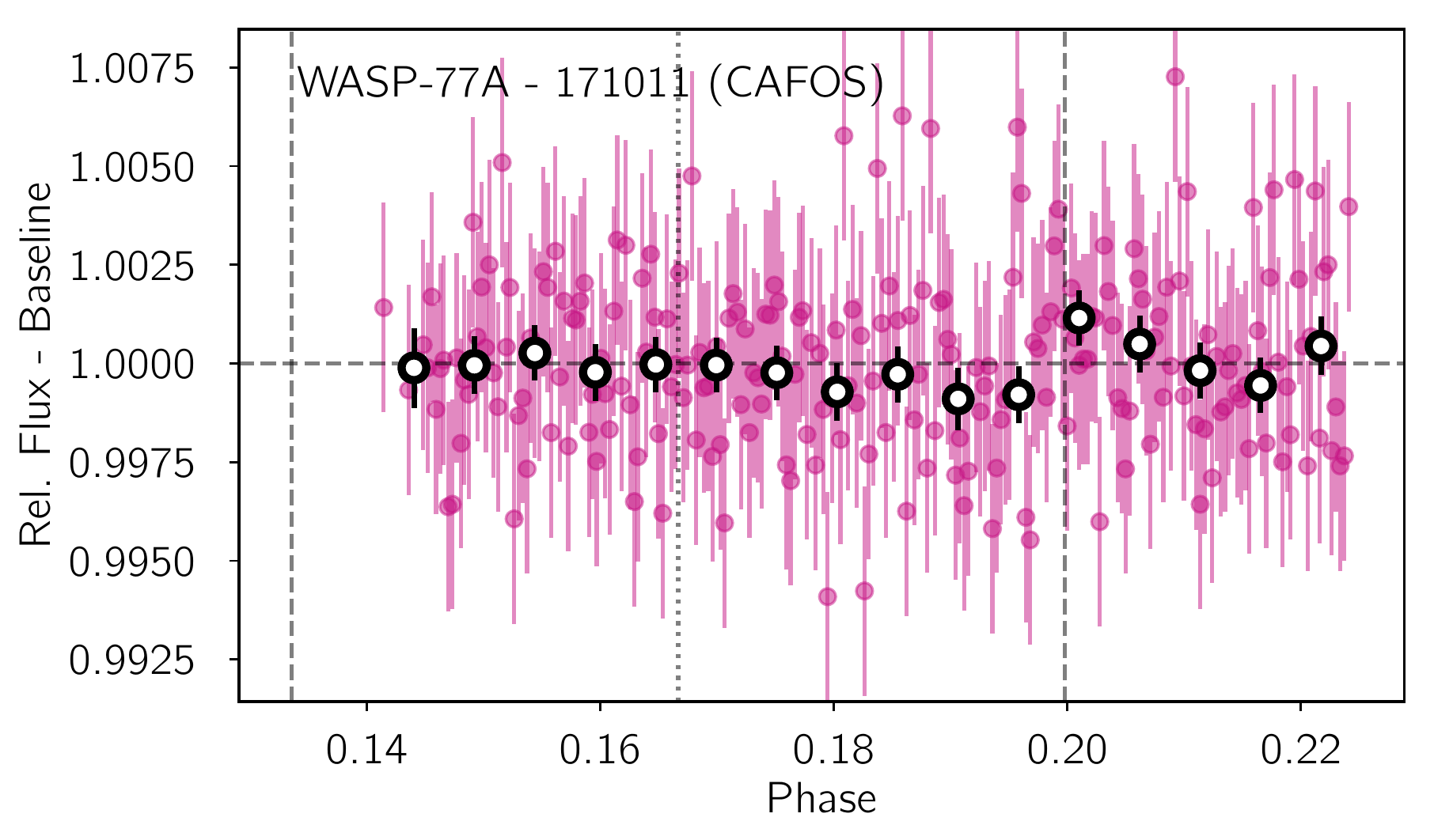}
\includegraphics[width=0.48\textwidth]{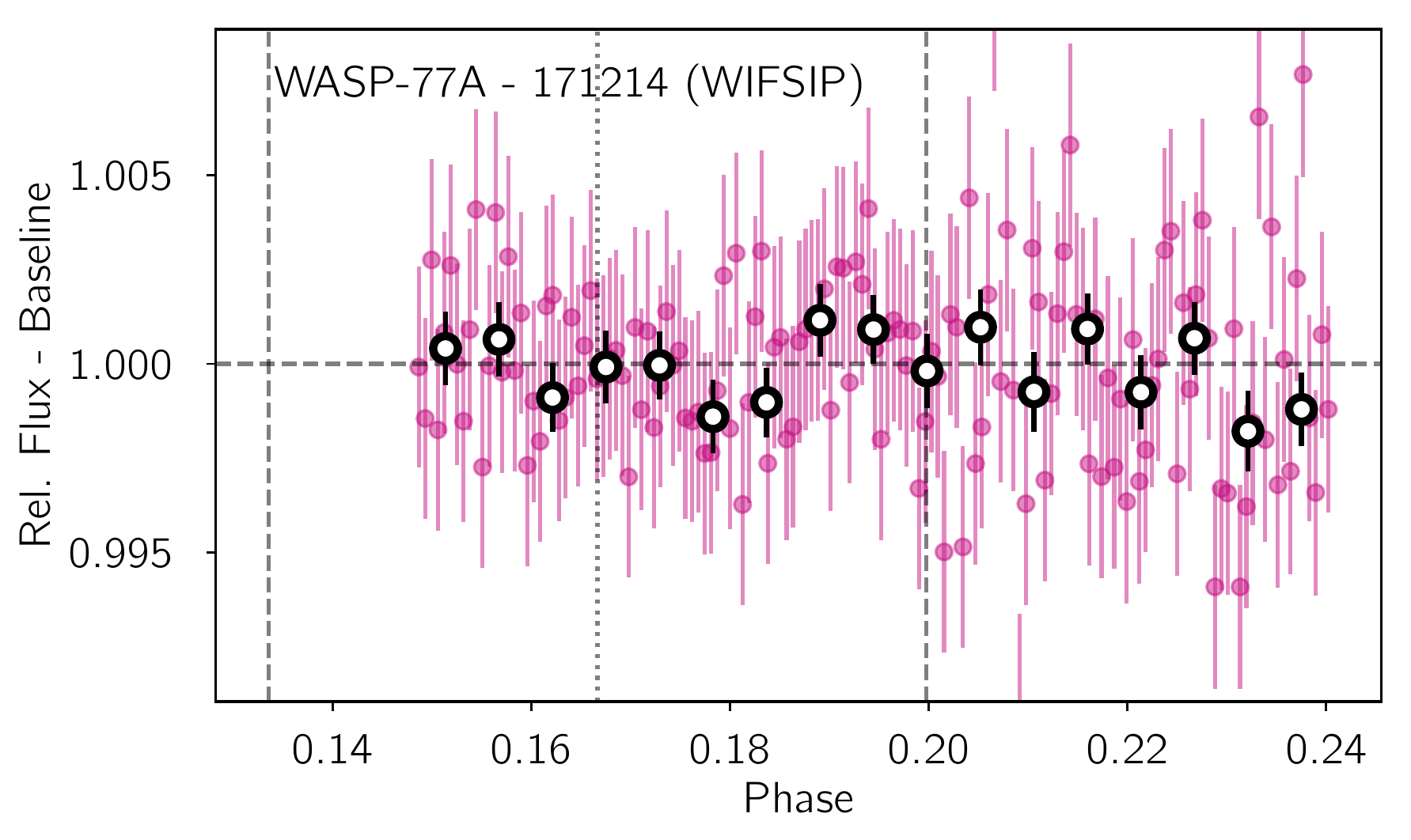}

\includegraphics[width=0.48\textwidth]{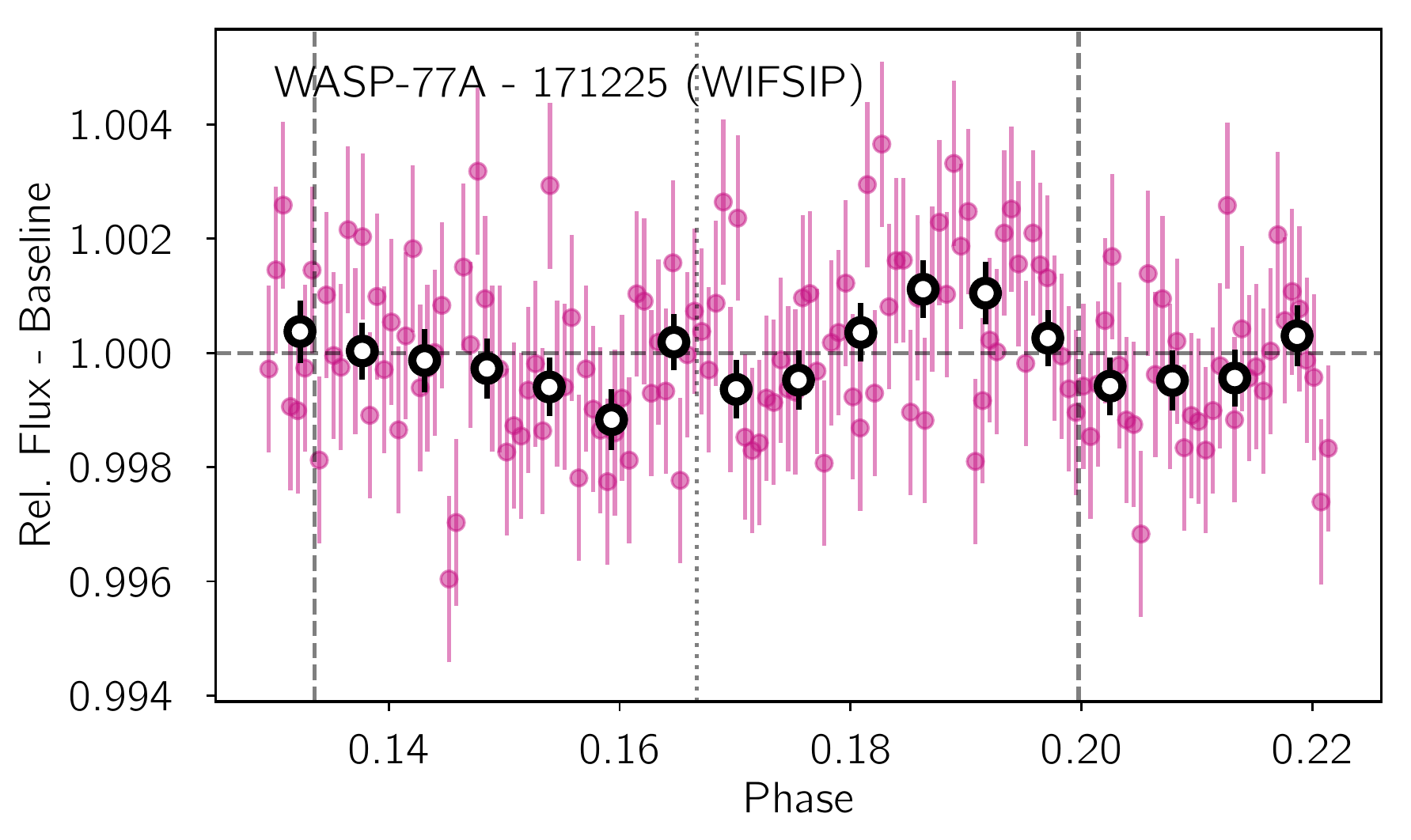}

\caption{Continuation of Fig~\ref{fig:LCfitting1}.}
\label{fig:LCfitting2}
\end{figure*}

% ========= Lib Ampl vs Trojan mass

\begin{figure*}[ht]
\centering
\includegraphics[width=0.32\textwidth]{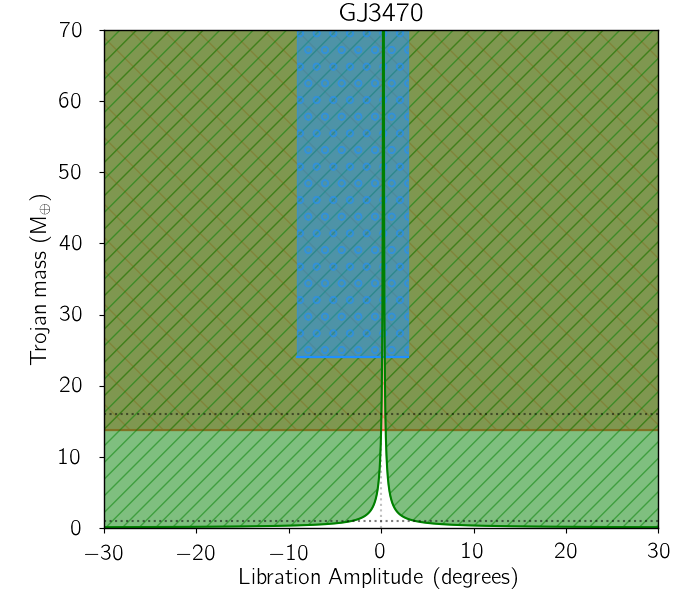}
\includegraphics[width=0.32\textwidth]{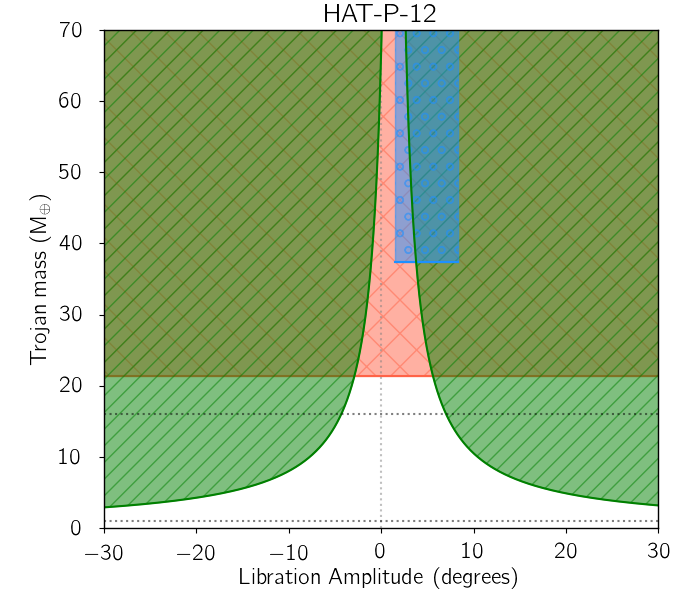}
\includegraphics[width=0.32\textwidth]{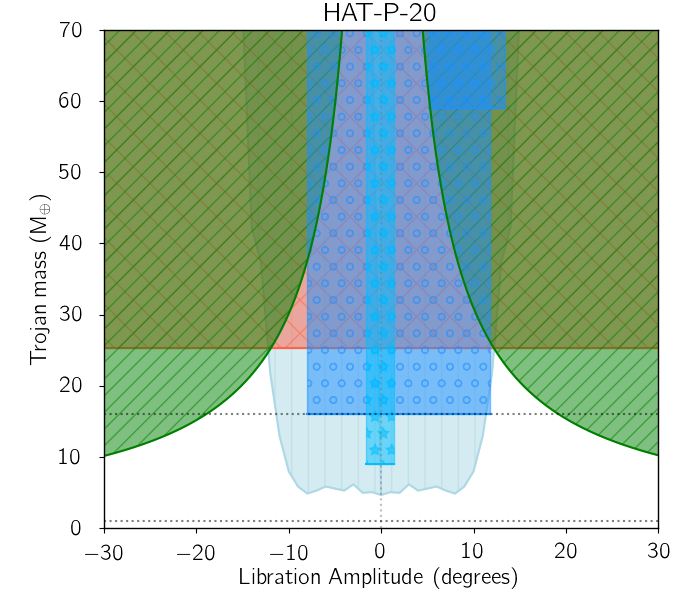}

\includegraphics[width=0.32\textwidth]{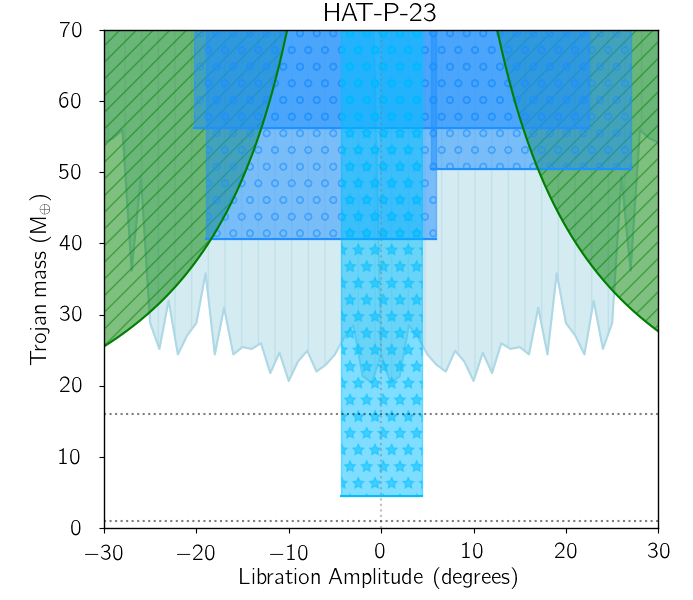}
\includegraphics[width=0.32\textwidth]{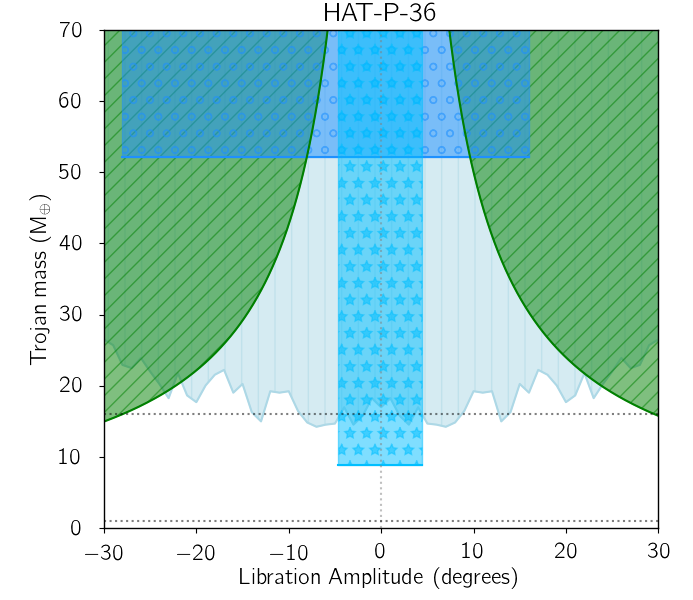}
\includegraphics[width=0.32\textwidth]{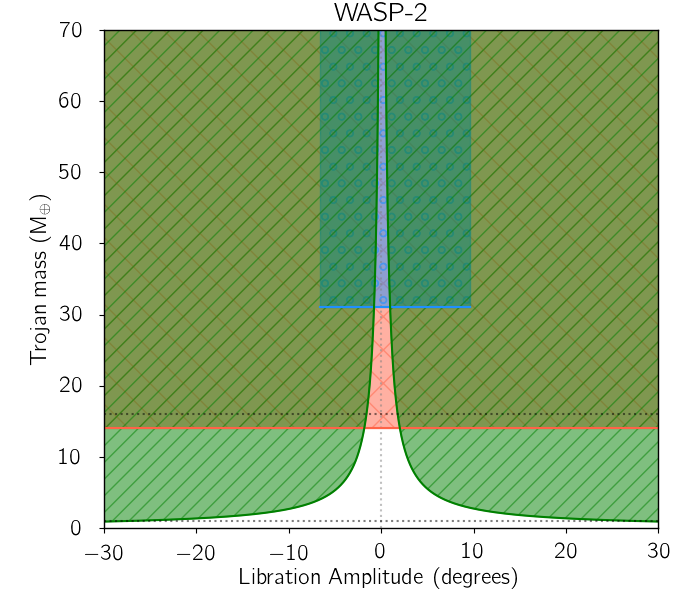}

\includegraphics[width=0.32\textwidth]{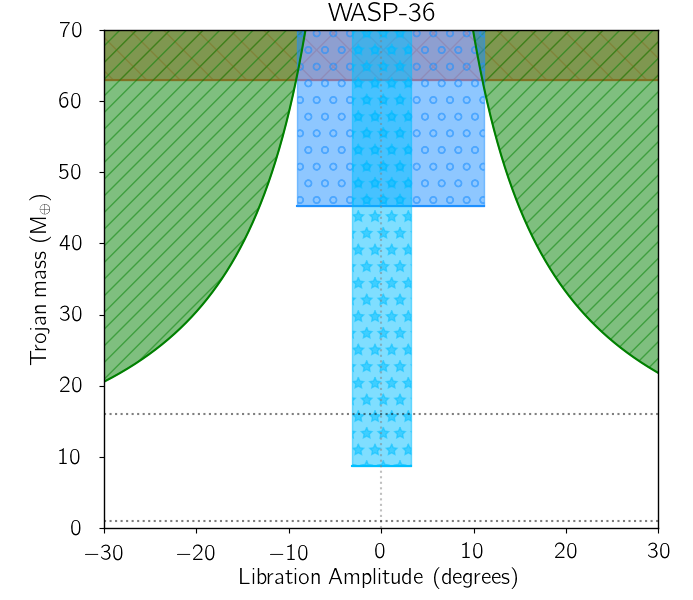}
\includegraphics[width=0.32\textwidth]{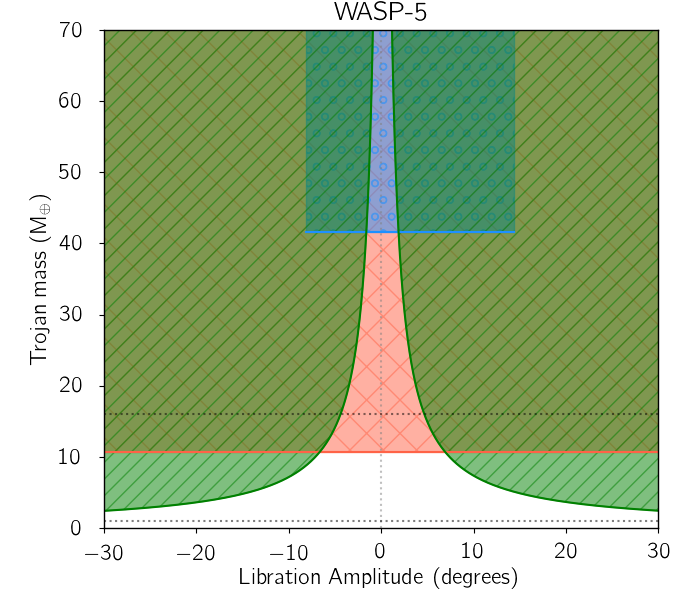}
\includegraphics[width=0.32\textwidth]{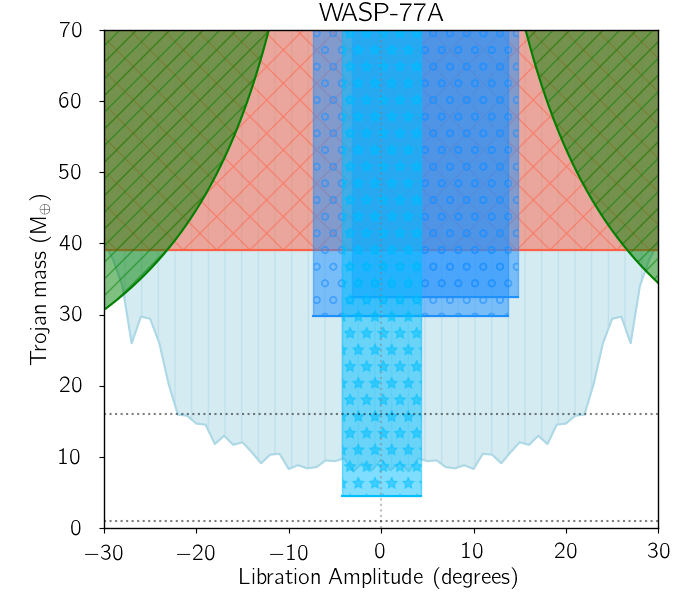}

\caption{Constrain of the trojan mass versus libration amplitude parameter space based on the three techniques used in this paper. The constrain from individual transits is shown as blue shaded regions with open circles, the constrain from the combined light curve is shown in light blue with open star symbols, the constrain from TTVs is shown as green diagonally striped shaded region, and the RV constrain is shown as a red vertically striped shaded region. The two horizontal dotted lines represent the Earth and Neptune and are shown to guide the eye.}
\label{fig:mtlib}
\end{figure*}

%----------------------------------------------------------------------------------------
%	TABLES
%----------------------------------------------------------------------------------------

\newpage

\section{Tables}
\label{app:tables}

\begin{table}
\scriptsize
\setlength{\extrarowheight}{5pt}
\caption{New radial velocities obtained for GJ\,3470. \label{tab:rv1}}
\begin{tabular}{c c c c }
%\hline
 Julian date & RV (km/s) & $\sigma_{\rm RV}$ (km/s) & Instrument \\
\hline \hline
2457791.50172 &         26.5240 &          0.0029 &    HARPS-N \\
2457791.58217 &         26.5266 &          0.0025 &    HARPS-N \\
2457792.36992 &         26.5200 &          0.0023 &    HARPS-N \\
2457792.55721 &         26.5118 &          0.0022 &    HARPS-N \\
2457793.40268 &         26.5123 &          0.0032 &    HARPS-N \\
2457793.53939 &         26.5186 &          0.0023 &    HARPS-N \\ \hline

\end{tabular}
\end{table}

% =====================================

\begin{table}
\scriptsize
\setlength{\extrarowheight}{5pt}
\caption{New radial velocities obtained for HAT-P-12. \label{tab:rv2}}
\begin{tabular}{c c c c }
%\hline
 Julian date & RV (km/s) & $\sigma_{\rm RV}$ (km/s) & Instrument \\
\hline \hline
2457784.64979 &        -41.1009 &          0.0026 &   CARMENES \\
2457784.75095 &        -41.0845 &          0.0023 &   CARMENES \\
2457785.63389 &        -41.0332 &          0.0045 &   CARMENES \\
2457785.73902 &        -41.0413 &          0.0039 &   CARMENES \\
2457785.76369 &        -41.0403 &          0.0037 &   CARMENES \\
2457791.66003 &        -40.4358 &          0.0067 &   HARPS-N \\
2457791.72821 &        -40.4217 &          0.0065 &   HARPS-N \\
2457792.67559 &        -40.4414 &          0.0051 &   HARPS-N \\
2457792.76510 &        -40.4491 &          0.0039 &   HARPS-N \\
2457793.68221 &        -40.4957 &          0.0050 &   HARPS-N \\
2457793.77300 &        -40.4957 &          0.0044 &   HARPS-N \\ \hline
\end{tabular}
\end{table}

% =====================================

\begin{table}
\scriptsize
\setlength{\extrarowheight}{5pt}
\caption{New radial velocities obtained for HAT-P-20. \label{tab:rv3}}
\begin{tabular}{l c c c c }
%\hline
 Julian date & RV (km/s) & $\sigma_{\rm RV}$ (km/s) & Instrument \\
\hline \hline
2457778.73074 &        -19.3379 &          0.0030 &     HARPS \\
2457779.62713 &        -17.5785 &          0.0021 &     HARPS \\
2457779.66825 &        -17.4749 &          0.0020 &     HARPS \\
2457779.70548 &        -17.3942 &          0.0029 &     HARPS \\
2457783.29928 &        -17.3966 &          0.0071 &   CARMENES \\
2457783.33735 &        -17.4780 &          0.0028 &   CARMENES \\
2457783.37978 &        -17.6001 &          0.0017 &   CARMENES \\
2457784.29131 &        -19.8829 &          0.0014 &   CARMENES \\
2457784.33715 &        -19.9465 &          0.0017 &   CARMENES \\
2457784.35189 &        -19.9118 &          0.0017 &   CARMENES \\
2457784.36669 &        -19.9133 &          0.0019 &   CARMENES \\
2457784.42444 &        -19.9156 &          0.0016 &   CARMENES \\
2457784.47847 &        -19.9112 &          0.0011 &   CARMENES \\
2457785.29185 &        -18.1693 &          0.0021 &   CARMENES \\
2457785.33989 &        -18.0342 &          0.0019 &   CARMENES \\
2457785.39013 &        -17.9107 &          0.0026 &   CARMENES \\
2457785.47954 &        -17.7162 &          0.0011 &   CARMENES \\
2457785.54400 &        -17.5504 &          0.0018 &   CARMENES \\
2457785.56634 &        -17.4910 &          0.0014 &   CARMENES \\
2457791.35401 &        -17.0477 &          0.0024 &   HARPS-N \\
2457791.37720 &        -17.0165 &          0.0025 &   HARPS-N \\
2457791.41835 &        -16.9499 &          0.0025 &   HARPS-N \\
2457791.46151 &        -16.8990 &          0.0020 &   HARPS-N \\
2457791.52634 &        -16.8384 &          0.0025 &   HARPS-N \\
2457792.40987 &        -18.2047 &          0.0017 &   HARPS-N \\
2457792.45106 &        -18.3115 &          0.0019 &   HARPS-N \\
2457792.47385 &        -18.3708 &          0.0021 &   HARPS-N \\
2457792.61402 &        -18.7004 &          0.0024 &   HARPS-N \\ \hline
\end{tabular}
\end{table}

% =====================================

\begin{table}
\scriptsize
\setlength{\extrarowheight}{5pt}
\caption{New radial velocities obtained for HAT-P-36. \label{tab:rv4}}
\begin{tabular}{l c c c c }
%\hline
 Julian date & RV (km/s) & $\sigma_{\rm RV}$ (km/s) & Instrument \\
\hline \hline
2457784.62650 &        -17.0151 &          0.0024 &   CARMENES \\
2457784.67574 &        -17.0888 &          0.0020 &   CARMENES \\
2457784.72818 &        -17.1169 &          0.0021 &   CARMENES \\
2457785.65525 &        -16.5881 &          0.0025 &   CARMENES \\
2457785.67649 &        -16.6234 &          0.0028 &   CARMENES \\
2457785.69869 &        -16.6133 &          0.0037 &   CARMENES \\
2457785.71784 &        -16.6445 &          0.0038 &   CARMENES \\
2457791.63596 &        -16.5054 &          0.0052 &   HARPS-N \\
2457791.70608 &        -16.4135 &          0.0057 &   HARPS-N \\
2457791.76242 &        -16.3341 &          0.0058 &   HARPS-N \\
2457792.72138 &        -16.5978 &          0.0037 &   HARPS-N \\
2457793.62341 &        -16.0329 &          0.0032 &   HARPS-N \\
2457793.70754 &        -16.1395 &          0.0035 &   HARPS-N \\
2457793.75023 &        -16.2006 &          0.0028 &   HARPS-N \\ \hline
\end{tabular}
\end{table}

% =====================================

\begin{table}
\scriptsize
\setlength{\extrarowheight}{5pt}
\caption{New radial velocities obtained for WASP-5.\label{tab:rv5}}
\begin{tabular}{l c c c c }
%\hline
 Julian date & RV (km/s) & $\sigma_{\rm RV}$ (km/s) & Instrument \\
\hline \hline
2457673.57366 &         19.7310 &          0.0051 &     HARPS \\
2457673.59153 &         19.7426 &          0.0042 &     HARPS \\
2457673.61023 &         19.7280 &          0.0040 &     HARPS \\
2457673.62602 &         19.7517 &          0.0029 &     HARPS \\
2457673.64561 &         19.7499 &          0.0028 &     HARPS \\
2457709.53130 &         19.7889 &          0.0034 &     HARPS \\
2457709.58860 &         19.8153 &          0.0029 &     HARPS \\
2457709.68416 &         19.8803 &          0.0036 &     HARPS \\
2457710.54488 &         20.1298 &          0.0021 &     HARPS \\
2457710.56189 &         20.1091 &          0.0022 &     HARPS \\
2457710.58478 &         20.0869 &          0.0020 &     HARPS \\
2457710.60295 &         20.0778 &          0.0022 &     HARPS \\
2457765.52618 &         20.2705 &          0.0040 &     HARPS \\
2457765.55328 &         20.2697 &          0.0038 &     HARPS \\
2457765.58348 &         20.2734 &          0.0054 &     HARPS \\
2457766.52305 &         19.7537 &          0.0033 &     HARPS \\
2457766.56604 &         19.7764 &          0.0040 &     HARPS \\
2457766.59011 &         19.7991 &          0.0043 &     HARPS \\ \hline
\end{tabular}
\end{table}

% =====================================

\begin{table}
\scriptsize
\setlength{\extrarowheight}{5pt}
\caption{New radial velocities obtained for WASP-77\,A. \label{tab:rv6} }
\begin{tabular}{l c c c c }
%\hline
 Julian date & RV (km/s) & $\sigma_{\rm RV}$ (km/s) & Instrument \\
\hline \hline
2458100.35100 &          1.3746 &          0.0036 &   CARMENES \\ 
2458101.29327 &          0.8239 &          0.0024 &   CARMENES \\
2458101.30958 &          0.8209 &          0.0023 &   CARMENES \\
2458101.35856 &          0.8380 &          0.0016 &   CARMENES \\
2458101.37478 &          0.8458 &          0.0014 &   CARMENES \\
2458101.42477 &          0.8961 &          0.0011 &   CARMENES \\
2458101.44026 &          0.9101 &          0.0013 &   CARMENES \\
           
\hline
\end{tabular}
\end{table}% =====================================

\begin{table}
\scriptsize
\setlength{\extrarowheight}{5pt}
\caption{New radial velocities obtained for WASP-36. \label{tab:rv7}}
\begin{tabular}{l c c c c }
%\hline
 Julian date & RV (km/s) & $\sigma_{\rm RV}$ (km/s) & Instrument \\
\hline \hline
2457778.76807 &        -12.8833 &          0.0107 &     HARPS \\ 
2457778.80797 &        -12.9182 &          0.0068 &     HARPS \\
2457778.85385 &        -12.9576 &          0.0069 &     HARPS \\
2457779.60030 &        -13.4827 &          0.0061 &     HARPS \\
2457779.73384 &        -13.3094 &          0.0088 &     HARPS \\
2457779.77421 &        -13.2352 &          0.0070 &     HARPS \\
2457779.82374 &        -13.1700 &          0.0108 &     HARPS \\
2457779.86853 &        -13.0910 &          0.0177 &     HARPS \\
2457783.44083 &        -13.5112 &          0.0146 &   CARMENES \\
2457783.45054 &        -13.4601 &          0.0542 &   CARMENES \\
2457784.44115 &        -13.7575 &          0.0061 &   CARMENES \\
2457784.45624 &        -13.7384 &          0.0057 &   CARMENES \\
2457784.53293 &        -13.5657 &          0.0051 &   CARMENES \\
2457784.55202 &        -13.5744 &          0.0057 &   CARMENES \\
2457784.59006 &        -13.5012 &          0.0103 &   CARMENES \\
2457784.60601 &        -13.5290 &          0.0120 &   CARMENES \\
2457785.41944 &        -14.0861 &          0.0106 &   CARMENES \\
2457785.52423 &        -14.1569 &          0.0052 &   CARMENES \\
2457785.59082 &        -14.1437 &          0.0071 &   CARMENES \\
2457791.55299 &        -13.4939 &          0.0096 &   HARPS-N \\
2457791.60835 &        -13.5219 &          0.0097 &   HARPS-N \\
2457792.50103 &        -12.8269 &          0.0059 &   HARPS-N \\
2457792.53141 &        -12.8277 &          0.0062 &   HARPS-N \\
2457792.57894 &        -12.8443 &          0.0102 &   HARPS-N \\
2457792.59412 &        -12.8832 &          0.0105 &   HARPS-N \\
2457792.64469 &        -12.9135 &          0.0087 &   HARPS-N \\
2457793.44677 &        -13.4685 &          0.0083 &   HARPS-N \\
2457793.50910 &        -13.4019 &          0.0060 &   HARPS-N \\
2457793.56663 &        -13.3037 &          0.0065 &   HARPS-N \\
2457793.59534 &        -13.2685 &          0.0060 &   HARPS-N \\
2457793.65484 &        -13.1750 &          0.0081 &   HARPS-N \\
2458100.54522 &        -14.0508 &          0.0137 &   CARMENES \\
2458100.56529 &        -14.1231 &          0.0129 &   CARMENES \\
2458100.67273 &        -14.1769 &          0.0106 &   CARMENES \\
2458100.68894 &        -14.1711 &          0.0117 &   CARMENES \\
2458100.73091 &        -14.2054 &          0.0104 &   CARMENES \\
2458100.74677 &        -14.2196 &          0.0101 &   CARMENES \\
2458101.54573 &        -13.4606 &          0.0078 &   CARMENES \\
2458101.56070 &        -13.4320 &          0.0078 &   CARMENES \\
2458101.66642 &        -13.5792 &          0.0060 &   CARMENES \\
2458101.68185 &        -13.5828 &          0.0054 &   CARMENES \\
2458101.69731 &        -13.6015 &          0.0053 &   CARMENES \\
2458101.71265 &        -13.6024 &          0.0060 &   CARMENES \\
2458101.75226 &        -13.6371 &          0.0069 &   CARMENES \\ \hline
\end{tabular}
\end{table}

\begin{table*}
\small
\setlength{\extrarowheight}{5pt}
\caption{Summary of phtometric data obtained for the 10 systems analyzed in this paper. \label{tab:photometry}}
\begin{tabular}{|l c c l c c c c c c| }
\hline
  \multicolumn{1}{|l}{System} &
  \multicolumn{1}{c}{Date} &
  \multicolumn{1}{c}{Instrument} &
%  \multicolumn{1}{c}{Defocus.} &
  \multicolumn{1}{l}{Filter} &
  \multicolumn{1}{c}{Lpoint} &
  \multicolumn{1}{c}{Coverage} &
  \multicolumn{1}{c}{Span} &
  \multicolumn{1}{c}{Texp} &
  \multicolumn{1}{c}{\#images} & 
  \multicolumn{1}{c|}{$\hat{\sigma}_{\rm LC}$\tablefootmark{(a)}} \\
  %\multicolumn{1}{c|}{rms\tablefootmark{a}} \\

           &             &         &            & (L4/L5) & (phase)     & (hours) & (s)   &  & (mmag)  \\ \hline \hline
  GJ 3470  & 2017-dec-30 & CAFOS   & SDSSz      & L4      & 0.813-0.859 & 3.63    & 2-20  & 330 &	0.93	\\ \hline
  HAT-P-12 & 2018-feb-08 & CAFOS   & SDSSi      & L4      & 0.830-0.864 & 2.64    & 20-180& 128 &	0.41	\\ \hline
  HAT-P-20 & 2017-apr-03 & CAFOS   & SDSSz      & L5      & 0.133-0.187 & 3.72    & 5-20  & 298 &	0.97	\\ 
           & 2017-nov-16 & CAFOS   & SDSSi      & L5      & 0.138-0.206 & 4.74    & 4-8   & 399 &	0.42	\\
           & 2017-dec-09 & CAFOS   & SDSSi      & L5      & 0.175-0.211 & 2.50    & 5-9   & 210 &	0.38	\\ \hline
  HAT-P-23 & 2017-jul-13 & CAFOS   & SDSSz      & L4      & 0.833-0.927 & 2.73    & 10-28 & <67 &	0.44	\\
           & 2017-jul-24 & CAFOS   & SDSSz      & L4      & 0.761-0.914 & 4.45    & 7-16  & 351 &	0.49	\\
           & 2017-nov-09 & WiFSIP  & rp         & L4      & 0.766-0.869 & 3.0     & 60    & 105 &	0.31	\\ \hline
  HAT-P-36 & 2017-02-23  & CAFOS   & SDSSi      & L5      & 0.064-0.256 & 5.00    & 10-38 & 231 &	0.50	\\ 
   		   & 2018-05-15  & WiFSIP  & rp      	& L5      & 0.123-0.217 & 2.98    & 60    & 102 &	0.41	\\ 
   		   & 2018-05-20  & WiFSIP  & rp      	& L5      & 0.064-0.256 & 2.99    & 10-38 & 104 &	0.40	\\ \hline
  WASP-2   & 2017-jul-20 & CAFOS   & SDSSi      & L4      & 0.806-0.869 & 3.21    & 7-12  & 236 &	0.28	\\ \hline
  WASP-36  & 2017-dec-09 & WiFSIP  & rp         & L5      & 0.159-0.228 & 2.54    & 60    & 88 &	0.76	\\ 
           & 2018-jan-01 & CAFOS   & SDSSz      & L5      & 0.128-0.209 & 2.99    & 25-60 & 101 &	0.46	\\ \hline
  WASP-5   & 2016-oct-26 & FORS2   & z\_SPECIAL & L4      & 0.792-0.885 & 3.64    & 12    & 174 &	0.18	\\ \hline
  WASP-77  & 2017-sep-26 & CAFOS   & SDSSi      & L5      & 0.090-0.241 & 4.93    & 4-12  & 376 &	0.80	\\
           & 2017-oct-11 & CAFOS   & SDSSi      & L5      & 0.141-0.224 & 2.70    & 5-13  & 228 &	0.40	\\
           & 2017-dec-14 & WiFSIP  & rp         & L5      & 0.149-0.240 & 2.99    & 10    & 145 &	0.26	\\ 
           & 2017-dec-25 & WiFSIP  & rp         & L5      & 0.130-0.221 & 3.0     & 10    & 148 &	0.24	\\ %\hline
\hline
\end{tabular}
\tablefoot{
\tablefoottext{a}{Mean uncertainty per 15-min bin.}
}
\end{table*}

\begin{table*}
\small
\setlength{\extrarowheight}{5pt}
\caption{Constraints for the transit time (phase in the case of the combined transits) and depth from the individual fitting to each 
epoch assuming no transit is found and a linear baseline model.}
\begin{tabular}{|l c c c c c|}
\hline
  \multicolumn{1}{|l}{System} &
  \multicolumn{1}{c}{Date} &
  \multicolumn{1}{c}{Inst.} &
  \multicolumn{1}{c}{JD\_in} &
  \multicolumn{1}{c}{JD\_end} &
  \multicolumn{1}{c|}{$R_t^{\rm max}$} \\

   &  &  &  \multicolumn{2}{c}{BJD-2457000 (days)}  & $R_{\oplus}$\\ \hline \hline
  GJ 3470   & 2017-dec-30 	& CAFOS 	& 8118.5564 & 8118.6678 & 3.73 \\ \hline
  HAT-P-12  & 2018-feb-08 	& CAFOS 	& 8158.6747 & 8158.7361 & 4.19 \\ \hline
  HAT-P-20  & 2017-apr-03 	& CAFOS 	& 7847.3845 & 7847.5008 & 8.28\\ 
   			& 2017-nov-16 	& CAFOS 	& 8074.5465 & 8074.7055 & 2.94\\ 
   			& 2017-dec-09 	& CAFOS 	& 8097.6551 & 8097.7208 & 6.38\\
   			& Combined 	  	& - 		& 0.140 	& 0.200 	&  2.02  \\ \hline
  HAT-P-23 	& 2017-jul-13 	& CAFOS 	& 7948.5736	& 7948.6464	& 5.78\\
   			& 2017-jul-24 	& CAFOS 	& 7959.4030	& 7959.5472	& 6.20\\
   			& 2017-nov-09 	& WiFSIP	& 8067.3540 & 8067.4378 & 5.13\\
   			& Combined 		& - 		& 0.775 	& 0.925 	& 1.39 \\ \hline
  HAT-P-36 	& 2018-feb-23 	& CAFOS 	& 8173.5360 & 8173.6985 & 5.97 \\ 
   		   	& 2018-may-15 	& WiFSIP 	& 8254.5726 & 8254.6509 & 9.64  \\ 
   		   	& 2018-may-20 	& WiFSIP 	& 8258.5943 & 8258.6731 & 13.3  \\ 
   			& Combined 		& - 		& 0.154  	& 0.179   	& 2.04  \\ \hline
  WASP-2   	& 2017-jul-20 	& CAFOS 	& 7955.5091 & 7955.6060 & 4.32 \\ \hline
  WASP-36 	& 2017-dec-09 	& WiFSIP 	& 8097.5313 & 8097.5992 & 8.02 \\ 
   			& 2018-mar-22 	& WiFSIP 	& 8200.4865 & 8200.5731 & 5.44 \\ 
   			& Combined 		& - 		& 0.158  	& 0.175   	& 2.01  \\ \hline
  WASP-5 	& 2016-oct-26 	& FORS2   	& 7687.5323 & 7687.6344 & 5.15 \\ \hline
  WASP-77A 	& 2017-sep-26 	& CAFOS  	& 8023.5241 & 8023.6846 & 7.81 \\ 
   			& 2017-oct-11 	& CAFOS  	& 8038.5539 & 8038.6216 & 4.48 \\
   			& 2017-dec-14 	& WiFSIP 	& 8102.4852 & 8102.5647 & 7.06 \\
   			& 2017-dec-25 	& WiFSIP 	& 8113.3395 & 8113.4193 & 4.27 \\
   			& Combined 		& - 		& 0.100  	& 0.230  	&  1.39 \\ %\hline
\hline\end{tabular}
\end{table*}

\end{document}